\documentclass[twocolumn,twocolappendix]{aastex63}

\usepackage[figuresright]{rotating}
\usepackage{amsmath}
\usepackage{amsfonts}
\usepackage{graphicx}
\usepackage{rotating}
\usepackage{natbib}
\usepackage{txfonts}
\usepackage{color}
\usepackage{wrapfig}
\usepackage{amssymb}
\usepackage{color}
\usepackage{xcolor}
\usepackage{ulem}
\usepackage{lineno}

\shorttitle{Low mass star-forming region and COMs}
\shortauthors{Bhat et al.}

\begin{document}
\title{Chemical evolution of some selected complex organic molecules in low-mass star-forming regions}

 \author[0000-0002-5224-3026]{Bratati Bhat}
 \affiliation{Institute of Astronomy Space and Earth Science, 1 Bidhan Sishu Sarani, Kolkata 700054, India}
 \author{Rumela Kar}
 \affiliation{Department of Metallurgical Engineering and Material Sciences, IIT Bombay, Powai, Mumbai, Maharashtra - 400076}
\author[0000-0002-7657-1243]{Suman Kumar Mondal}
  \affiliation{S. N. Bose National Center for Basic Sciences, JD-Block, Salt Lake, Kolkata, 700098, India}
\author[0000-0003-1745-9718]{Rana Ghosh}
 \affiliation{Indian Centre for Space Physics, 43 Chalantika, Garia Station Road, Kolkata 700084, India}
 \author[0000-0003-1602-6849]{Prasanta Gorai}
\affiliation{Department of Space, Earth \& Environment, Chalmers University of Technology, SE-412 96 Gothenburg, Sweden}
\author[0000-0002-0095-3624]{Takashi Shimonishi}
\affiliation{Center for Transdisciplinary Research,
Niigata University, Nishi-ku, Niigata 950-2181, Japan}
\affiliation{Environmental Science Program, Department of Science, Faculty
of Science, Niigata University, Nishi-ku, Niigata 950-2181, Japan}
\author{Kei E. I. Tanaka}
\affiliation{Department of Earth and Planetary Sciences, Tokyo Institute of Technology, Meguro, Tokyo, 152-8551, Japan}
\author{Kenji Furuya}
\affiliation{Center for Computational Sciences, University of Tsukuba, Tsukuba, 305-8577, Japan \& National Astronomical Observatory of Japan, Tokyo 181-8588, Japan}
 \author[0000-0003-4615-602X]{Ankan Das}
 \email{ankan.das@gmail.com}
 \affiliation{Institute of Astronomy Space and Earth Science, 1 Bidhan Sishu Sarani, Kolkata 700054, India}

\begin{abstract}
 The destiny of complex organic molecules (COMs) in star-forming regions is interlinked with various evolutionary phases. Therefore, identifying these species in diversified environments of identical star-forming regions would help to comprehend their physical and chemical heritage. We identified multiple COMs utilizing the Large Program ‘Astrochemical Surveys At IRAM’ (ASAI) data, dedicated to chemical surveys in Sun-like star-forming regions with the IRAM 30 m telescope. It was an unbiased survey in the millimeter regime, covering the prestellar core, protostar, outflow region, and protoplanetary disk phase. Here, we have reported some transitions of seven COMs, namely, methanol (CH$_3$OH), acetaldehyde (CH$_3$CHO), methyl formate (CH$_3$OCHO), ethanol (C$_2$H$_5$OH), propynal (HCCCHO), dimethyl ether (CH$_3$OCH$_3$), and methyl cyanide (CH$_3$CN) in some sources L1544, B1-b, IRAS4A,  and SVS13A. We found a trend among these species from the derived abundances using the rotational diagram method and MCMC fit. We have found that the abundances of all of the COMs, except for HCCCHO, increase from the L1544 (prestellar core) and peaks at IRAS16293-2422 (class 0 phase). It is noticed that the abundance of these molecules correlate with the luminosity of the sources. The obtained trend is also visible from the previous interferometric observations and considering the beam dilution effect.
\end{abstract}


\section{Introduction} \label{sec:intro}
One of the most compelling topics of contemporary Astrophysics is understanding matter's chemical origins and evolution during the formation of stars \citep{case12}. Interstellar matter consists of molecules and dust produced during a stellar cycle. These molecules play a dominant role in gas cooling and initiate gravitational collapse, giving birth to many stars. It is thus vital to understand and characterize the evolutionary anatomies of our solar system. This study is particularly inquisitive about forming low-mass stars concerning different evolutionary stages. It begins with the fragmentation of a molecular cloud into several gravitationally bound cores supported against gravity by thermal, magnetic, and turbulent pressures. The prestellar core has a temperature of $\leqslant$30K \citep{garr06,garr09}. Nowadays, it is understood that many of the complex molecules or ancestors could have been formed by the active grain catalysis process\citep{hase92,das08a,das10,das11,das16,sriv22,sil21,bhat22,ghos22}. Unfortunately, due to the low temperature of this region, it is buried under interstellar ice, which makes it challenging to observe. However, due to the refinement of existing observation facilities, many COMs were identified in prestellar cores L1544, L183, L1512, L1498, etc. \citep{latt20,case17,vast18}. 
Once the prestellar cores become unstable and gravitational collapse ensues, the gravitational energy freely radiates away, and the collapsing fragment stays isothermal. A robust central concentration of matter is formed initially due to this isothermal collapse. Hence, an opaque, hydrostatic protostellar core in the center of a thick envelope remains wherefrom it accretes matter. In the beginning, the central object is obscured due to the thickness of the envelope. The cold outer regions of the envelope govern the spectral energy distribution (SED). This phase is represented by the Class 0 phase of the star formation \citep{andr93}. While the major parts of the envelope are glacial, there seems to be a depletion of the heavy element-bearing molecules frozen into grain mantles, precisely as in prestellar cores. But, the presence of the central core is fuelled by gravitational energy, which causes heating of the inner-most region of the envelope. It causes the evaporation of grain mantles formed during prestellar cores, and the molecules trapped in the ice are introduced to the gas
phase, where they may undergo further reactions. Expanding over 100 au, they would form the hot corino. Hot corinos share some similarities with the hot cores, but they would not be considered scaled versions of each other and differs chemically \citep{bott07}. Hot cores are rich in complex organic molecules. Many COMs are detected in high-mass star-forming regions \citep{gor20a,gor20b,mond21,mond23}. The most notable hot Corinos are IRAS 16293-2422, IRAS4A IRAS4B, IRAS2A, HH212, L483, and B335 \citep{sahu18,jorg04,jabe14,sant15,jaco19}. A Class 0 core starts to evolve into class I after the accretion of more than half of its envelope onto the central core. After a million years, with the onset of thermonuclear fusion in this core, the mighty stellar wind is produced, which restricts the in-fall of new masses. Obtained evidence suggests that the chemical composition of minor components of our solar system (comets, asteroids, etc.) is partially obtained from the early stages of solar-type protostar formation \citep{bock00}. Class I sources would act as a bridge between class 0 and the prominent disk present in class II and III phases of star formation. Comparatively more complex molecules (HCOOCH$_3$, $\rm{CH_3OCH_3}$, CH$_3$CHO, CH$_3$OH, HCOOH, $\rm{CH_3CH_2OH}$, NH$_2$CHO, CH$_3$CN, etc.) were observed in the hot corino phase than the protoplanetary disk. Radio astronomy and far-infrared observatories play a spectacular role in modern-day Astrophysics. Radio telescopes that operate in the millimeter wavebands probe the cold universe around, making it possible for detailed observations of systems at different evolutionary stages, which ultimately sheds light on the most decisive chemical processes determining evolution. One of the most important examination tools to comprehensively study the evolution of star-forming regions is the systematic spectral line surveys, especially in the millimeter range. 
Recent remarkable progress in observational facilities in radio and far-infrared regimes opens a new molecular detection era in star-forming regions. Dedicated line surveys of different astronomical sources at diverse evolutionary stages of star formation will be beneficial in understanding the chemical evolution along the star formation stages. Here, we have analyzed the large survey of IRAM 30 m data for L1544, B1-b, IRAS4A, SVS13A to understand the chemical evolution through the different stages of low-mass star-forming regions. 
 
This paper is organized as follows. In section \ref{sec:observation}, we discuss the observational detail. 
Results and discussions are presented in section \ref{sec:results}, and finally, in section \ref{sec:conclusion}, we conclude.

\section{Observations}\label{sec:observation}
We used the archival data of the Large Program of Astrochemical Surveys At IRAM (ASAI, PI: Bertrand Lefloch and Rafael Bachiller). This systematic line survey was dedicated to understanding solar-type protostar's chemical and dynamical evolution. The observation was carried out from September 2012 to March 2015 using the EMIR receivers of the IRAM 30 m. This work considers four different sources L1544, B1-b, IRAS4A, and SVS13A (belonging to four distinct stages from the prestellar phase to class I). The survey contained ten template sources at different stages of evolution. The 3 mm (80-116 GHz), 2 mm (130-170 GHz), and 1.3 mm (200-276 GHz) bands were covered by this line survey. 
While all these bands were covered for B1b, IRAS4A, and SVS3A, in the case of L1544, only the 3 mm (80-116 GHz) band was covered. The observational details were already discussed elaborately in \cite{lefl18}.
The beam size or half power beam width (HPBW) of the IRAM 30 m telescope would be determined by HPBW($\arcsec$) = $2460$/frequency(GHz) relation. Here, the antenna temperature ($T_A^*$), was converted to the main beam temperature ($T_{MB}$), by $T_{MB}$ = $T_A^*$/$\eta_{MB}$,  where $\eta_{MB}$ is the antenna efficiency. All the intensities in tables and figures are shown in terms of the main beam temperature. 
The details of the sources considered here are discussed in the section below. The targetted positions and required information about these sources are summarized in Table \ref{table:source}.

\begin{table*}
\centering
\caption{Targeted positions and relevant information of the sample sources. \label{table:source}}
\begin{tabular}
{|c|c|c|c|c|c|}
\hline
Source name&Stage of the source&Coordinates&Distance\footnote{See running text of Section \ref{sec:observation}}&Luminosity\footnote{See running text of Section \ref{sec:observation}}&{$V_{LSR}$}\\
&&(J2000)&(pc)&(L$_\odot$)&(km s$^{-1}$)\\
\hline
L1544&Evolved prestellar core&$05^{h}04^{m}17.^{s}21$+$25^{\circ}10^{'}42^{\arcsec}.8$&140 
 (171.7$^{*}$)&1.0&7.2 \citep{jime16}\\
B1-b&First hydrostatic core (FHSC)&$03^{h}33^{m}20.^{s}80$+$31^{\circ}07^{'}34^{\arcsec}.0$&230&0.77&6.5 \citep{lope15}\\
IRAS4A&Class 0&$03^{h}29^{m}10.^{s}42$+$31^{\circ}13^{'}32^{\arcsec}.2$&260 (293$^{*}$)&9.1&7.2 \citep{geri09}\\
SVS13A&Class I&$03^{h}29^{m}03.^{s}73$+$31^{\circ}16^{'}03^{\arcsec}.8$&260 (300$^{*}$)&34.0&8.6 \citep{chen09}\\
\hline
\end{tabular}\\
{$^{*}$Updated distance.}
\end{table*}

\subsection{Sources considered}
\subsubsection{L1544}
In the Taurus molecular cloud (TMC), L1544 is a dense starless core that is in an early phase of star formation before collapsing \citep{tafa98,ciol00}. This ideal proto-type evolved prestellar core is situated at 140 pc distance from the sun \citep{cern87}. Recenty \cite{gall19} using GAIA data obtained a distance of 171.7 pc for L1544. The central part of the prestellar core (up to a few thousand au scale) was unexplored until \cite{case19,case22} used the ALMA high angular resolution band 6 continuum emission data. They named the compact region of 0.1 \(\textup{M}_\odot\)  (10$\arcsec$ radius $\sim$ 1400 au) as the `kernel'. Besides the possible local density enhancement, this kernel has an average number density of  $\sim 10^6$ cm$^{-3}$. Though this kernel is fragmented, non-ideal magnetohydrodynamic simulations and synthetic interferometric observations suggest a smooth kernel with a peak number density of $10^7$ cm$^{-3}$. In the prestellar core phase, matters accumulated in the center of the cloud yield a drop in temperature and growth in density. As a result, the atoms and molecules in the gas phase freeze on the dust and form icy grain mantles. A central density of $\sim$ 10$^6$ cm$^{-3}$ is reached along with a very low temperature $\sim$ 7 K. Due to heavy depletion, a high degree of deuterium fractionation of $\rm{N_2H^+}$ compared to HCO$^+$ is observed \citep{caseb02,rade19}. Many COMs are detected in L1544 \citep{vast14}, ranging from numerous sulfur-related molecules to pre-biotic molecules \citep{vast18,lope15}. Detailed modeling of L1544 by \cite{keto10b} found that they needed to consider a high dust opacity to reproduce the measured temperature drop inside 2000 au. \cite{case99} indicated that it could mean ﬂuﬀy grains in the core center, where CO is heavily frozen, and volume densities become greater than 10$^6$ cm$^{-3}$. Numerous observations well constrain the physical structure of this source. The high volume densities and centrally concentrated form make it a unique target to study possible variations in the opacity.

\subsubsection{B1-b}
Barnard 1 (B1) belongs to the Perseus molecular cloud complex situated at a distance of 230 pc \citep{fuen17}. It is an important source from the chemical and kinematic point of view because it contains several active sources which are in different stages of star formation. For example, B1-a and B1-c are in the class 0 stage, associated with outflows. B1-b is the main core divided into three parts: two young stellar objects, B1b-N and B1b-S, and one more evolved entity, B1b-W \citep{huan13,fuen16}. B1b-N and B1b-S are separated from each other by approximately 18$^{''}$, which are in the first hydrostatic core (FHSC) stages. The FHSC is categorized based on the spectral energy distribution and outflow present in it \citep{geri15,pezz12}. The most investigated part is B1-b because of its rich molecular spectrum.
The total luminosity of the source B1-b is 0.77 L$_\odot$ \citep{lefl18}.  
Numerous species like CH$_3$O \citep{cern12}, $\rm{NH_3D^+}$ \citep{cern13}, HCNO \citep{marc09} were identified in this source. The detection of D$_2$CS, ND$_2$H, and ND$_3$ confirms a high deuteration observed towards this source. Recent observation using ALMA interferometer has identified various complex organic molecules like CH$_3$OCOH, CH$_3$CHO, and many COMs ($\rm{NH_2CN, NH_2CHO, CH_3CH_2OH, CH_2OHCHO,}$ ${\rm CH_3CH_2OCOH}$) is tentatively detected towards B1-b \citep{marc18b}. 

\subsubsection{NGC 1333-IRAS4A}
IRAS4A is a proto-binary system located in the NGC1333 reﬂection nebula in the Perseus cloud. 
The recent result from GAIA found that it is situated at a distance of $\sim$293$\pm$22 pc (\cite{orti18}; \cite{zuck18}).
The total mass of the gas envelope of IRAS4A is 3.5 M$_\odot$, and the total luminosity is 9.1 L$_\odot$ \citep{lefl18}. Another component, IRAS4B, is separated by an angular distance of 31$^{''}$ from IRAS4A \citep{marv08}. Till date many COMs like CH$_3$OH, $\rm{CH_3OCH_3, C_2H_5CN, CH_2OHCHO}$ are detected towards IRAS4A. It is noticed that a very high collimated outflow is associated with IRAS4A from CO, CS, and SiO emission \citep{blak95,lefl98}.  The infall motion was detected in this source by \cite{fran01} and \cite{choi99} with an estimated accretion rate of $1.1\times10^{-4}$ M$_\odot$ per year, an inner mass of 0.7 M$_\odot$ and age of $\sim$ 6500 yr (see also \cite{mare03}). Previous observations with the IRAM 30 m classify IRAS4A as a hot corino protostar. Recently, a high-resolution interferometric data reveals that IRAS4A consists of two components \citep{choi10,choi11}: IRAS4A1, and IRAS4A2, separated by an angular distance of 1.8$^{''}$ ($\sim$527 au; \citealt{des20a}) from one another. A striking chemical difference is observed between IRAS4A1 and IRAS4A2. \cite{sant15} and \cite{lope17} conﬁrmed that IRAS4A2 is a hot corino protostar, but more analysis is needed to confirm it for IRAS4A1. Recently \cite{des20a} concluded that IRAS4A1 is a hot corino, but the lack of iCOMS detected towards IRAS4A1 is due to large dust optical depth towards the center. 
\cite{sahu19} depicted two possible scenarios of IRAS4A1: a) The observed absorption features are probably arising from a hot-corino-like atmosphere against a very compact ($\leq$ 36 au) disk in A1 b) the absorption may arise from different layers of a temperature-stratified dense envelope in A1. 



\subsubsection{SVS13A}
SVS13A is a relatively evolved protostar already in the class I phase. The luminosity of the source is 34 L$_\odot$ and located at a distance of $\sim$ 260 pc \citep{schl14}. Recently using GAIA data the distance of SVS13A is 300 pc \citep{diaz22}. It is a part of the NGC 1333-SVS13 system. It consists of three sources, A, B, and C. The angular separation of SVS13A from its two counterparts, SVS13B and SVS13C are 15$^{''}$ and 20$^{''}$, respectively \citep{bach98,loon07}. SVS13A has an extended outflow associated with it, also attached with well-known Herbig-Haro chain (HH) objects 7-11 \citep{reip93}. The VLA observation revealed SVS13A as a closed binary system, VLA4A and VLA4B, separated by 0.3$^{''}$ ($\sim$70 au; \citealt{angl00}). A recent observation by  \cite{diaz22} reported a distance of $\sim$ 300 pc for SVS13A and separation of VLA4A and VLA4B of $\sim$ 90 au. Many COMs, such as acetaldehyde (CH$_3$CHO), methyl formate (HCOOCH$_3$), dimethyl ether (CH$_3$OCH$_3$), ethanol (CH$_3$CH$_2$OH), and formamide (NH$_2$CHO) were already identified from the ASAI data, in this source \citep{bian19}.

\subsection{Line identification}\label{sec:data-analysis}
The line identification is carried out using CASSIS\footnote{\url{http://cassis.irap.omp.eu/?page=cassis}} software. For the spectroscopic details, we have used Cologne Database for
Molecular Spectroscopy (CDMS\footnote{\url{https://cdms.astro.uni-koeln.de/}}; \citealt{mull01,mull05}), and Jet Propulsion Laboratory (JPL\footnote{\url{https://spec.jpl.nasa.gov/}}; \citealt{pick98}) databases. 
We consider only the unblended lines with a signal-to-noise ratio (SNR) greater than the 3$\sigma$ limit. The observed transitions of various complex molecules together with their quantum numbers, upper state energies (E$_{up}$), $V_{LSR}$, line parameters such as line width (FWHM), and the integrated intensity ($\int$T$_{mb}$dV) are noted in Table. \ref{tab:observation_1}, \ref{tab:observation_2}, \ref{tab:observation_3}, \ref{tab:observation_4}. The line parameters are obtained using a single Gaussian fit to the observed spectral profile of each unblended transition. The identified lines are plotted in black color in Fig. \ref{fig:ch3oh_l1544mcmc}, \ref{fig:ch3oh_barnardmcmc}, \ref{fig:ch3oh_irasmcmc}, \ref{fig:ch3oh_svsmcmc}, \ref{fig:ch3cho_mcmc}, \ref{fig:ch3ocho_mcmc}, \ref{fig:c2h5oh_mcmc}, \ref{fig:hcccho_mcmc}, \ref{fig:ch3och3_mcmc}, \ref{fig:ch3cn_mcmc1}, \ref{fig:ch3cn_mcmc2}.
\\
\\

\subsection{H$_2$ column density} \label{sec:H2_col}

Due to the lack of the continuum observation data, we used the H$_{2}$ column density from the literature to derive the abundances of species. 
For the L1544, we used a H$_{2}$ column density of $8.9 \times 10^{22}$ cm$^{-2}$ (calculated by \citealt{hily22}) for a beam size of $26\arcsec$ in L1544. 
\cite{dani13} obtained an average H$_2$ column density (1.2 mm observation with IRAM) of $7.6 \times 10^{22}$ cm$^{-2}$  within the 30$\arcsec$ beam in B1. \cite{john10} estimated a H$_{2}$ column density $\sim 8.2 \times 10^{22}$ cm$^{-2}$ for the same beam. Following \cite{lope15}, we used an average H$_2$ column density $\sim 7.9 \times 10^{22}$ cm$^{-2}$ for B1-b in estimating the abundances. For IRAS 4A, \cite{mare02} obtained a H$_2$ column density of $2.9 \times 10^{22}$ cm$^{-2}$ for a 30$\arcsec$ beam, whereas with a 0.5 $\arcsec$ beam it was $2.5 \times 10^{24}$ cm$^{-2}$. Since we analyzed the data obtained from the IRAM 30 m telescope, we use $2.9 \times 10^{22}$ cm$^{-2}$ for abundance estimation. For SVS13A, we used the H$_2$ column density of $10^{23}$ cm$^{-2}$ estimated by \cite{lefl98} for 20$\arcsec$ beam. Due to this uncertainty in the H$_2$ column density, the derived abundances and the chemical trend have an uncertainty also. To minimize this effect, we used the values of H$_2$ column density in different sources for $\sim$ 30$\arcsec$ beam. The H$_2$ column density used in deriving the abundances is summarized in Table \ref{tab:rotdiag}.

\clearpage
\begin{table*}
\centering
{\scriptsize
 \caption{Comparison of column densities obtained using rotational diagram method and MCMC fitting with the values obtained from the literature for COMs observed in different sources. \label{table:comparison}}
\begin{tabular}{|c|c|c|c|c|c|}
  \hline
  \hline
  Species&Column Density&L1544&B1-b&IRAS4A&SVS13A\\
  \hline
  \hline  &Previous&$ (2.70\pm0.60)\times10^{13a}$&$(2.50\pm1.30)\times10^{14b}$&$(5.10\pm1.00)\times10^{14c}$&$(1.00\pm0.20)\times10^{17d}$\\
  CH$_3$OH&Rotational Diagram&$(7.14^{+0.70}_{-0.60})\times10^{12}$&$(9.73^{+0.05}_{-0.04})\times10^{13}$&$(1.42^{+0.03}_{-0.02})\times10^{14}$&$(1.21^{+0.03}_{-0.02})\times10^{14}$\\
  &&&&$(1.41^{+0.04}_{-0.04})\times10^{14}$&\\
&MCMC&$(4.60\pm0.92)\times10^{12}$&$(1.13\pm0.16)\times10^{14}$&$(1.80\pm0.29)\times10^{14}$(C1)&$(2.43\pm1.07)\times10^{14}$\\
  &&&&$(2.00\pm0.28)\times10^{14}$(C2)&\\
 \hline
&Previous&$ 1.20\times10^{12h}$&$1.50\times10^{12f}$&$2.60\times10^{12g}$&$(1.20\pm0.70)\times10^{16d}$\\
  CH$_3$CHO&Rotational Diagram&$(1.25^{+0.40}_{-0.30})\times10^{12}$&$(3.48^{+0.5}_{-0.50})\times10^{12}$&$(9.47^{+0.6}_{-0.50})\times10^{12}$&$(5.89^{+1.6}_{-1.30})\times10^{12}$\\
  &&&&$(9.53^{+1.00}_{-0.90})\times10^{12}$&\\
&MCMC&$(6.10\pm3.55)\times10^{11}$&$(4.40\pm0.56)\times10^{12}$&$(1.30\pm0.21)\times10^{13}$(C1)&$(7.20\pm4.89)\times10^{12}$\\
  &&&&$(1.10\pm0.21)\times10^{13}$(C2)&\\
\hline
&Previous&$(4.40\pm4.00)\times10^{12h}$&$3.00\times10^{12f}$(A+E)&$(5.50\pm2.70)\times10^{16i}$(A)&$(1.30\pm0.10)\times10^{17d}$\\
&&&&$(5.80\pm1.10)\times10^{16i}$(E)&\\
 CH$_3$OCHO&Rotational Diagram&$3.70\times10^{12}x$&$(6.41^{+2.30}_{-5.00})\times10^{12}$&$(3.49^{+0.50}_{-0.40})\times10^{13}$&$(7.78^{+1.30}_{-1.00})\times10^{13}$\\
  &MCMC&-&$(1.20\pm1.57)\times10^{13}$&$(3.65\pm1.63)\times10^{13}$&$(7.10\pm3.52)\times10^{13}$\\
 \hline
&Previous&&$\leqslant4.80\times10^{12j}$&$(4.40\pm1.40)\times10^{16k}$&$(1.10\pm0.50)\times10^{17d}$\\
  C$_2$H$_5$OH&Rotational Diagram&-&$1.00\times10^{13x}$&$(1.69^{+0.30}_{-0.20})\times10^{13}$&$(1.37^{+1.80}_{-1.60})\times10^{13}$\\
 &MCMC&-&-&$(2.10\pm1.10)\times10^{13}$&$(1.60\pm1.22)\times10^{13}$\\
 \hline
 &Previous&$1.80-6.30\times10^{11h}$&$5.50\times10^{11f}$&-&$\leqslant1.00\times10^{16d}$\\
  HCCCHO&Rotational Diagram&$(3.16^{+0.80}_{-0.70})\times10^{12}$&$2.56\times10^{12x}$&$3.90\times10^{12x}$&$4.00\times10^{14x}$\\
 &MCMC&$(4.00\pm1.33)\times10^{11}$&-&-&-\\
 \hline
&Previous&$ (1.50\pm0.20)\times10^{12h}$&$3.00\times10^{12f}$&$\leqslant4.50\times10^{16i}$&$(1.40\pm0.40)\times10^{17d}$\\
 CH$_3$OCH$_3$&Rotational Diagram&$1.72\times10^{12y}$&$6.12\times10^{12y}$&$1.20^{+0.40}_{-0.30}\times10^{13}$&$1.41\times10^{13y}$\\ &MCMC&$(2.20\pm1.61)\times10^{12}$&$(8.50\pm5.52)\times10^{12}$&$(2.10\pm1.03)\times10^{13}$&\\
\hline &Previous&$(6.1\pm1.8)\times10^{11l}$&$(3.2^{+1.6}_{-1.4})\times10^{14v}$(B1-b S)&$(6.5\pm2.9)\times10^{15m}$&$2.0\times10^{16n}$\\
 &&&$< 8.5\times10^{13v}$(B1-b N)&&\\
  CH$_3$CN&Rotational Diagram&$4.85\times10^{11y}$&$4.95\times10^{11y}$&$(1.78^{+0.17}_{-0.15})\times10^{12}$&$(4.57^{+0.23}_{-0.22})\times10^{12}$\\
 &MCMC&-&-&$(1.34\pm0.35)\times10^{12}$&$(2.8\pm0.57)\times10^{12}$\\
  \hline
  \hline
 \end{tabular}}\\
 {\scriptsize \noindent $^x$Upper limits, $^y$Lte fitting,$^a$\cite{bizz14}, $^b$\cite{ober09}, $^c$\cite{mare05}, $^d$\cite{bian19} (Interferometric observation), 
 $^f$\cite{cern12}, $^g$\cite{hold19}, $^h$\cite{jime16}, $^i$\cite{bott04} (Interferometric observation), $^j$\cite{ober10}, $^k$\cite{taqu15} (Interferometric observation), $^l$\cite{nagy19}, $^m$\cite{taqu15} (Interferometric observation), $^n$\cite{bian22b} (Interferometric observation), $^v$\cite{yang21} (Interferometric observation)}
 \end{table*}

\begin{figure*}
\begin{minipage}{0.49\textwidth}
\includegraphics[width=\textwidth]{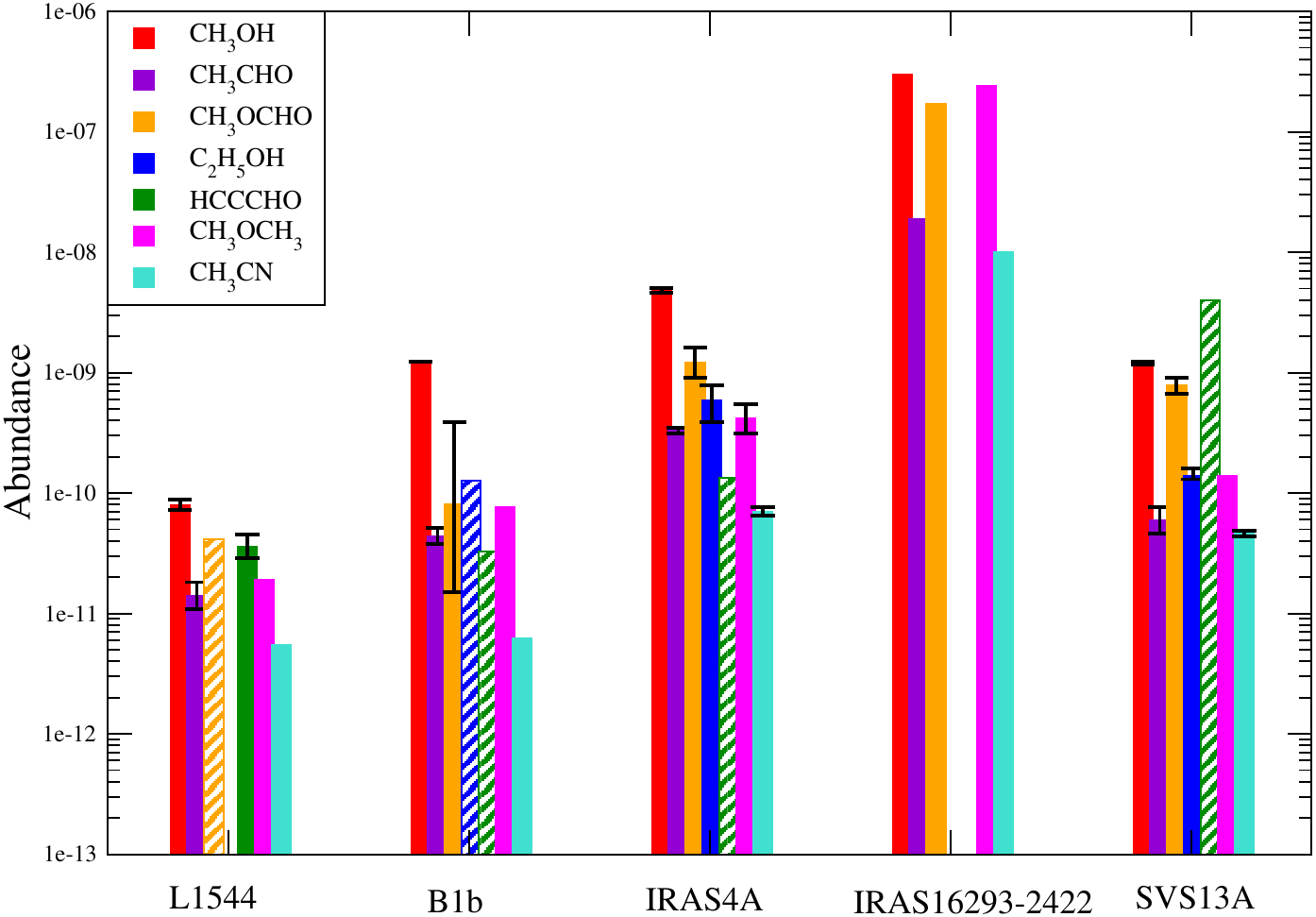}
\end{minipage}
\begin{minipage}{0.49\textwidth}
\includegraphics[width=\textwidth]{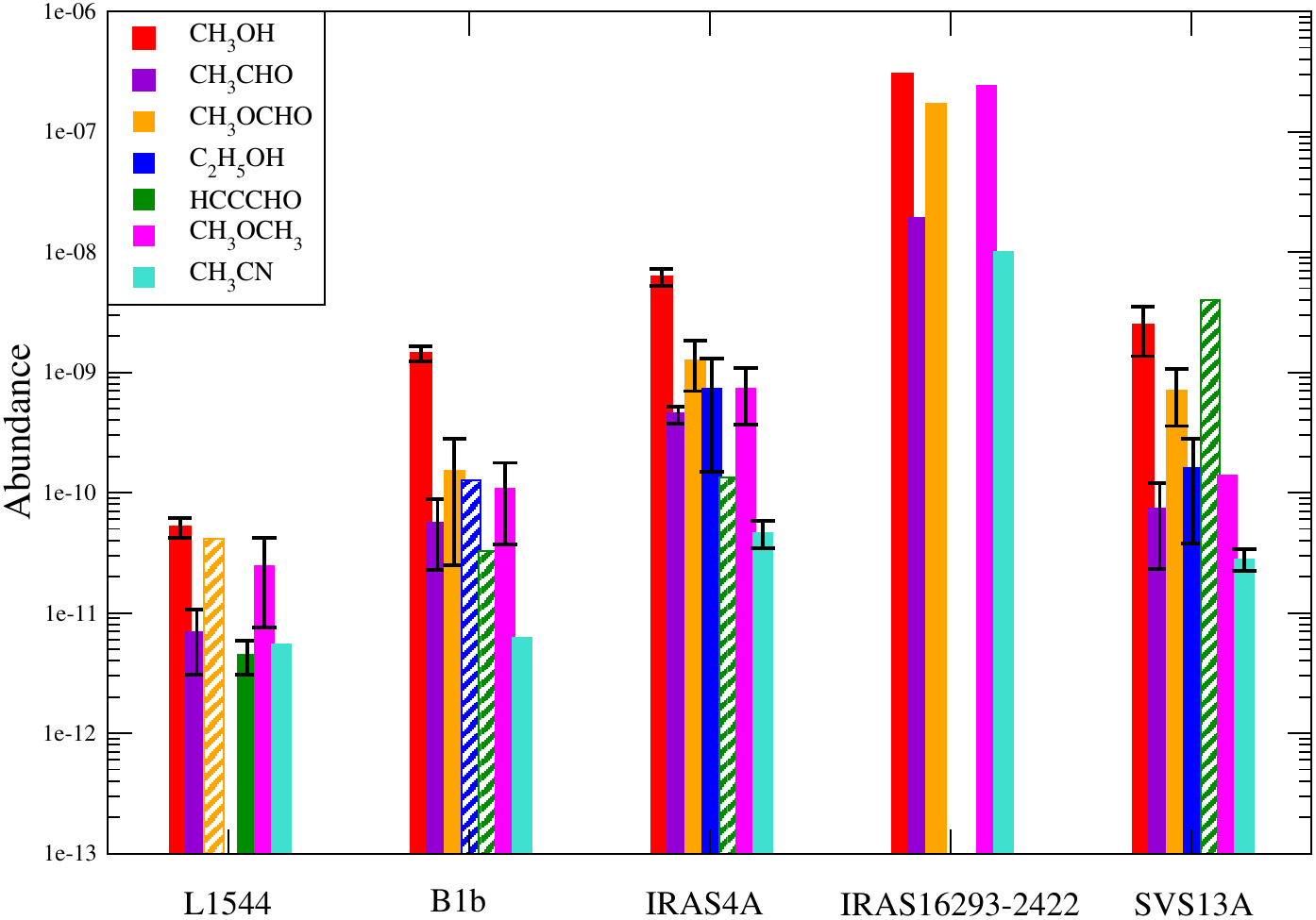}
\end{minipage}
\caption{The evolution of abundances in different stages of low-mass star-forming regions is shown. The left panel shows the abundances obtained from the rotational diagram method (dashed parts are the upper limits), and the right panel depicts the same obtained with the MCMC fit. Black vertical lines represent the error bars. For CH$_3$OCH$_3$ in L1544 and B1b, the lines have the same upstate energy, so a rotation diagram cannot be performed, and the column density is calculated using simple LTE fitting. For CH$_3$OCH$_3$ in SVS13A, the column density is calculated using the value from \cite{bian19} scaling it for 30$\arcsec$ beam. A new class 0 source, IRAS16293-2422 (22 L$_\odot$), is included where CH$_3$OH, CH$_3$CHO, CH$_3$OCHO, CH$_3$OCH$_3$ and CH$_3$CN is observed \citep{caza03}.}
\label{fig:clmdensity}
\end{figure*}

\section{Results \& Discussions} \label{sec:results}
The main goal of this work is to find the correlation between the observed COMs in various evolutionary phases of star-forming regions. Observing all the instances of a single low-mass star-forming region is impossible. Instead, it is beneficial to selectively follow some evolutionary stages of similar types of the ongoing process before the formation of stars to have an overview of the complete process. Though it is not expected that all the targeted regions would follow a similar footprint, it gives an overall idea of the general low-mass star-formation process. 
We were able to identify several transitions of interstellar COMs, i.e methanol (CH$_3$OH), acetaldehyde (CH$_3$CHO), methyl formate (CH$_3$OCHO), ethanol (C$_2$H$_5$OH), propynal (HCCCHO), dimethyl ether (CH$_3$OCH$_3$) and methyl cyanide (CH$_3$CN) in different sources.

We apply two robust LTE methods namely rotational diagram (RD) and Monte Carlo Markov Chain (MCMC) fitting, to derive the column density and excitation temperature of a species. The details description of RD analysis and MCMC method are discussed in the section \ref{sec:RD} and \ref{sec:mcmc}, respectively. The obtained column densities and temperature from rotational diagram analysis are summarized in Table \ref{tab:rotdiag}. The rotation diagram plots are shown in \ref{fig:rotational_diag_ch3oh_single}, \ref{fig:rotational_diag_ch3cho}, \ref{fig:rotational_diag_hcooch3}, \ref{fig:rotational_diag_C2H5OH}, \ref{fig:rotational_diag_HCCCHO}, \ref{fig:rotational_diag_CH3OCH3}, \ref{fig:rotational_diag_CH3CN} . The variable parameters and the ranges used for the MCMC method and the best-fitted values are summarized in Table \ref{table:mcmc_lte}. In addition, Table \ref{table:comparison} compares the abundances obtained through MCMC fitting and rotational diagram analysis with earlier findings. It depicts that our results are consistent with the earlier results. The values obtained using both methods differ from the previously obtained values within a factor of 0.17-5.02. Also, the abundances obtained from rotational diagram analysis and MCMC fitting are almost similar and differ within a minimal range, which justifies the accuracy of the results obtained from the two methods. In the left panel of Figure \ref{fig:clmdensity}, we show how the abundances obtained from the rotational diagram method varies in different sources, whereas the right panel shows similar results obtained with the MCMC fitting method. The results obtained from our analysis are discussed here with individual species. The column density values of H$_2$ in different sources are used from literature to calculate the abundances. The values are noted in Table \ref{tab:rotdiag} and in section \ref{sec:H2_col}.
\clearpage

\subsection{Observed species}
\subsubsection{CH$_3$OH}
Since methanol is a major chemical constituent of the various phases of star formation, its analysis has become of utmost importance. Previously, \cite{vast14} identified CH$_3$OH emission in L1544 from the ASAI data and obtained a column density $3 \times 10^{13}$ cm$^{-2}$ 
using non-LTE approximation. They concluded that methanol lines are likely originated from $\sim$ 8000 au, where the temperature is $\sim$ 10 K. Methanol was previously detected in B1-b using IRAM 30m telescope by \cite{ober09} with a column density of $2.5\times10^{14}$ cm$^{-2}$. In IRAS4A, \cite{mare05} obtained a column density $5.1\times10^{14}$ cm$^{-2}$. 
Here, we have identified several transitions of CH$_3$OH in all four sample sources (see Table \ref{tab:observation_1}). The obtained values of column densities and excitation temperatures are noted from rotational diagram are prsented in Table \ref{tab:rotdiag} and same for MCMC fitting, it is noted in Table \ref{table:mcmc_lte}. We have performed MCMC fitting considering a single component of the detected CH$_3$OH transitions in L1544, B1-b, and SVS13A, and we get a good fit. 
In the case of class 0 protostar IRAS4A, we found that a two-component fit is required for the rotational diagram of methanol: (1) having E$_{up}$ $>$ 50 K (hot component), (2) having E$_{up}$ $<$ 50 K (cold component). Similarly in the MCMC method, we obtain a good fit when we are considering two components. The chemical evolution of CH$_3$OH with respect to the evolutionary stages of the star formation is represented by red bars in the left (RD) and right (MCMC) panels of Fig. \ref{fig:clmdensity}. It shows that the methanol abundance gradually increases from a minimum value in L1544 to a maximum value in IRAS16293-2422. The increase in abundance might be attributed due to the gradual rise in temperature. Since a lower temperature (10-20 K) is efficient for the ice phase hydrogenation reactions, methanol formation by successive hydrogen additions to CO could be possible during the prestellar core phase. 
$$
\rm{CO \rightarrow HCO\rightarrow H_2CO \rightarrow CH_2OH/CH_3O\rightarrow CH_3OH} 
$$

However, at this stage, the thermal desorption is not efficient enough \citep[having a binding energy of 5264 K]{das18} to transfer the ice phase methanol contents to the gas phase. The non-thermal desorptions \citep{ober09,garr07}
are mainly responsible for the observed gas-phase abundance of methanol at this stage. Furthermore, the cloud evolves into a first hydrostatic core and then a protostar phase. In the protostar phase, methanol would also be formed by  radical-radical surface reaction, $\rm{CH_3+OH \rightarrow CH_3OH}$. As a result, temperature increases, eventually enhancing the chance of production and their release by thermal desorption.
We get a less abundance of methanol in SVS13A (class I object). 
A decrease in the abundance of methanol for the class I object may occur due to the lower methanol formation rate in the class I phase than in the class 0 phase because of the competition between the reaction and thermal desorption of the reactants.

 \cite{scib20} surveyed the presence of complex organic molecules in starless and prestellar cores within the Taurus Molecular Cloud by ARO 12m telescope on Kitt Peak (beam size $\sim 62.3 ^{''}$). They obtained a methanol abundance $(0.53-3.36) \times 10^{-9}$, and excitation temperature ranging from $6.79-8.66$ K for all 31 cores. From the rotational diagram analysis, we obtained a comparatively lower abundance of methanol $ \sim 8.02 \times 10^{-11}$, in L1544 and an excitation temperature of $8.6$ K. \cite{scib20} found that the less evolved regions have comparatively higher methanol abundance than more evolved regions. Since the L1544 is more evolved than the sources studied by \cite{scib20}, a comparatively lower abundance of methanol is consistent with the observed trend.
 
\subsubsection{CH$_3$CHO}
Acetaldehyde, is an asymmetric top molecule with a widespread presence in the various evolutionary phases of star formation. We have identified multiple transitions of CH$_3$CHO in all the selected sources. Fitted line parameters of this species are summarised in Table \ref{tab:observation_2}. The upper state energy of the observed lines ranges from $9$ to $121$ K. 
\cite{jime16} obtained a column density 1.2 $\times$ 10$^{12}$ cm$^{-2}$ towards the dense, highly extinguished continuum peak within the inner 2700 au and $3.2 \times 10^{12}$ cm$^{-2}$ towards a low-density shell located at 4000 au from the core center. However, we obtain a very similar column density of $1.25 \times 10^{12}$ cm$^{-2}$ for acetaldehyde in our work. In B1-b, a column density of $1.5 \times10^{12}$ cm$^{-2}$ was obtained by \cite{cern12} with the MADEX code assuming the excitation temperature of 10 K. From our rotation diagram analysis, we found the column density $3.5\times10^{12}$ cm$^{-2}$ of this species. \cite{hold19} and \cite{bian19} observed CH$_3$CHO towards IRAS4A and SVS13A and obtained a column density of $2.6\times10^{12}$ cm$^{-2}$ and $1.2\times10^{16}$ cm$^{-2}$, respectively using single-dish telescope. From our rotation diagram analysis, we yield the column densities of $9.5\times10^{12}$ cm$^{-2}$ and $5.9\times10^{12}$ cm$^{-2}$ for IRAS4A and SVS13A, respectively. Similar to the CH$_3$OH, for CH$_3$CHO, two components are observed in the rotational diagram analysis (see Figure \ref{fig:rotational_diag_ch3cho}) in IRAS4A, (1) E$_{up}$ $>$ 50 K (hot component) and (2) cold component having  E$_{up}$ $<$ 50 K. We obtain a temperature of $22.1$ K for the low excitation lines and $64.5$ K for the high excitation lines from RD analysis. Similarly, a two-component MCMC fit is performed and yields a temperature of $71.2$K and $11.1$ K for components 1 and 2, respectively. It is noticed that the results derived from  the two methods  are very similar. The violet-colored bar lines in Figure \ref{fig:clmdensity}  represent the variation of its abundance and it shows the same behavior as $\rm{CH_3OH}$ throughout the evolutionary stages of star formation. \cite{scib20} obtained acetaldehyde abundance $(0.6-3.9) \times 10^{-10}$, and excitation temperature 3.57 K. From the rotational diagram analysis, we obtained a comparatively lower abundance of acetaldehyde $\sim 1.4 \times 10^{-11}$ in L1544 and an excitation temperature of 6 K. Similar to CH$_3$OH since the L1544 is more evolved than the sources studied by \cite{scib20}, a comparatively lower abundance of acetaldehyde is consistent with the observed trend.

\subsubsection{CH$_3$OCHO}
Methyl formate is a simple asymmetric top complex organic species observed in most star-forming regions. Its isomers, acetic acid (CH$_3$COOH) and glycolaldehyde (HCOCH$_2$OH), are comparatively less abundant. Observing complex molecules in a prestellar core is always challenging because of their low temperature. \cite{jime16} identified methyl formate in L1544 with IRAM 30 m at two positions; one at the center of the core and another at the same place where the peak abundance of methanol arises. They found a column density of $(4.4\pm 4.0)\times 10^{12}$ cm$^{-2}$  and $(2.3\pm 1.4)\times 10^{12}$ cm$^{-2}$ at these two position respectively. \cite{cern12} observed this species towards B1-b, and derived a total column density of $3\times10^{12}$ cm$^{-2}$. \cite{bott04} observed CH$_3$OCHO towards the low-mass protostar IRAS4A and derived a column densities of $5.5\times10^{16}$ cm$^{-2}$ and $5.8\times10^{16}$ cm$^{-2}$ for A-CH$_3$OCHO and E-CH$_3$OCHO, respectively. \cite{bian19} identified CH$_3$OCHO transitions towards SVS13A, and obtained a column density of $1.3\times10^{17}$ cm$^{-2}$. In the prestellar core L1544, one transition at 90.22765 GHz is tentatively detected. An upper limit of column density $3.7\times10^{12}$ cm$^{-2}$ is estimated using LTE method (see Table  \ref{tab:upper limit}).
For B1-b, IRAS4A, and SVS13A, many transitions are detected, and the results from RD analysis and MCMC method are shown in Table \ref{tab:rotdiag} and \ref{table:mcmc_lte}, respectively. The RD plot for CH$_3$OCHO is shown in Figure \ref{fig:rotational_diag_hcooch3} and the MCMC fitting is shown in Figure \ref{fig:ch3ocho_mcmc}. \\
We have found that $\rm{CH_3OCHO}$ follow a similar pattern as methanol and acetaldehyde in its evolution from the prestellar core phase to the class I stage, as represented by an orange-colored bar line in Figure \ref{fig:clmdensity}.
 Unlike methanol and acetaldehyde, no ice phase hydrogenation reactions are directly involved in forming $\rm{CH_3OCHO}$. Instead, it is the radical-radical reaction between $\rm{CH_3O}$ and HCO that can form ice phase $\rm{CH_3OCHO}$:
$$
{\rm CH_3O + HCO \rightarrow CH_3OCHO}.
$$
The activities are restricted at low temperatures due to the high binding energy of these radicals i.e., 4400 K for CH$_3$O \citep{wake17} and 2206 K for HCO \citep{das18}.
However, it plays an active role in the warmer region. Moreover, the binding energy of $\rm{CH_3OCHO}$ (6295 K, \url{https://kida.astrochem-tools.org}) is comparatively higher than the methanol and acetaldehyde \citep[5264 and 4573, respectively]{das18}. 
The decline in the CH$_3$OCHO abundance at the class I stage would be attributed to the favorable sublimation rate over the reaction rate of radicals and its subsequent gas phase destruction after desorption.

\subsubsection{$\rm{C_2H_5OH}$}
Ethanol (C$_2$H$_5$OH) emission was detected for the first time in 1975 toward Sagittarius B2. It was observed with the help of one of the pioneering radio telescopes of the twentieth century, the 11 m radio telescope of the National Radio Astronomy Observatory (NRAO, \cite{zuck75}). It has been observed towards low and high mass in both the star-forming regions. Due to the orientation of the OH group, ethanol consists of two different conformers, one is anti, and another one is gauche. A gauche$^+$ and gauche $^-$ state forms when the tunnelling between the two equivalent gauche conformers lifts the degeneracy. Recently, \cite{bian19} performed ab initio quantum chemical calculations to characterize the geometry and energy of ethanol conformers with high accuracy. In this work, no clear C$_2$H$_5$OH lines are detected in L1544. In IRAS4A and SVS13A, multiple transitions of ethanol are clearly identified. Three possible transitions of C$_2$H$_5$OH are detected toward B1-b. However, the transition at 135.989923 GHz is below 3$\sigma$, and the other transition at 270.444085 GHz is blended. Only one unblended transition at 131.502781 GHz is detected. Therefore, we consider ethanol is tentatively detected in this source and derived upper limit of column density  $\sim 1.0 \times 10^{13}$ cm$^{-2}$ ( see Table \ref {tab:upper limit}). Figure \ref{fig:clmdensity} shows that the abundance is maximum in IRAS4A (class 0). It would form on grain surface by radical-radical reaction between $\rm{C_2H_5}$ and OH.  \cite{das18} estimated binding energy of $\rm{C_2H_5}$ and OH to be 2081 K and 3781 K, respectively. The absence of $\rm{C_2H_5OH}$ in the prestellar core may be due to the mobility of these two radicals being limited at low temperatures. Also, the binding energy of ethanol itself is high \citep[5400 K]{wake17} to release in the gas phase. A comparatively reduced abundance of $\rm{C_2H_5OH}$ in the class I stage would be due to the lack of production of it by the radical-radical reaction at such a high temperature and its subsequent destruction after its desorption.

\subsubsection{$\rm{HCCCHO}$}
Propynal ($\rm{HCCCHO}$) was discovered in a cold cloud, TMC-1, using the NRAO's 42.7 m radio telescope at Green Bank \citep{irvi88}. The first laboratory-based infrared spectra of amorphous and crystalline forms of propynal were presented by \cite{huds19} at multiple temperatures. Due to its band's intensity and spectral position, it becomes an attractive candidate for an astronomical search involving interstellar ices. 
Here, we identify only three transitions of HCCCHO in L1544. With the IRAM 30 m telescope, \cite{jime16} carried out high-sensitivity single-pointing 3 mm observations toward the dust-continuum peak in L1544. They detected a single transition for HCCCHO at  83.775842 GHz. We also identify the same transition (9$_{0,9}$ - 8$_{0,8}$) along with another two transitions. 
Our RD analysis yields a column density of $3.2\times10^{12}$ cm$^{-2}$ in L1544. We are not able to identify any clear transitions in other sources.
 However, we estimate an upper limit of HCCCHO for these sources (see Table \ref{tab:upper limit}). \cite{lois16} reported a single line at 83.775832 GHz (9$_{0,9}$ - 8$_{0,8}$) of propynal in B1-b, having the column density of 7.9 $\times$ 10$^{11}$ cm$^{-2}$ but we and \cite{marg20} did not observe this particular transition from the ASAI data. \cite{marg20} estimated an upper limit of HCCCHO based on 102.298 GHz transition, but in our observation, we have seen that this transition is shifted and is also blended with s-Propanal. 
Although the line at 93.043 GHz is slightly shifted from the peak but it is not blended with any other lines. Therefore, On the basis of this transition, we made an upper limit of column density ($\sim$ 2.56 $\times$ 10$^{12}$ cm$^{-2}$) in B1-b. 

The variation of abundance of HCCCHO in various phases of star formation is shown in Figure \ref{fig:clmdensity} with a green colored bar. 
We notice that the abundance is relatively low in L1544. It remains roughly the same in B1-b. In the case of IRAS4A, it increases a little. 
On the other hand, we get a high abundance in SVS13A. 

Given that the majority of the abundances provided here are based on the predicted upper limit of propynal, making a strong statement regarding the evolution related to propynal would not be justified. However, in the ice phase, the formation of HCCCHO can process by the following reactions:
$$
{\rm O + CH_2CCH \rightarrow HCCCHO + H}
$$
$$
{\rm C_2H + H_2CO \rightarrow HCCCHO + H}
$$
The binding energy value of the reactants obtained from \cite{das18} are 770 K, 3238 K, 3315 K, and 3851 K for O, $\rm{CH_2CCH}$, C$_2$H, and H$_2$CO, respectively. So at the low temperature, HCCCHO formation could process by oxygen addition, whereas in the warmer region, by C$_2$H and H$_2$CO. The highest column density obtained at SVS13A indicates that warm chemistry would be efficient.

\subsubsection{$\rm{CH_3OCH_3}$}
Dimethyl ether (DME) is an asymmetric top molecule with two CH$_3$ groups, which undergoes a large amplitude motion along CO-bond. The two internal rotations cause the splitting of a rotational level into four sub-states AA, EE, AE, and EA. DME was detected in the Orion nebula by \citealt{snyd74}. A gas phase formation route for DME was proposed by \citealt{blak87}. Its presence in the high-mass star-forming regions \citep{turn99, sutt95,numm00} and low-mass binary system \citep{caza03,kuan04} have been observed. From the experiments, observations, and theoretical perspectives, \cite{peet06} discussed the fate of DME in various astrophysical environments. We have identified only one transition (3$_{1,3}$ - 2$_{0,2}$) of DME with four sub-states AA, AE, EA, and EE in L1544. The EE and AA sub-states are clearly resolved in our study, whereas the EA and AE are overlapped with each other. It is not possible to perform RD analysis as those sub-states have the same upper-state energy. We estimate a column density of $1.6 \times 10^{12}$ cm$^{-2}$ from LTE fitting. 
\cite{jime16} also observed the same transition in their work and mentioned a column density of $1.5 \times 10^{12}$ cm$^{-2}$.
In B1-b, we have identified four similar sub-states identical to those observed in L1544. We derive a column density of $6.0 \times 10^{12}$ cm$^{-2}$ in B1-b from LTE fitting. In IRAS4A, we have observed four unblended transitions, and each of them consists of four sub-states, but they are overlapped with each other. The integrated intensity is obtained with Gaussian fitting and then divided according to their $S\mu^2$ values \citep{shim16}. The transitions with maximum intensity among these sub-states are considered for the rotation diagram analysis to
calculate the rotation temperature and column density. Only these transitions are noted in Table \ref{tab:observation_3}. We get a rotational temperature of $61.1$ K and column density of $1.2 \times 10^{13}$ cm$^{-2}$ for this species in this source.
The observed transitions of DME in SVS13A have asymmetric line profiles (transitions are not mentioned here; see \citealt{bian19}), and hence Gaussian fit is not possible to perform for these types of transitions. \cite{bian19} have calculated the column density of DME for asymmetric line profile (see method described in Appendix A1 in \citealt{bian19}). They used a source size of $0.3\arcsec$ to derive the column density ($1.4 \times 10^{17}$ cm$^{-2}$) of this species. We use their value after scaling by beam filling factor\footnote{$ff$ = $\frac{\theta_s^2}{\theta_s^2 + \theta_b^2}$} considering the source size $30\arcsec$ and the obtained value is  $\sim 1.4 \times 10^{13}$ cm$^{-2}$. We also perform LTE fitting and get a good fit with a column density a few times higher than the scaled value. We use the above-mentioned scaled value in this work.


The variation of the abundance of DME in different sources which are in different phases of star-formation is shown in Figure \ref{fig:clmdensity} using a magenta-colored bar. We also observed a rising tendency from the L1544 to the IRAS4A and IRAS16293-2422 for DME, similar to what was shown with methanol, acetaldehyde, and methyl formate. Whereas its value gets decreased in SVS13A.

 DME is considered to be formed either in the grain surface \citep{cupp17} or gas phase \citep{balu15}. Both the hydrogenation and radical-radical pathways were involved in the ice phase formation of DME:
$$
{\rm CH_3CHO+H\rightarrow CH_3OCH_2},
$$
$$
{\rm CH_3OCH_2+H\rightarrow CH_3OCH_3},
$$
$$
{\rm CH_3 + CH_3O \rightarrow CH_3OCH_3}.
$$
Both the hydrogenation and the radical-radical route form dimethyl ether. Like methanol, acetaldehyde, and methyl formate, its abundance gradually increases throughout the evolutionary phase up to the class 0 phase and then decreases in the class I phase.

\subsection{$\rm{CH_3CN}$}\label{sec:ch3cn}
Methyl cyanide or CH$_3$CN is a symmetric-top molecule with a high dipole moment of $\sim$ 3.91 debyes. The K-ladders in the rotational level of CH$_3$CN can be excited only by collisional excitation. Hence CH$_3$CN can be a very good tracer to calculate the kinetic temperature of the molecular clouds. We have detected several transitions of CH$_3$CN in all the selected sources. Similar to the CH$_3$OH and CH$_3$CHO, for CH$_3$CN, a two components fit is required for the rotational diagram of methyl cyanide (see Figure. \ref{fig:rotational_diag_CH3CN}) in IRAS4A, (1) E$_{up}$ $>$ 50 K (hot component) and (2) E$_{up}$ $<$ 50 K (cold component). RD analysis yields a temperature of 61.2 K for the high excitation lines and 25.8 K for the low excitation lines. Similarly, a two-component MCMC fit is performed and yields a temperature of 70.1K and 21.0 K for hot component and cold component, respectively.\\
Taking the ratio between two transitions of the same frequency band can nullify various uncertainties obtained from observation. Here we used the line ratios of different CH$_3$CN transitions observed to calculate the kinetic temperature. Different K$_a$ ladders are connected by collisional excitation. The relative population of two K$_a$ ladders follows the Boltzmann equation at kinetic temperature. Considering the selection rules mentioned in \cite{mang93}, we calculated the line ratio between two transitions. Details of the selection rule and the method are described in \cite{das19,mond23}. Some selected line ratios are calculated ($\frac{J1_{k_{a}}-J2_{k_{a}}}{J3_{k_{a'}}-J4_{k_{a'}}}$). Using LTE approximation, the ratio (R) between two transitions satisfying the conditions mentioned in \cite{mang93} is,
R = S$_{R}$ exp(D/T$_{K}$), where $D = E(J3,k_{a'}) - E(J1, k_{a})$ and $S_{R} =\frac{S_{J1k_{a}}}{S_{J3k_{a'}}}$. 

Under the LTE approximation, as the kinetic temperature is considered equal to the excitation temperature, we calculated the kinetic temperatures for different CH$_3$CN k-ladder transitions for the sources (L1544, Barnard1 b, IRAS4A, SVS13A) using the above-mentioned formula. Calculated values of kinetic temperatures for all the CH$_3$CN transitions are mentioned in Table \ref{table:kladder}. In Figure \ref{fig:temp}, we plotted the average kinetic temperature obtained from this calculations for the transitions having E$_{up}>50$ K with the solid blue line, and the dashed blue line in Figure \ref{fig:temp} is the same for transitions having E$_{up}<50$ K.

\begin{table*}
\centering
{\scriptsize
 \caption{Calculation of kinetic temperature using line ratio of observed CH$_3$CN transitions.} \label{table:kladder}
\begin{tabular}{|c|c|c|c|c|c|c|c|c|c|}
  \hline
  \hline
  Source&Frequency&Quantum No.&E$_{up}$&$\int$T$_{mb}$dv&S$_{ij}$&R&T$_{k}$&Average T$_k$&\\
  &(GHz)&&(K)&(K.km.s$^{-1}$)&&&(K)&(K)&\\
  \hline
  \hline  
L1544&91.985314&$5_1$ - $4_1$&20.4 &0.05&6.61439&$\frac{5_0-4_0}{5_1-4_1}$&5.12&5.12&\\
&91.987087&$5_0$ - $4_0$&13.2 &0.068&2.2051&&&&\\
\hline
B1-b&91.985314&$5_1$ - $4_1$&20.4&0.034&6.61439&$\frac{5_0-4_0}{5_1-4_1}$&4.41&4.41&\\
&91.987087&$5_0$ - $4_0$&13.2&0.058&2.2051&&&&\\
\hline
IRAS4A&128.75703&$7_3$ – $6_3$&89&0.089&15.74921&$\frac{7_2-6_2}{7_3-6_3}$&33.79&&\\
&128.769436&$7_2$ - $6_2$&53.3&0.144&8.85861&&&&\\
\cline{2-8}
&128.769436&$7_2$ - $6_2$&53.3&0.144&8.85861&$\frac{7_1-6_1}{7_2-6_2}$&46.24&&\\
&128.776881&$7_1$ – $6_1$&31.9&0.244&9.44898&&&&\\
\cline{2-8}
&128.75703&$7_3$ – $6_3$&89&0.089&15.74921&$\frac{7_1-6_1}{7_3-6_3}$&37.58&&\\
&128.776881&$7_1$ – $6_1$&31.9&0.244&9.44898&&&&\\
\cline{2-8}
&128.75703&$7_3$ – $6_3$&89&0.089&15.74921&$\frac{7_0-6_0}{7_3-6_3}$&23.19&&\\
&128.779363&$7_0$ - $6_0$&24.7&0.285&3.1501&&&&\\
\cline{2-8}
&128.769436&$7_2$ - $6_2$&53.3&0.144&8.85861&$\frac{7_0-6_0}{7_2-6_2}$&16.66&&\\
&128.779363&$7_0$ - $6_0$&24.7&0.285&3.1501&&&&\\
\cline{2-8}
&147.163244&$8_2$ – $7_2$&60.4&0.098&10.33654&$\frac{8_1-7_1}{8_2-7_2}$&26.44&48.60&E$_{up} >$ 50 K\\
&147.171751&$8_1$ – $7_1$&38.9&0.232&10.85164&&&&\\
\cline{2-8}
&147.163244&$8_2$ – $7_2$&60.4&0.098&10.33654&$\frac{8_0-7_0}{8_2-7_2}$&14.17&&\\
&147.174588&$8_0$ – $7_0$&31.8&0.258&3.61771&&&&\\
\cline{2-8}
&165.540377&$9_3$ – $8_3$&104&0.131&22.05132&$\frac{9_2-8_2}{9_3-8_3}$&106.31&&\\
&165.556321&$9_2$ - $8_2$&68.3&0.098&11.7908&&&&\\
\cline{2-8}
&165.556321&$9_2$ - $8_2$&68.3&0.098&11.7908&$\frac{9_1-8_1}{9_2-8_2}$&28.45&&\\
&165.565891&$9_1$ – $8_1$&46.9&0.216&12.25078&&&&\\
\cline{2-8}
&165.540377&$9_3$ – $8_3$&104&0.131&22.05132&$\frac{9_1-8_1}{9_3-8_3}$&52.49&&\\
&165.565891&$9_1$ – $8_1$&46.9&0.216&12.25078&&&&\\
\cline{2-8}
&165.540377&$9_3$ – $8_3$&104&0.131&22.05132&$\frac{9_0-8_0}{9_3-8_3}$&29.16&&\\
&165.569081&$9_0$ – $8_0$&39.7&0.22	&4.08322&&&&\\
\cline{2-8}
&165.556321&$9_2$ - $8_2$&68.3&0.098&11.7908&$\frac{9_0-8_0}{9_2-8_2}$&15.30&&\\
&165.569081&$9_0$ – $8_0$&39.7&0.22&4.08322&&&&\\
\cline{2-10}
&73.588799&$4_1$ – $3_1$&16&0.127&5.16717&$\frac{4_0-3_0}{4_1-3_1}$&5.20&&\\
&73.590218&$4_0$ – $3_0$&8.8&0.169&1.72263&&&&\\
\cline{2-8}
&91.979994&$5_2$ – $4_2$&41.8&0.082&5.78723&$\frac{5_1-4_1}{5_2-4_2}$&33.07&&\\
&91.985314&$5_1$ – $4_1$&20.4&0.179&6.61439&&&&\\
\cline{2-8}
&91.985314&$5_1$ – $4_1$&20.4&0.179&6.61439&$\frac{5_0-4_0}{5_1-4_1}$&5.50&&\\
&91.987087&$5_0$ – $4_0$&13.2&0.221&2.2051&&&&\\
\cline{2-8}
&91.979994&$5_2$ – $4_2$&41.8&0.082&5.78723&$\frac{5_0-4_0}{5_2-4_2}$&14.62&&\\
&91.987087&$5_0$ – $4_0$&13.2&0.221&2.2051&&&&\\
\cline{2-8}
&110.381372&$6_1$ – $5_1$&25.7&0.233&8.03828&$\frac{6_0-5_0}{6_1-5_1}$&5.99&10.31&E$_{up} <$ 50 K\\
&110.383499&$6_0$ – $5_0$&18.5&0.258&2.6798&&&&\\
\cline{2-8}
&128.776881&$7_1$ – $6_1$&31.9&0.244&9.44898&$\frac{7_0-6_0}{7_1-6_1}$&5.74&&\\
&128.779363&$7_0$ - $6_0$&24.7&0.285&3.1501&&&&\\
\cline{2-8}
&147.171751&$8_1$ – $7_1$&38.9&0.232&10.85164&$\frac{8_0-7_0}{8_1-7_1}$&5.89&&\\
&147.174588&$8_0$ – $7_0$&31.8&0.258&3.61771&&&&\\
\cline{2-8}
&165.565891&$9_1$ – $8_1$&46.9&0.216&12.25078&$\frac{9_0-8_0}{9_1-8_1}$&6.45&&\\
&165.569081&$9_0$ – $8_0$&39.7&0.22&4.08322&&&&\\
\hline
SVS13A&110.364353&$6_3$ - $5_3$&82.8&0.08&12.40333&$\frac{6_2-5_2}{6_3-5_3}$&42.87&&\\
&110.374989&$6_2$ - $5_2$&47.1&0.109&7.34927&&&&\\
\cline{2-8}
&110.364353&$6_3$ - $5_3$&82.8&0.08&12.40333&$\frac{6_1-5_1}{6_3-5_3}$&72.47&&\\
&110.381372&$6_1$ – $5_1$&25.7&0.114&8.03828&&&&\\
\cline{2-8}
&110.364353&$6_3$ - $5_3$&82.8&0.08&12.40333&$\frac{6_0-5_0}{6_3-5_3}$&38.21&&\\
&110.383499&$6_0$ – $5_0$&18.5&0.093&2.6798&&&&\\
\cline{2-8}
&220.709016&$12_3$ – $11_3$&133.2&0.265&31.00599&$\frac{12_2-11_2}{12_3-11_3}$&47.07&&\\
&220.73026&$12_2$ - $11_2$&97.4&0.294&16.07897&&&&\\
\cline{2-8}
&220.74301&$12_1$ - $11_1$&76&0.278&16.4203&$\frac{12_0-11_0}{12_1-11_1}$&24.76&63.90&E$_{up} >$ 50 K\\
&220.747261&$12_0$ - $11_0$&68.9&0.373&16.53946&&&&\\
\cline{2-8}
&220.709016&$12_3$ – $11_3$&133.2&0.265&31.00599&$\frac{12_1-11_1}{12_3-11_3}$&83.68&&\\
&220.74301&$12_1$ - $11_1$&76&0.278&16.4203&&&&\\
\cline{2-8}
&220.709016&$12_3$ – $11_3$&133.2&0.265&31.00599&$\frac{12_0-11_0}{12_3-11_3}$&66.27&&\\
&220.747261&$12_0$ - $11_0$&68.9&0.373&16.53946&&&&\\
\cline{2-8}
&220.73026&$12_2$ - $11_2$&97.4&0.294&16.07897&$\frac{12_0-11_0}{12_2-11_2}$&135.87&&\\
&220.747261&$12_0$ - $11_0$&68.9&0.373&16.53946&&&&\\
\hline
\hline
 \end{tabular}}
 \end{table*}

\begin{table*}
\centering
{\scriptsize
 \caption{Calculation of kinetic temperature using line ratio of observed CH$_3$CN transitions.} \label{table:kladder}
\begin{tabular}{|c|c|c|c|c|c|c|c|c|c|}
  \hline
  \hline
  Source&Frequency&Quantum No.&E$_{up}$&$\int$T$_{mb}$dv&S$_{ij}$&R&T$_{k}$&Average T$_k$&\\
  &(GHz)&&(K)&(K.km.s$^{-1}$)&&&(K)&(K)&\\
  \hline
\hline
&91.979994&$5_2$ – $4_2$&41.8&0.067&5.78723&$\frac{5_1-4_1}{5_2-4_2}$&184.38&&\\
&91.985314&$5_1$ – $4_1$&20.4&0.086&6.61439&&&&\\
\cline{2-8}
&91.985314&$5_1$ – $4_1$&20.4&0.086&6.61439&$\frac{5_0-4_0}{5_1-4_1}$&6.42&&\\
&91.987087&$5_0$ – $4_0$&13.2&0.088&2.2051&&&&\\
\cline{2-8}
&91.979994&$5_2$ – $4_2$&41.8&0.067&5.78723&$\frac{5_0-4_0}{5_2-4_2}$&23.11&&\\
&91.987087&$5_0$ – $4_0$&13.2&0.088&2.2051&&&43.58&E$_{up} <$ 50 K\\
\cline{2-8}
&110.381372&$6_1$ – $5_1$&25.7&0.114&8.03828&$\frac{6_0-5_0}{6_1-5_1}$&8.04&&\\
&110.383499&$6_0$ – $5_0$&18.5&0.093&2.6798&&&&\\
\cline{2-8}
&110.374989&$6_2$ - $5_2$&47.1&0.109&7.34927&$\frac{6_0-5_0}{6_2-5_2}$&33.64&&\\
&110.383499&$6_0$ – $5_0$&18.5&0.093&2.6798&&&&\\
\cline{2-8}
&128.7768817&$7_1$ – $6_1$&31.9&0.092&9.44898&$\frac{7_0-6_0}{7_1-6_1}$&5.85&&\\
&128.779363&$7_0$ - $6_0$&24.7&0.105&3.1501&&&&\\
  \hline
  \hline
 \end{tabular}}
 \end{table*}

\subsection{Abundance variation in different sources}
We found a steadily increasing abundance from L1544 to IRAS4A and peaked at IRAS16293-2422 for CH$_3$OH, CH$_3$CHO, CH$_3$OCHO, C$_2$H$_5$OH, CH$_3$OCH$_3$ and CH$_3$CN and a reduced abundance in SVS13A. 
In contrast, the abundance of HCCCHO shows an increasing trend up to the SVS13A. The trend obtained with HCCCHO is based on the upper limit derived for most of the sources except L1544. In the case of C$_2$H$_5$OH, we could not even estimate an upper limit in L1544. 

\cite{scib20} noticed that CH$_3$OH and CH$_3$CHO formed early and often in the starless prestellar stages and they proposed a chemical link between them at the early stages. However, this proposed linkage in the prestellar cores is yet to be observationally supported by the higher spatial resolution map. They pointed out the chemical linkage between CH$_3$OH and CH$_3$CHO by the gas phase reaction:
$\rm{CH + CH_3OH \rightarrow CH_3CHO + H}$ 
This reaction was experimentally studied by \cite{john00}, who found it to be barrierless. \cite{vasy17} considered this reaction for their chemical modeling of the prestellar core. However, in the UMIST database, this reaction is considered above 298 K only. In the KIDA database, this reaction yields,
$\rm{CH + CH_3OH \rightarrow CH_3 + H_2CO}$  in the 10-300 K range.

\begin{figure}
\centering
\begin{minipage}{0.50\textwidth}
\includegraphics[width=\textwidth]{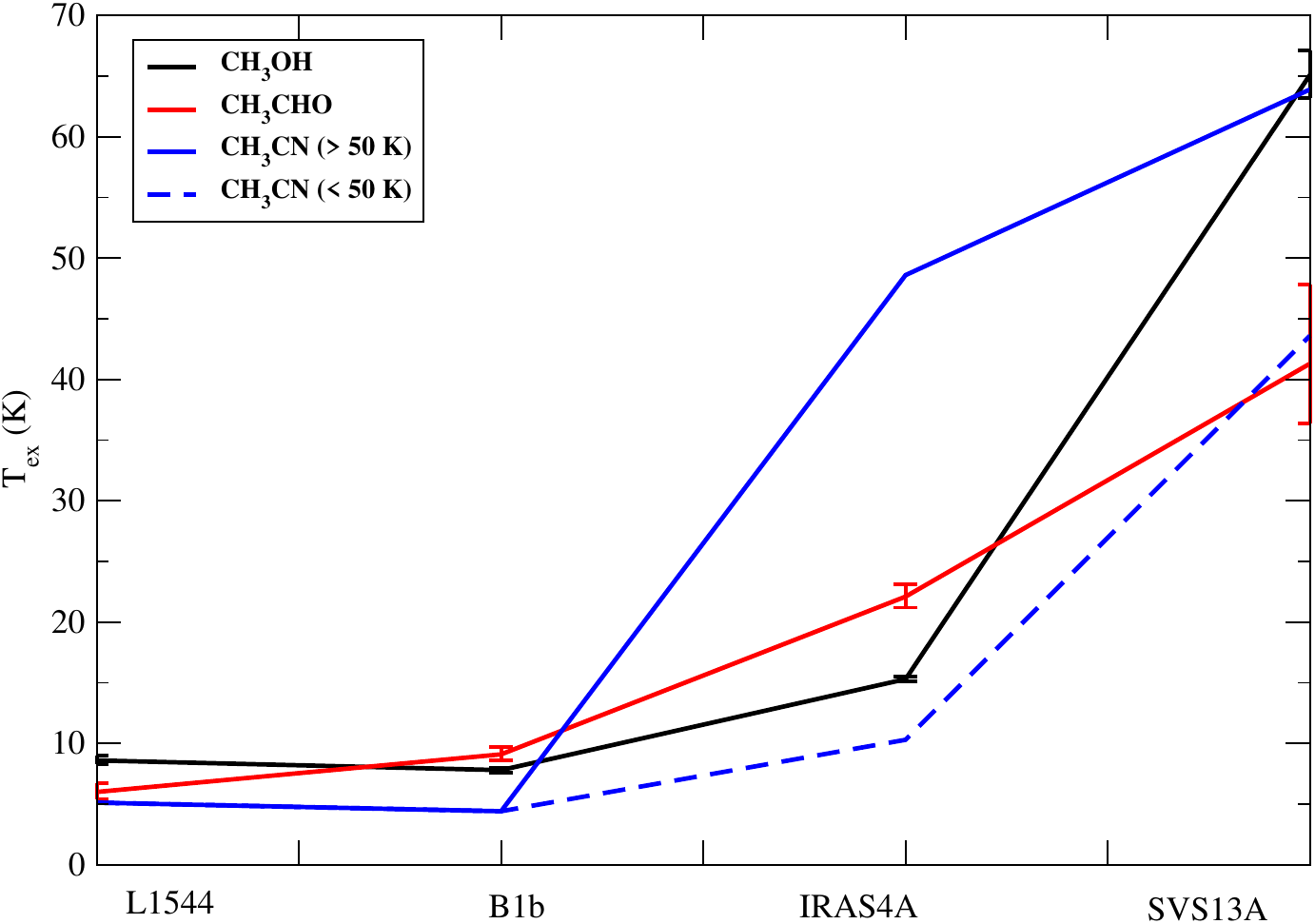}
\end{minipage}
\caption{Excitation Temperature derived from rotation diagram for methanol (in black) and acetaldehyde (in red), respectively. Vertical lines represent the corresponding errors. Kinetic temperature calculated from Table \ref{table:kladder} using CH$_3$CN transitions for high-temperature (solid blue) component and low-temperature (dashed blue) component present in IRAS4A and SVS13A.}
\label{fig:temp}
\end{figure}

\begin{figure}
\centering
\begin{minipage}{0.50\textwidth}
\includegraphics[width=\textwidth]{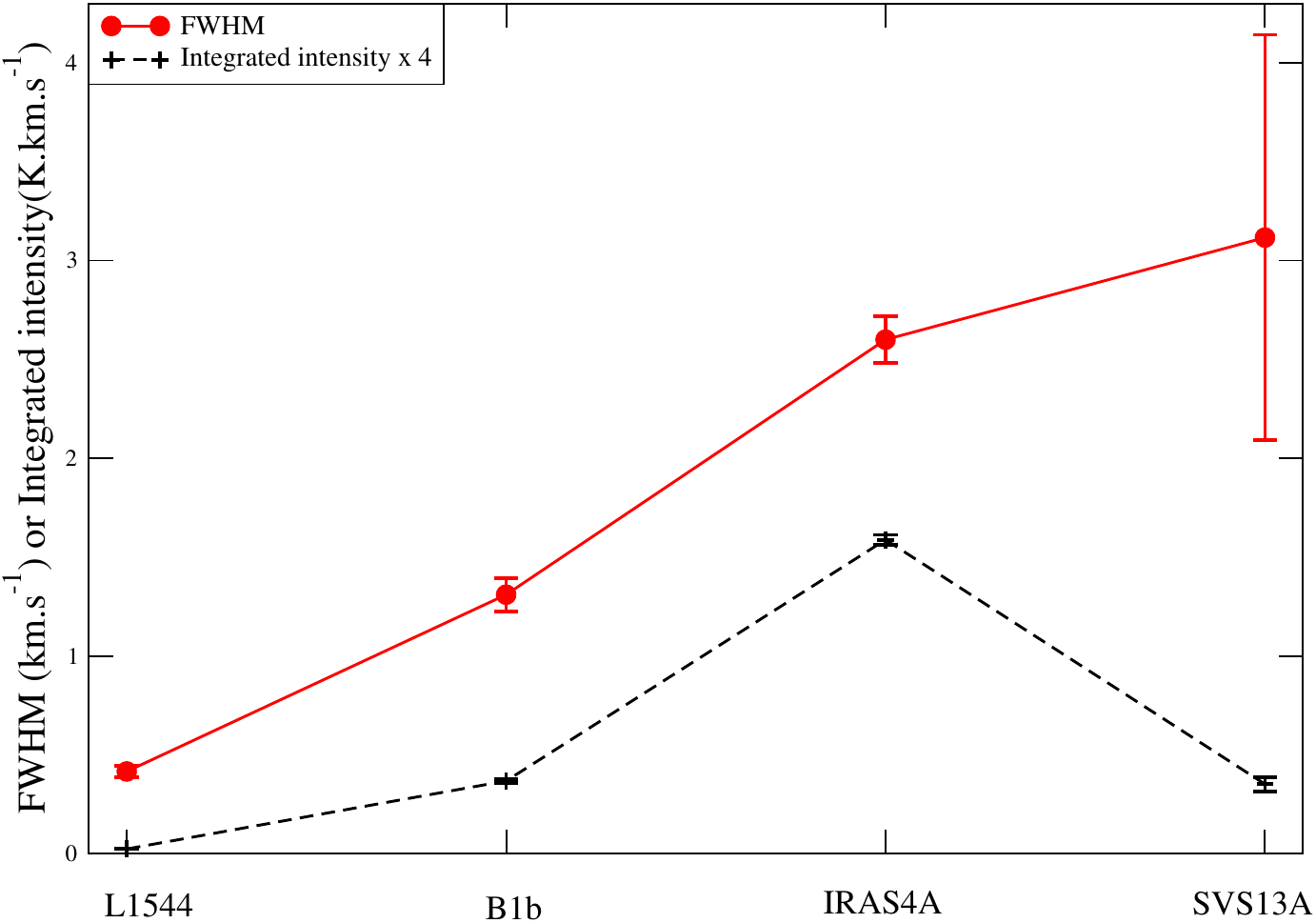}
\end{minipage}
\caption{FWHM (red line) and integrated intensity (black-dashed) for 96.755501 GHz transition of methanol. The vertical lines represent the error bars.}
\label{fig:fwhm}
\end{figure}

The hydrogen column density in each source is required to derive the abundances of the species and look out for their linkage with the various evolutionary phases, if any. For the reliability of the abundance derivation from the obtained column density, we use the hydrogen column density obtained from a beam size comparable to the source size of the molecule (see Section \ref{sec:H2_col}). 
As our beam size varies from 30$\arcsec$ to 9$\arcsec$ in our frequency range, we could not resolve the molecular emission originating from the core. To have an idea, we plot the rotational temperature of two species and the kinetic temperature obtained from k-ladder calculation for CH$_3$CN (details in section \ref{sec:ch3cn}) in all our sample sources. As usual, we found an increased temperature (see Figure \ref{fig:temp}) from L1544 to SVS13A. 
Moreover, we studied a particular transition of a species that is common in all the sources. Only one transition (96.755501 GHz) of CH$_3$OH is present in all the sources. We plot the integrated intensity of this transition of CH$_3$OH, and it is presented in Figure \ref{fig:fwhm}. It implies that the integrated intensity also shows a similar trend with the different sources at different stages as it was obtained for the abundances. 
Additionally, for acetaldehyde, we identified a transition at $211.2738$ GHz for all the sources except SVS13A. It shows an increasing trend. In IRAS4A and SVS13A, we identified a transition at $216.5819$ GHz and obtained a reduced integrated intensity in SVS13A compared to IRAS4A.
Thus the adopted N(H$_2$) may not only be responsible for the obtained abundance trend. 

Further, we check whether this transition is affected by any of the physical properties of the source. Hence, we plot the variation of FWHM of this transition (96.7555 GHz) for all the sources (see Figure \ref{fig:fwhm}). It depicts a steadily increasing trend from L1544 to SVS13A, and its value is comparable to the other observed transitions.

Since all these sources have different luminosity, in section \ref{sec:luminosity}, we discuss the linkage between the obtained abundances and source luminosity.

Except for the L1544, all our sample sources have multiple cores. The targeted position of all the sample sources in ASAI data is mentioned in Table \ref{table:source}. Our observed beam includes all the cores presented in each source, even with the highest frequency of our observation. So we need to check whether any chemical differentiation is present among the cores of each source and confirm that these cores follow the similar trend that we have obtained. Therefore, the high resolution interferometric data is required to verify the obtained trend for different sources in different stages. Since the analysis of the interferometric observation is out of scope for this paper, we discuss the interferometric observational results obtained by various authors in these sources in Section \ref{sec:INO}.


\begin{figure*}
\begin{minipage}{0.33\textwidth}
\includegraphics[width=\textwidth]{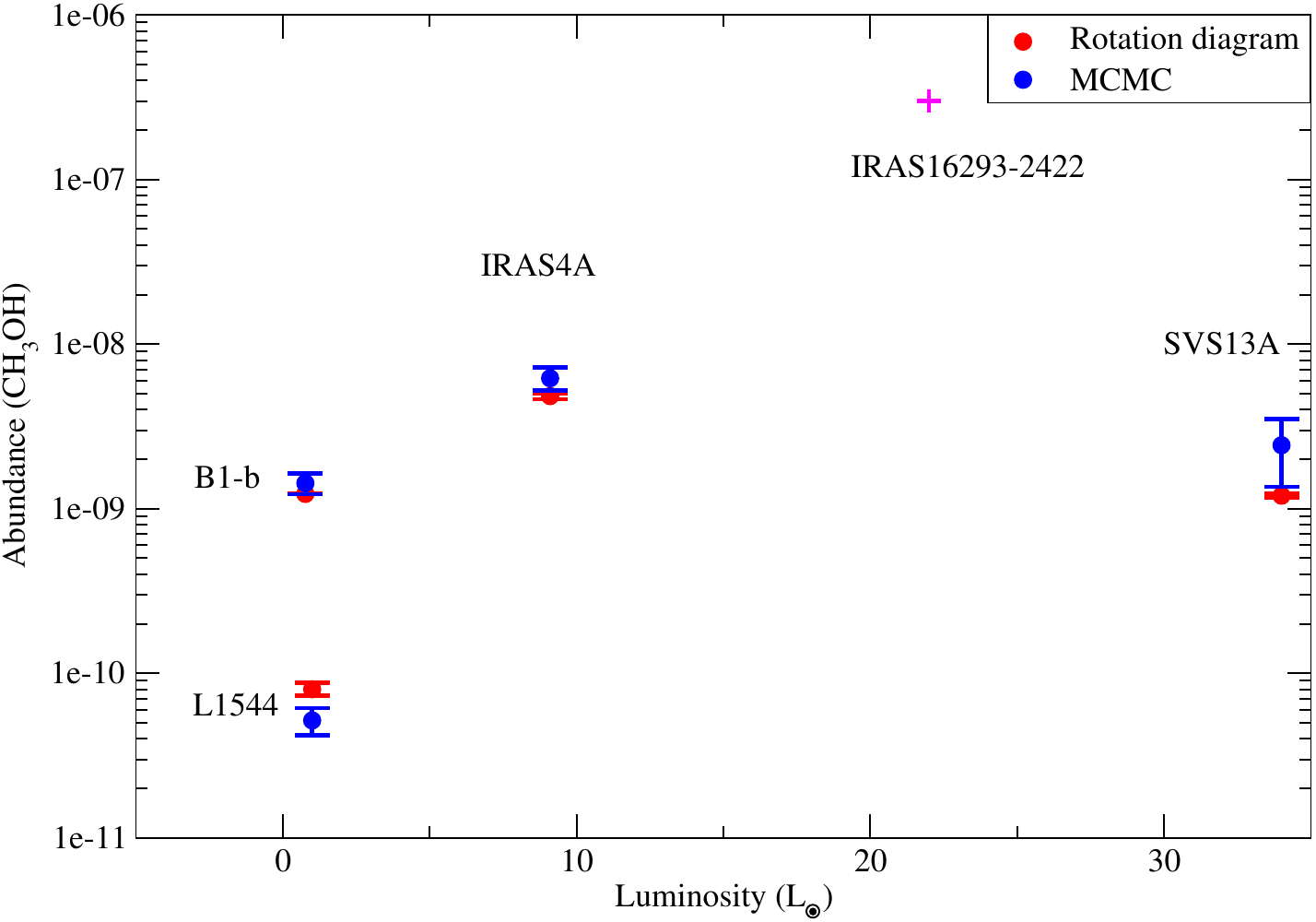}
\end{minipage}
\begin{minipage}{0.33\textwidth}
\includegraphics[width=\textwidth]{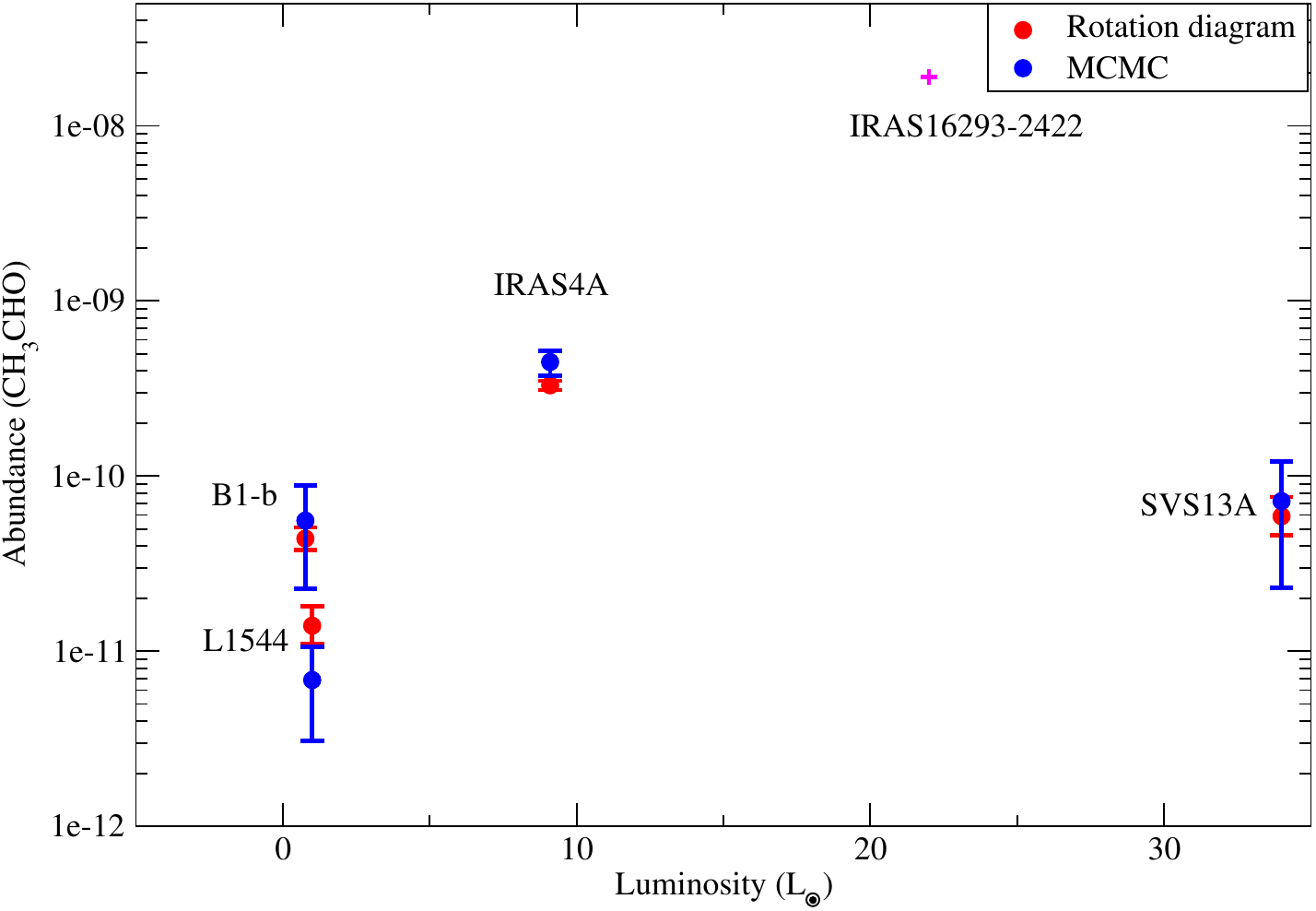}
\end{minipage}
\begin{minipage}{0.33\textwidth}
\includegraphics[width=\textwidth]{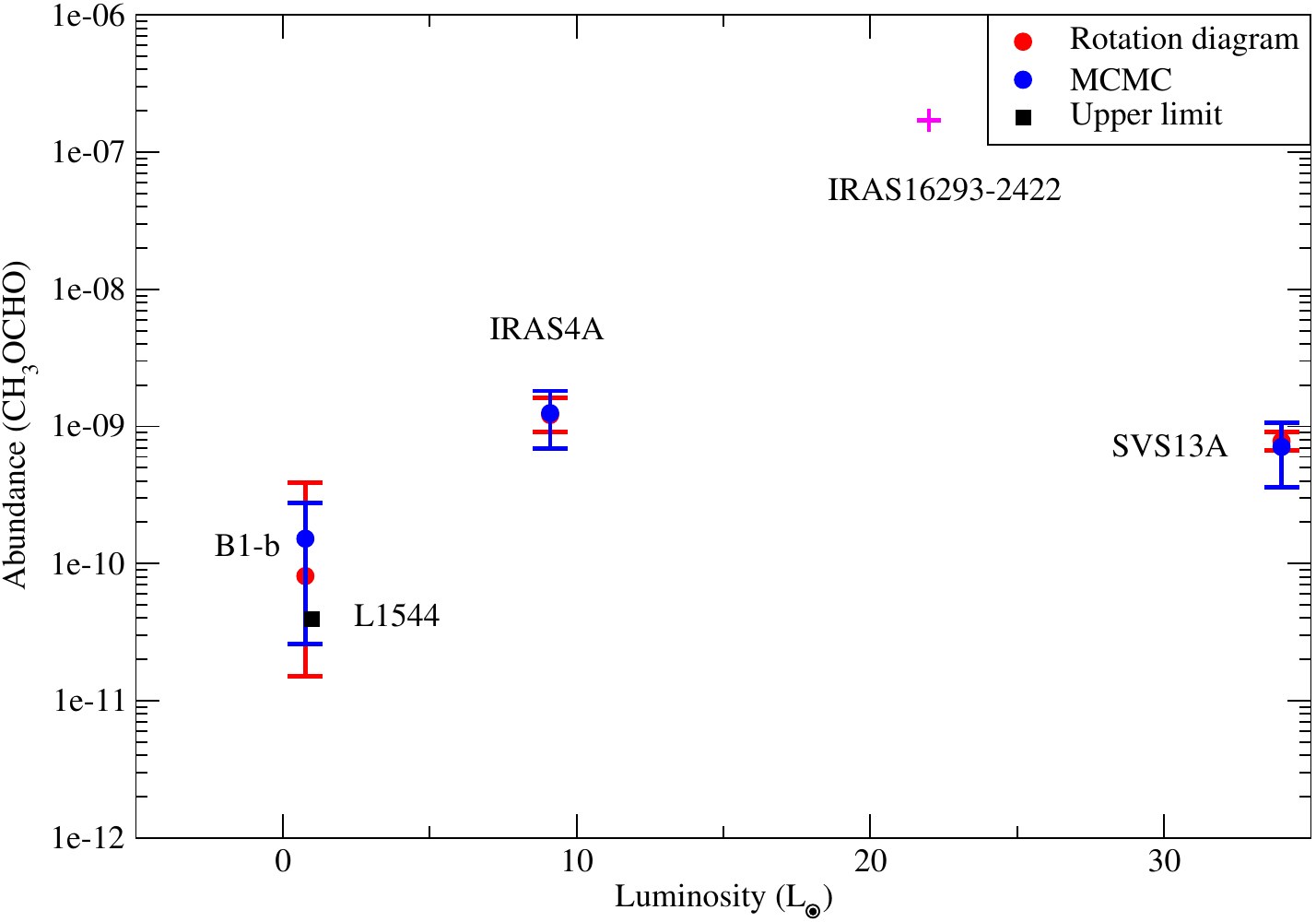}
\end{minipage}
\begin{minipage}{0.33\textwidth}
\includegraphics[width=\textwidth]{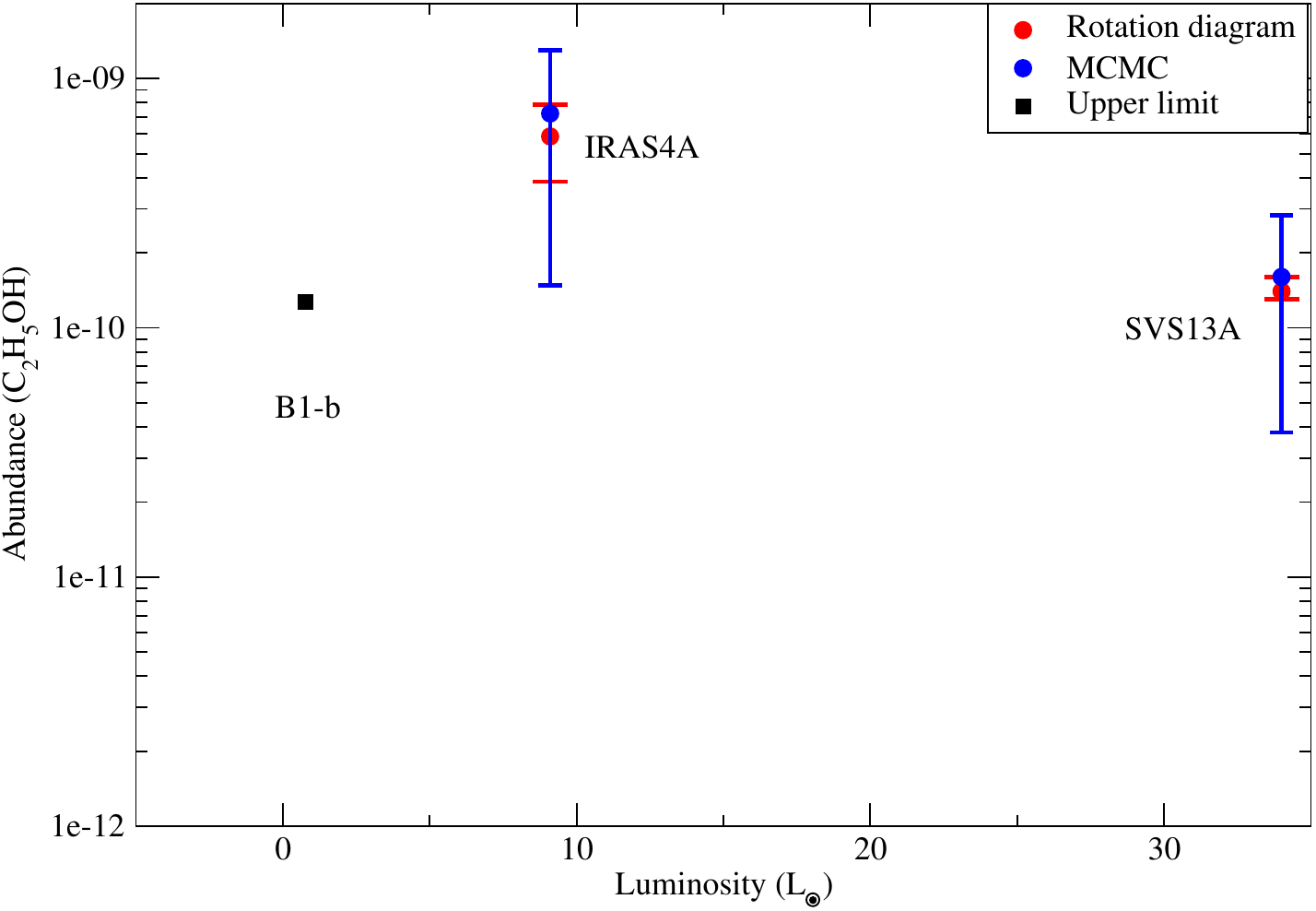}
\end{minipage}
\begin{minipage}{0.33\textwidth}
\includegraphics[width=\textwidth]{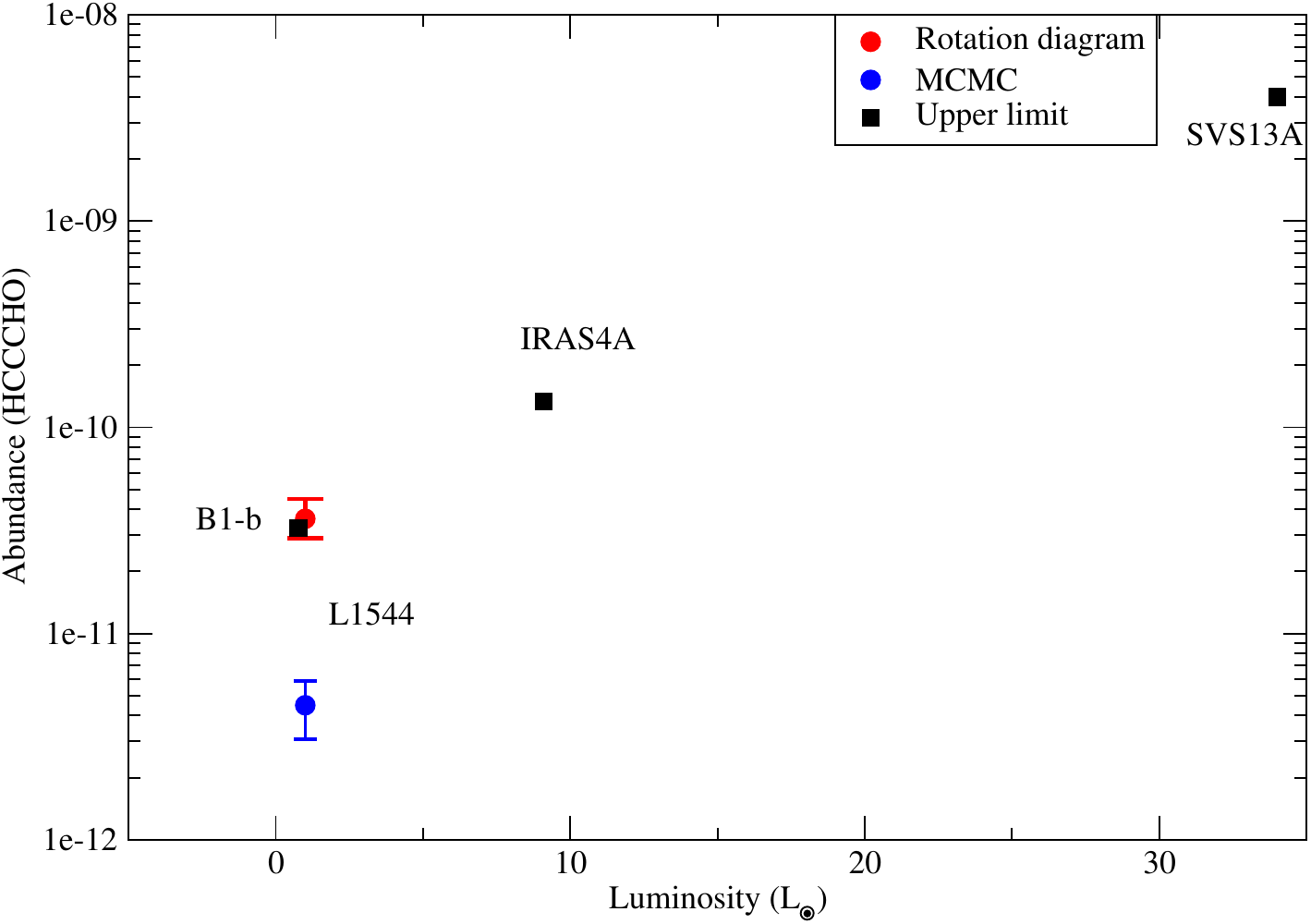}
\end{minipage}
\begin{minipage}{0.33\textwidth}
\includegraphics[width=\textwidth]{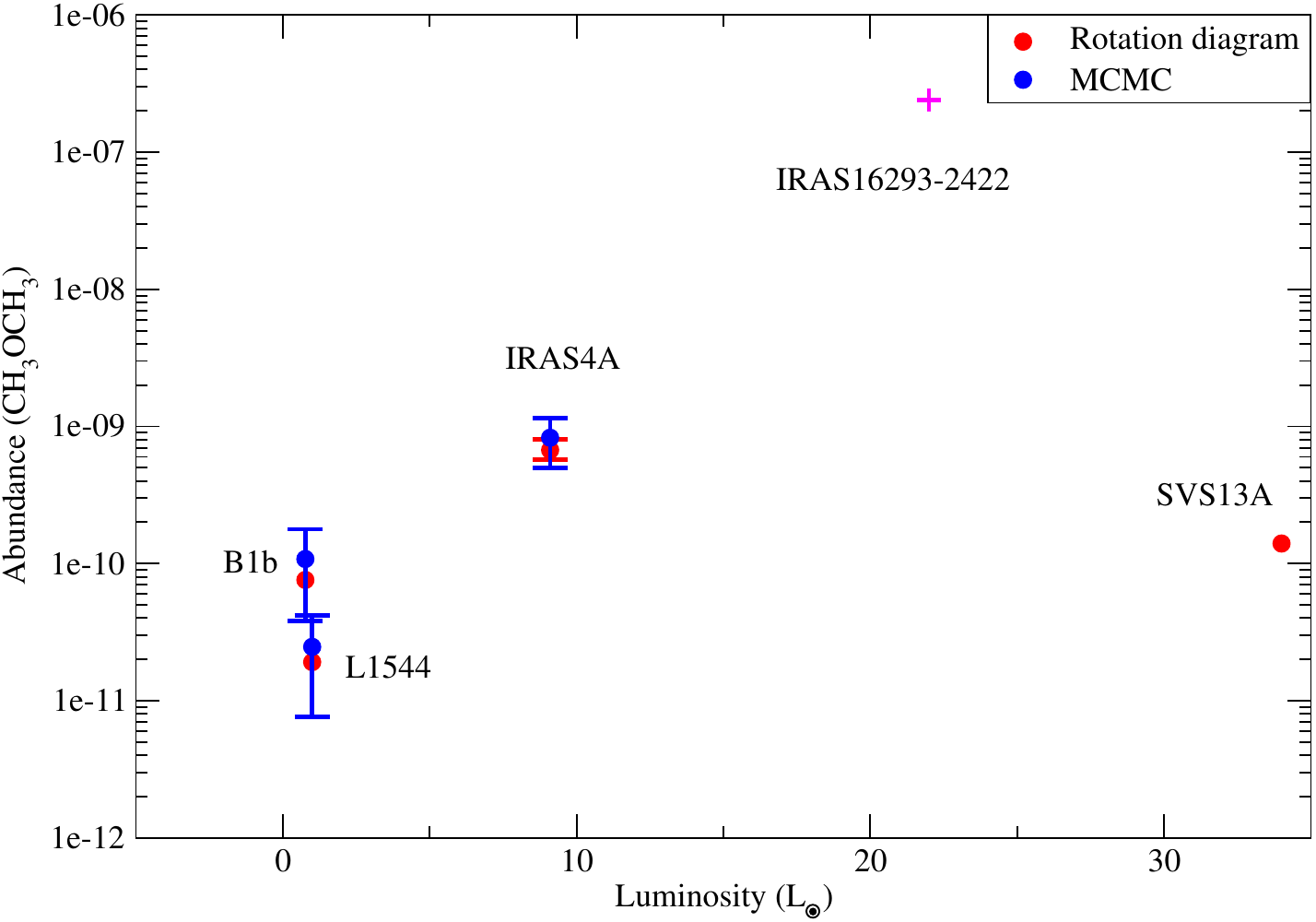}
\end{minipage}
\begin{minipage}{0.33\textwidth}
\includegraphics[width=\textwidth]{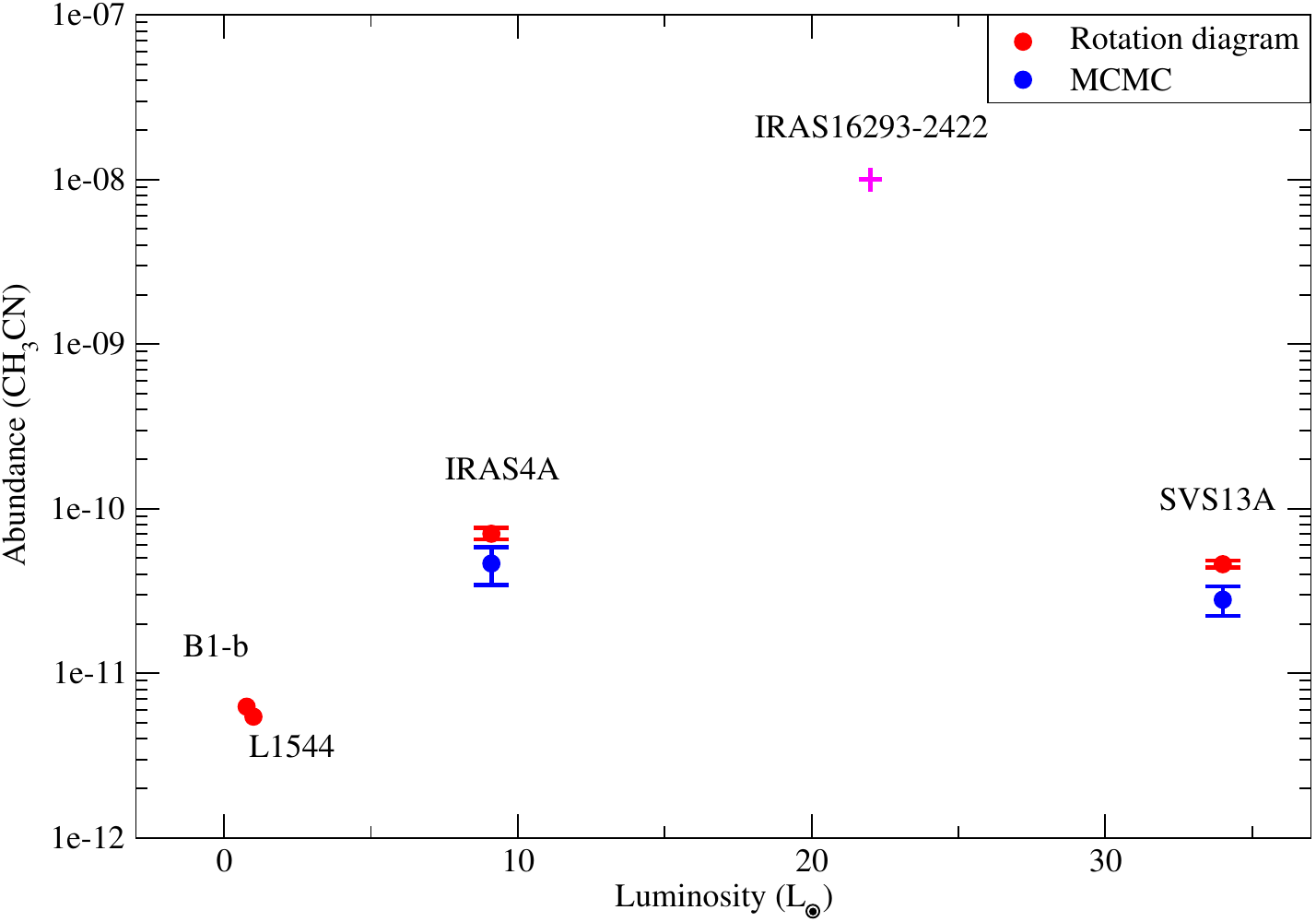}
\end{minipage}
\caption{Abundance variation of CH$_3$OH, CH$_3$CHO, CH$_3$O-CHO, C$_2$H$_5$OH, HCCCHO, CH$_3$OCH$_3$ and CH$_3$CN shown with source luminosity. The red circle represents the value obtained from the rotational diagram, and the blue circle represents the same obtained from MCMC. The solid black squares represent the same calculated using upper limits. 
The plus sign (magenta) represents the abundance for IRAS4A 16293-2422 (22 L$_\odot$) taken from \cite{caza03}.
The vertical lines represent the error bars.}
\label{fig:luminosity}
\end{figure*}

\begin{figure*}
\begin{minipage}{0.33\textwidth}
\includegraphics[width=\textwidth]{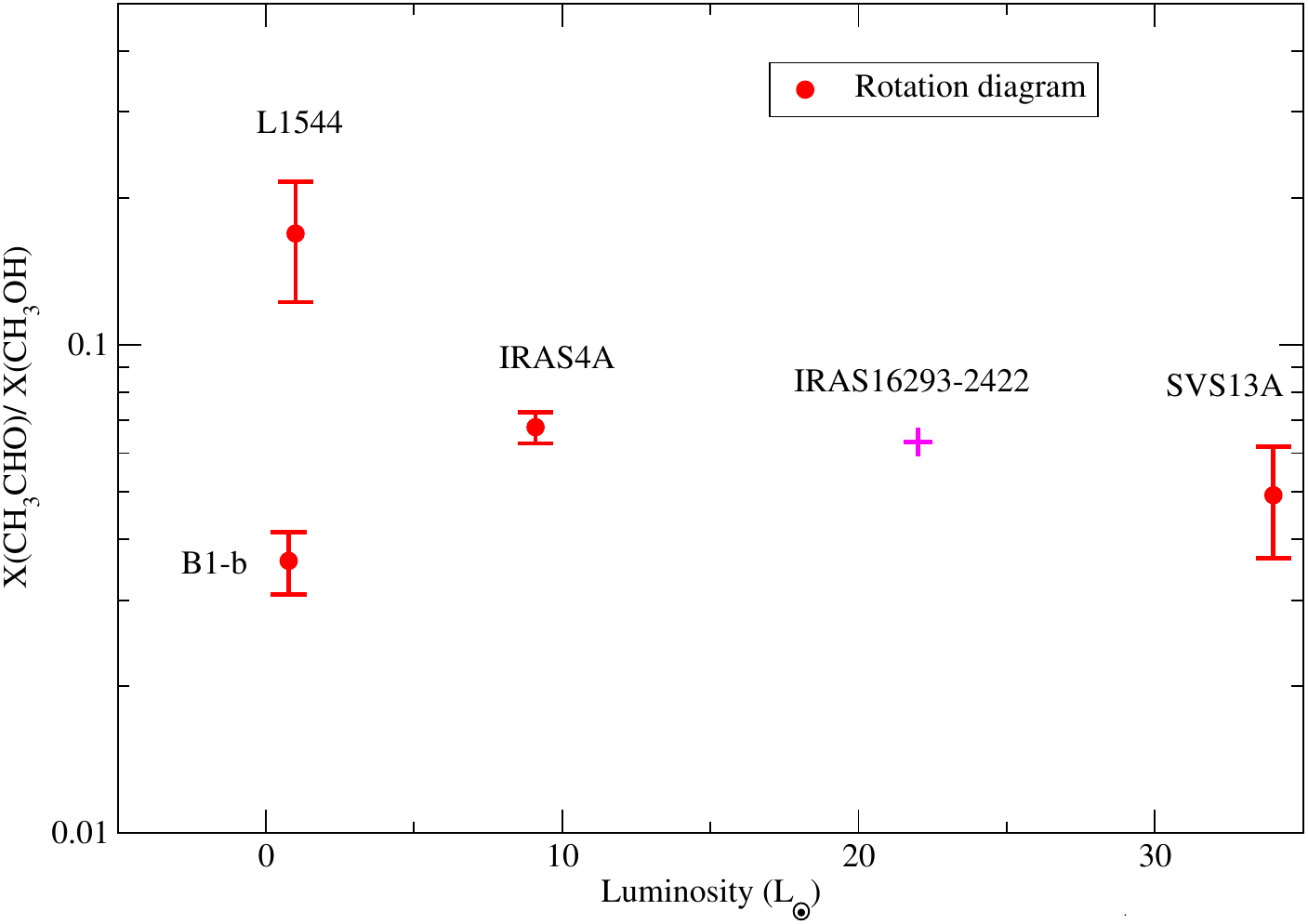}
\end{minipage}
\begin{minipage}{0.33\textwidth}
\includegraphics[width=\textwidth]{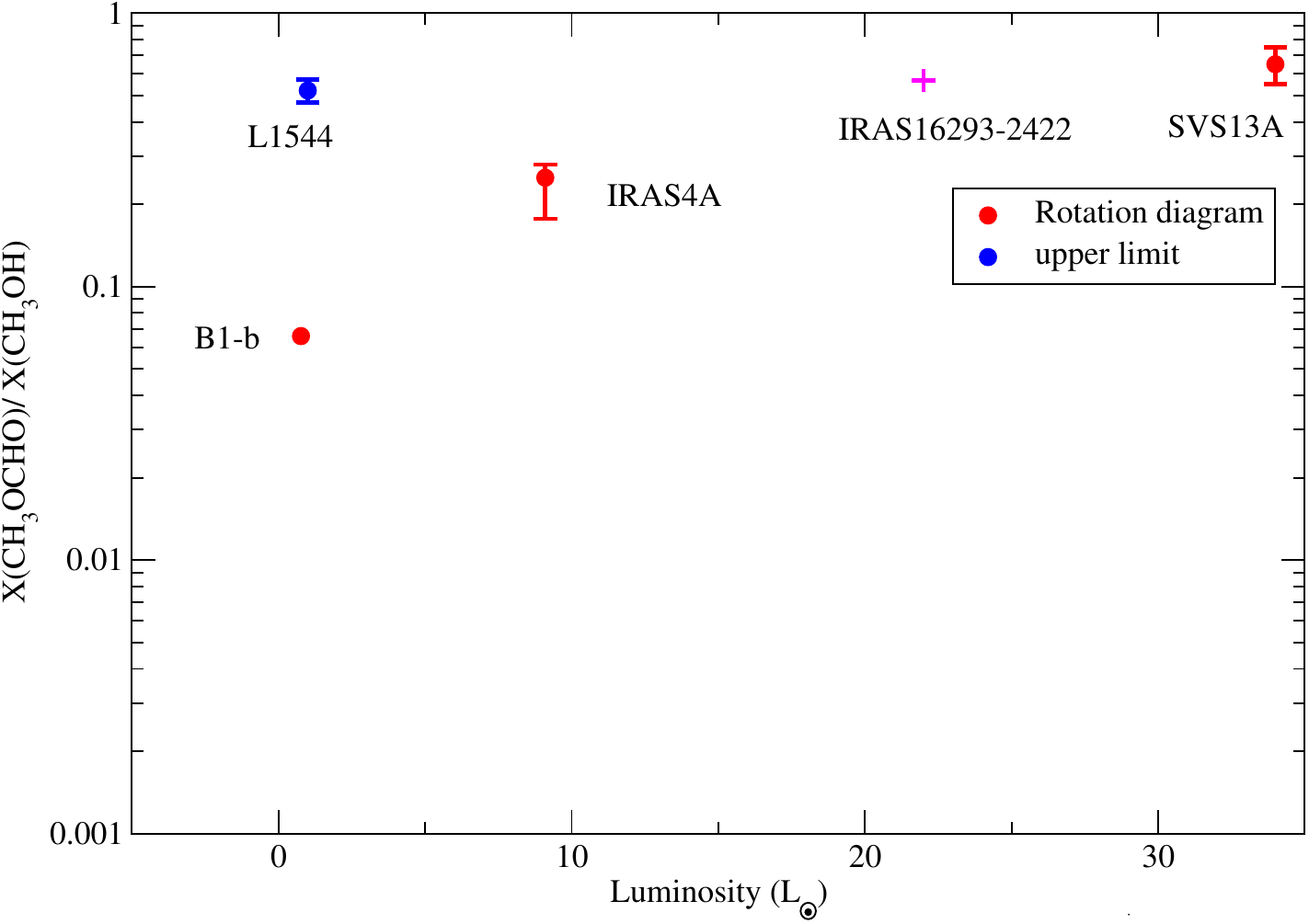}
\end{minipage}
\begin{minipage}{0.33\textwidth}
\includegraphics[width=\textwidth]{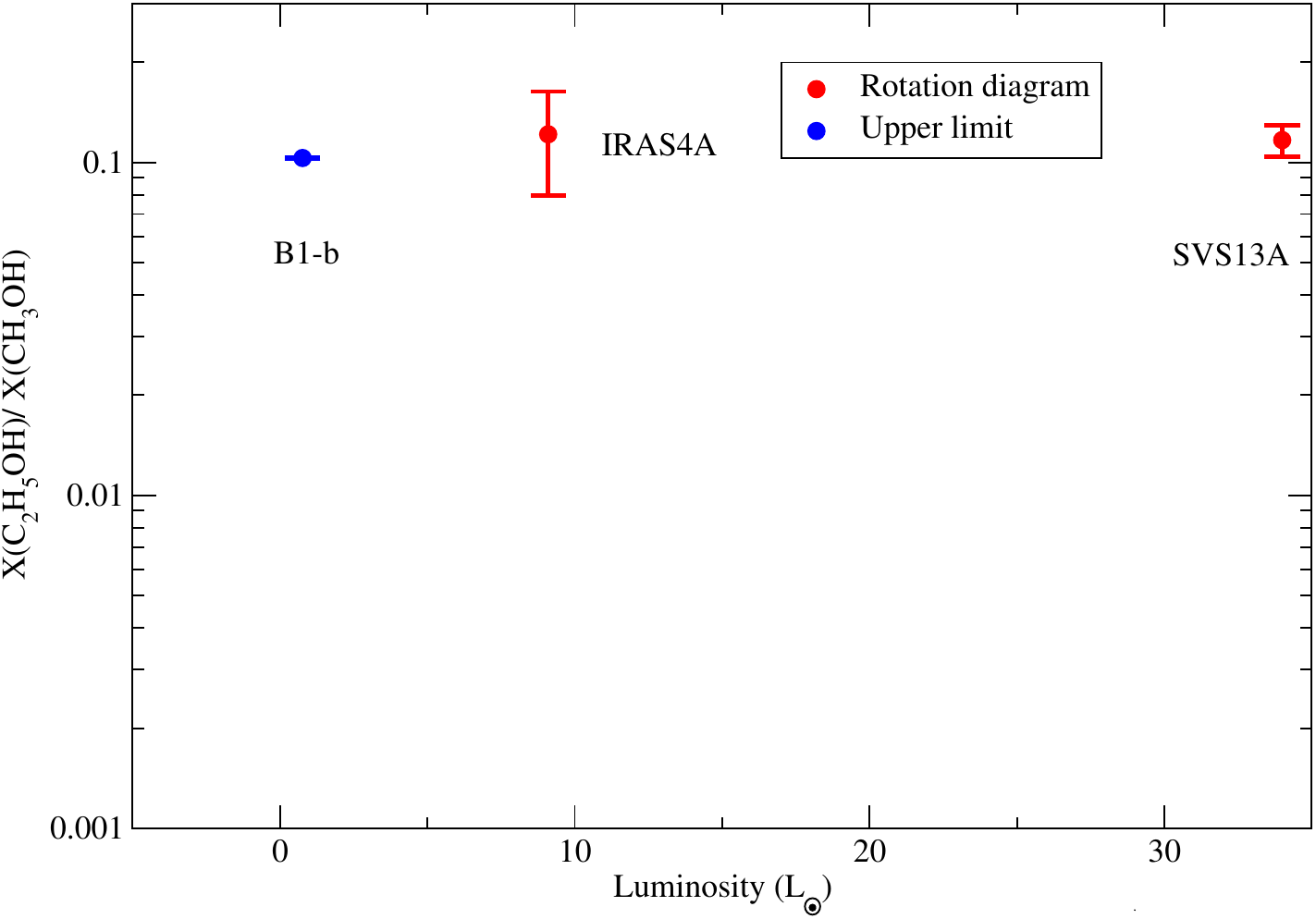}
\end{minipage}
\begin{minipage}{0.33\textwidth}
\includegraphics[width=\textwidth]{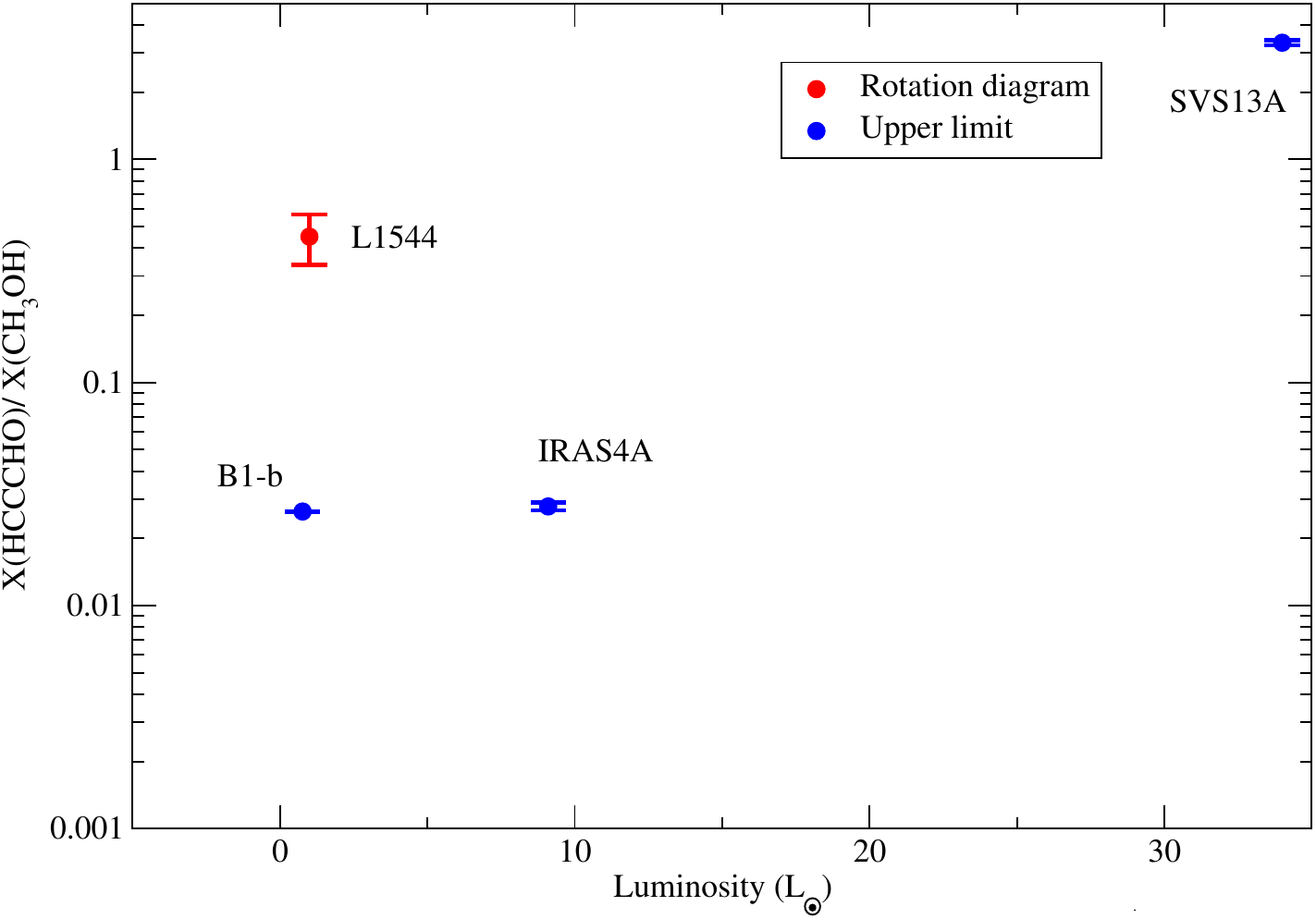}
\end{minipage}
\begin{minipage}{0.33\textwidth}
\includegraphics[width=\textwidth]{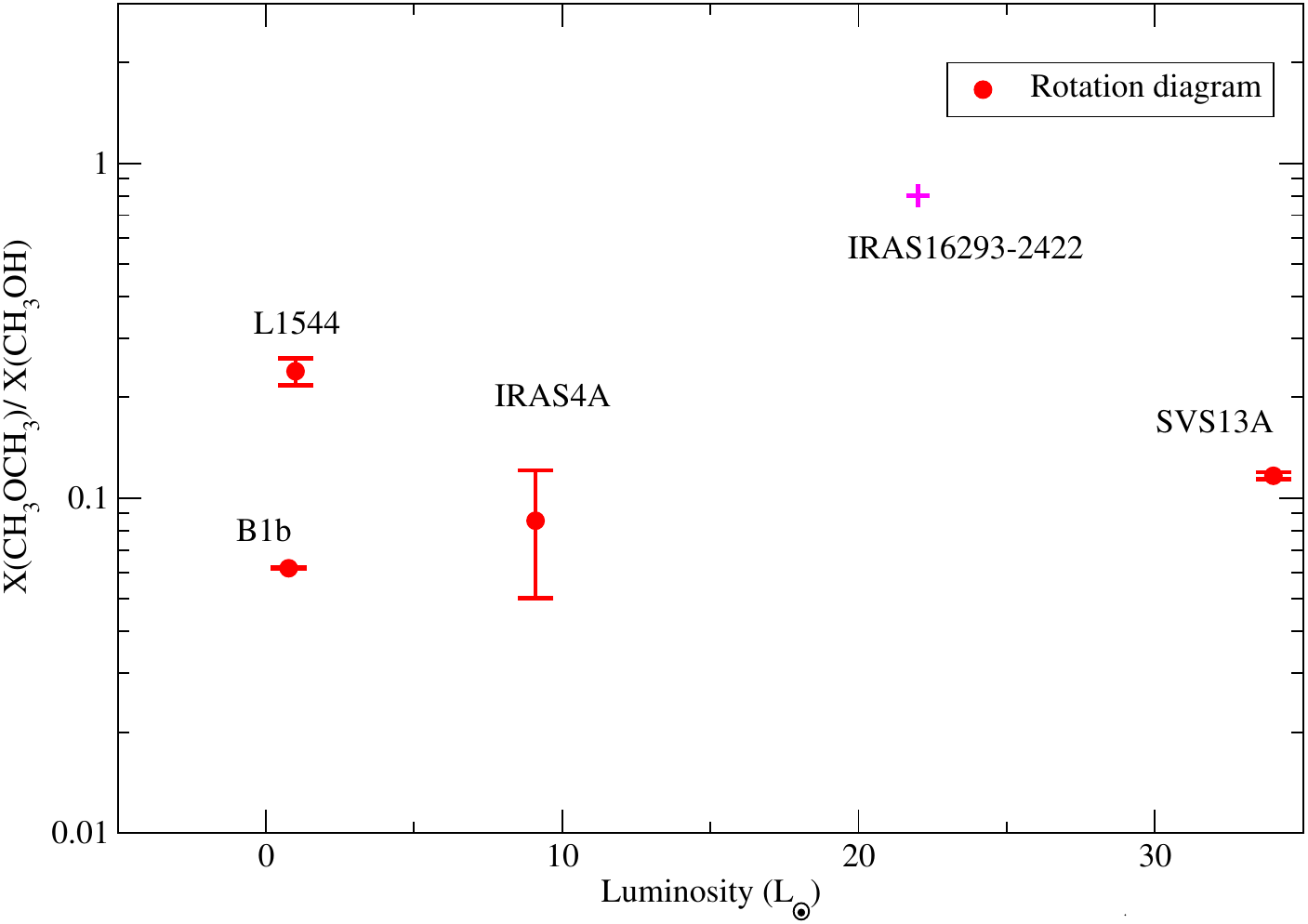}
\end{minipage}
\begin{minipage}{0.33\textwidth}
\includegraphics[width=\textwidth]{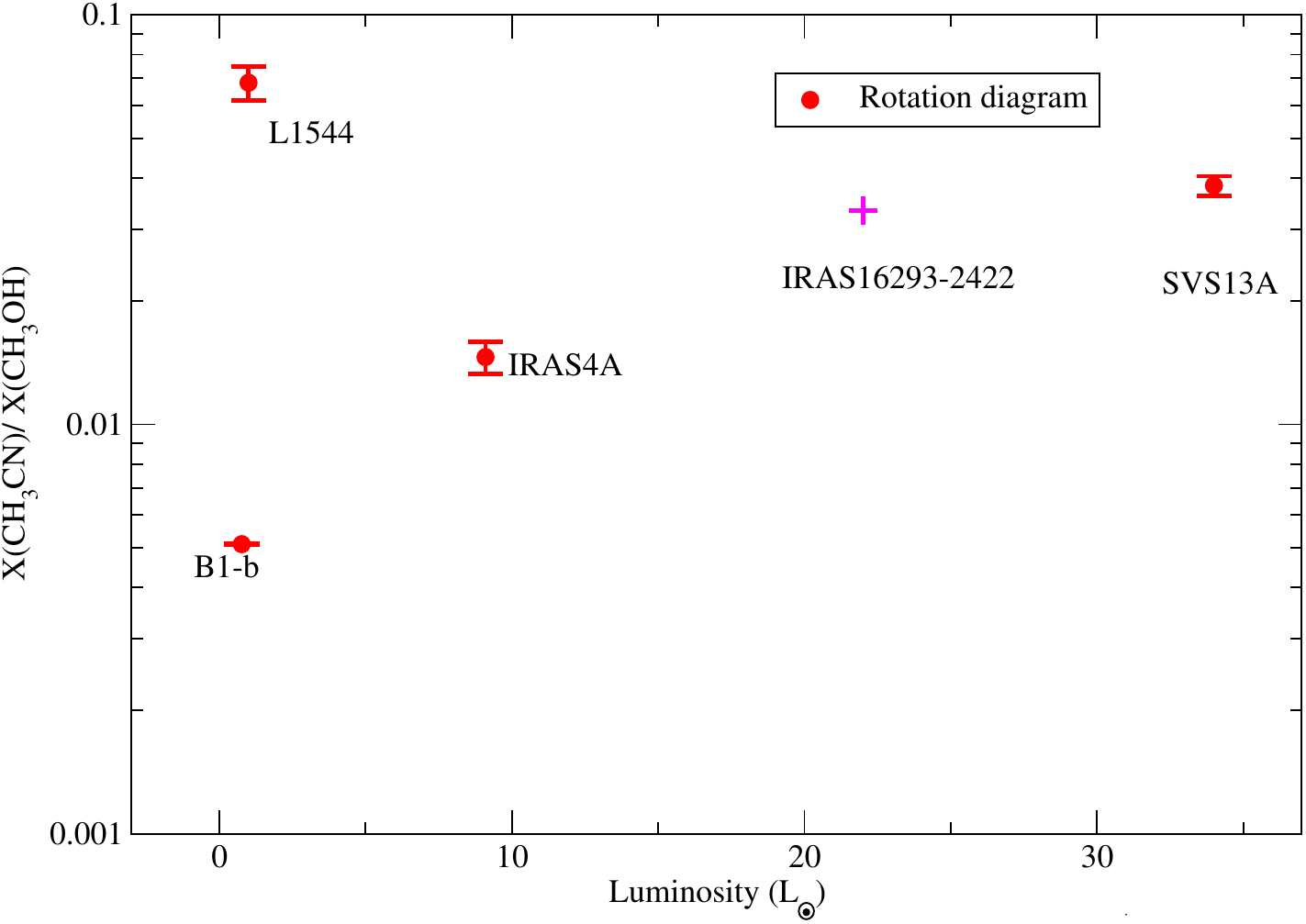}
\end{minipage}
\caption{The abundance ratios of CH$_3$CHO, CH$_3$OCHO, C$_2$H$_5$OH, HCCCHO, CH$_3$OCH$_3$ and CH$_3$CN w.r.t methanol (CH$_3$OH) plotted with source luminosity. The red circles represent the value obtained from the rotational diagram, and the blue circles represent the same obtained from the upper limits. The plus sign (magenta) represents the abundance obtained for IRAS4A 16293-2422 (22 L$_\odot$) taken from \cite{caza03}. Vertical lines represent the error bars.}
\label{fig:ratio_methanol}
\end{figure*}

\subsection{Luminosity effect \label{sec:luminosity}} 
Source luminosity or bolometric luminosities may indicate the different evolutionary stages of star formation \citep{myer98}. The luminosity considered for L1544, B1-b, IRAS4A, and SVS13A are 1 L$_{\odot}$, 0.77 L$_{\odot}$, 9.1 L$_{\odot}$, and 34 L$_{\odot}$, respectively \citep{doty05,lefl18}.
Since the luminosity of SVS13A is much higher than the other sources considered here, we consider another object, IRAS16293-2422, lying between the IRAS4A and SVS13A to infer the trend related to the luminosity. IRAS16293-2422 is a solar-type Class 0 protostar in the eastern part of the $\rho$ Ophiuchi star-forming region at a distance of 120 pc. It has a bolometric luminosity of 22 L$_{\odot}$. We consider the obtained abundance of methanol, acetaldehyde, methyl formate, dimethyl ether and methyl cyanide for this source from \cite{caza03}. The abundances of acetaldehyde, methyl formate, and dimethyl ether for this source were derived from the IRAM 30 m data, whereas methanol is from the JCMT.
Figure \ref{fig:luminosity} depicts the luminosity effect on the molecules considered in this study.
The abundances obtained from rotational diagram analysis are shown with solid red circles, the MCMC method with solid blue circles, and the data obtained from \cite[IRAS 16293-2422]{caza03} with magenta crosses in Figure \ref{fig:luminosity}.  
The black solid squares are the values obtained by using the upper limit. 
It shows a similar behavior as discussed in Figure \ref{fig:clmdensity}.
In the case of IRAS 16293-2422, compared to another class 0 object, IRAS4A, a higher column density of these species was obtained by \cite{caza03}. It may be due to a higher luminosity ($\sim 22$ L$_{\odot}$) for IRAS 16293-2422 compared to IRAS4A ($\sim$ 9.1 L$_{\odot}$).

Since there are considerable uncertainties in deriving the abundances, it is beneficial to see the molecular ratio of some specific pairs to constrain our understanding better.
Multi-line observations of methanol and several COMs toward the two low-mass protostars NGC 1333-IRAS2A and NGC 1333-IRAS4A with the PdBI at an angular resolution of 2 ${\arcsec}$ were carried out by \cite{taqu15}. They calculated the ratio in column density of CH$_3$OCH$_3$ to C$_2$H$_5$OH in low mass protostar IRAS4A is about 0.7 considering the source size 0.5 $\arcsec$. In our case, we are getting this ratio with a similar value of 0.7 from the rotation diagram analysis values of IRAS4A.
Figure \ref{fig:ratio_methanol} shows the abundance ratio with respect to methanol plotted with the luminosity of the sources. We did not find any specific trend by plotting the molecular ratio. But we found that  
methanol remains the most abundant among the species studied here (ratios are $<$1 except for HCCCHO). The first panel of Figure \ref{fig:ratio_methanol} displays the variation of the abundance ratio of acetaldehyde and methanol with luminosity. The ratio remains roughly invariant in IRAS4A, IRAS16293-2422 (class 0) and SVS13A (class I phase). The ratio between methyl formate and methanol in the second panel and between ethanol and methanol in the third panel of Figure \ref{fig:ratio_methanol} shows a similar nature.

\subsubsection{Interferometric observations} \label{sec:INO}
\begin{figure*}
    \centering
    \begin{minipage}{0.50\textwidth}
    \includegraphics[width=\textwidth]{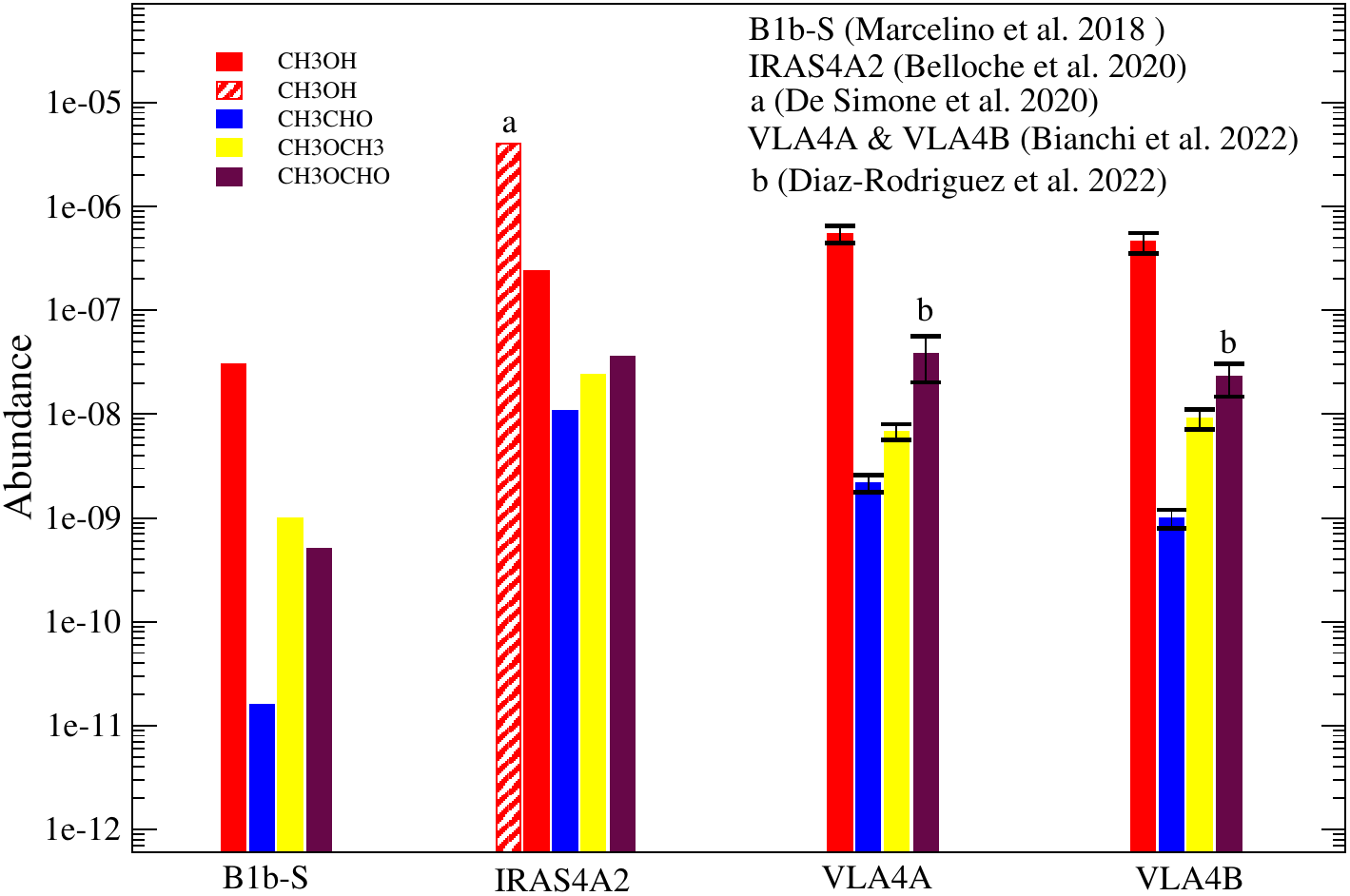}
    \end{minipage}
    \caption{Abundance variation of CH$_3$OH, CH$_3$CHO, CH$_3$OCH$_3$,and CH$_3$OCHO obtained from interferometric observation. Black vertical lines represent the error bar.}
    \label{fig:inter}
\end{figure*}

It is well known that the evolution of COMs is linked with various evolutionary stages of star formation. We use the large program ASAI data to examine this aspect. We also include the previous interferometric observation to consolidate the obtained trend. 
Figure \ref{fig:inter} shows the obtained abundance trend with the interferometric observation. In Table \ref{tab:inter}, we have noted the abundances and H$_2$ column density obtained with the interferometric observations.

There is considerable uncertainty in deriving the abundances of these molecules from the obtained column density. 
\cite{marc18b} presented the ALMA Band 6 spectral line observations at the angular resolution of $\sim 0.6$ $\arcsec$ towards B1-b. They extracted the spectra towards both protostars, B1b-S and B1b-N. However, it was found that B1b-S is rich in COMs, whereas B1b-N is free of COMs emission. They derived the column densities of  $^{13}$CH$_3$OH  for  source sizes of 0.35 $\arcsec$. We utilize a value of $^{12}$C/$^{13}$C = 60 in deriving the column density of $^{12}$CH$_3$OH from this. They also derived the column density of CH$_3$CHO, CH$_3$OCH$_3$, CH$_3$OCHO, and N(H$_2$) for a source size of 0.60 $\arcsec$.  
In the case of IRAS4A2, the column density is derived from the 0$\arcsec$.35 region \citep{bell20}. \cite{simo20} obtained a very high column density ($\sim 10^{19}$ cm$^{-2}$) for methanol with a source size of 0.24 $\arcsec$. In deriving the abundances of this region, we use a H$_2$ column density of 
$2.5 \times 10^{24}$ cm$^{-2}$
\citep{lope15} obtained from the 0.5 $\arcsec$ component.

For IRAS16293-2422 B the column density of these molecules are taken from \cite{jorg18} for a source size of 0.5$^{\arcsec}$. The column density of CH$_3$CN is taken from \cite{calc18}. For hydrogen column density, we used a value of $1.2\times10^{25}$ cm$^{-2}$ \citep{jorg16}.

For SVS13A, the column density of CH$_3$OH, CH$_3$CHO, and CH$_3$OCH$_3$ are taken from \cite{bian22}. They observed SVS13A with an angular resolution of 0.106 $\arcsec$.
The column density of CH$_3$OCHO is taken from  \cite{diaz22} with beam size regions 0.16$\arcsec$ $\times$ 0.08 $\arcsec$ toward each component. The H$_2$ column density of $\sim 1.1 \times 10^{25}$ cm$^{-2}$ is taken from  \cite{lope15} for compact component size of 1.0 $\arcsec$.

Figure \ref{fig:inter} clearly shows that the general trend (abundances gradually increased up to the class 0 stage and then decreased for the class I phase) obtained for the ASAI survey was also obtained with the interferometric observations. 

\begin{table*}
   \centering
   \caption{Column density of observed molecules obtained from interferometric observations. \label{tab:inter}}
    \begin{tabular}{|c|c|c|c|c|c|}
    \hline
         Sources&\multicolumn{4}{|c|}{Column density [abundance] in the form of \rm{a(b) = a $\times$ 10$^b$}}&N(H$_2$)\\
\cline{2-5}
&CH$_3$OH&CH$_3$CHO&CH$_3$OCH$_3$&CH$_3$OCHO&(cm$^{-2}$)\\
\hline
B1-b-S&3.0(17)$^a$[3.0(-8)]&1.6(14)$^a$[1.6(-11)]&1.0(16)$^a$[1.0(-9)]&5.0(15)$^a$[5.0(-10)]&1.1(25)$^a$\\
\hline
IRAS4A2&6.0(17)$^b$[2.4(-7)]/&2.7(16)$^b$[1.1(-8)]&6.0(16)$^b$[2.4(-8)]&8.9(16)$^b$[3.5(-8)]&2.5(24)$^f$\\
&1.0(19)$^c$[4.0(-6)]&&&&\\
\hline
IRAS16293-2422 B&1.0(19)$^g$[8.3(-7)]&1.2(17)$^g$[1.0(-8)]&2.4(17)$^g$[2.0(-8)]&2.6(17)$^g$[2.2(-8)]&1.2(25)$^h$\\
\hline
VLA4A&6.0(18)$^d$[5.5(-7)]&2.4(16)$^d$[2.2(-9)]&7.5(16)$^d$[6.8(-9)]&4.2(17)$^e$[3.8(-8)]&1.0(25)$^f$\\
\hline
VLA4B&5.0(18)$^d$[4.5(-7)]&1.1(16)$^d$[1.0(-9)]&1.0(17)$^d$[9.0(-9)]&2.5(17)$^e$[2.3(-8)]&1.0(25)$^f$\\
\hline
    \end{tabular}
    \\
    {\scriptsize \noindent $^a$\citep{marc18b},$^b$\citep{bell20},$^c$\citep{simo20},$^d$\citep{bian22},$^e$\citep{diaz22},$^f$\citep{lope15},$^g$\cite{jorg18},$^h$\cite{jorg16}}
\end{table*}

\section{Conclusion} \label{sec:conclusion}
Here, we analyze the ASAI large program data for five sources to understand the chemical and physical evolution of solar-type star-forming regions. We identify $\rm{CH_3OH, \ CH_3CHO, \ CH_3OCHO, \ C_2H_5OH, \ HCCCHO}$, $\rm{CH_3OCH_3}$, and CH$_3$CN in some of these sources we have considered. Observing these complex organic molecules in these four sources gave us a transparent view of evolutionary inheritance. The following are our initial findings in this article:

$\bullet$ An extensive study of the ASAI large program samples identifies many species with different transitions. We employ various LTE methods to measure excitation temperature and column density of a species. For the very first time in B1-b, we tentatively identify some transitions of $\rm{C_2H_5OH}$ with an upper limit of column density $1.0 \times 10^{13}$ cm$^{-2}$.

$\bullet$ We noticed a comparatively enhanced column density of these species at the first hydrostatic core phase compared to the prestellar core. Moreover, no significant difference between class 0 (IRAS4A) and class I phase (SVS13A) has been observed. The two-class 0 objects IRAS4A and  IRAS 16293-2422 show a significant difference (possibly because of the noticeable difference in luminosity). 
Abundances of the COMs gradually increase from the L1544 to IRAS16293-2422 and then decrease for SVS13A (Figure \ref{fig:clmdensity}), except HCCCHO (upper limit). A similar trend is obtained by considering the beam dilution effect (Figure \ref{fig:beamdil}) and interferometric data from the literature (Figure \ref{fig:inter}).

$\bullet$ 
We found a trend in the obtained FWHM for a particular transition of methanol. It seems to increase steadily from L1544 (prestellar core) to SVS13A (class I). It also indicates a steady increase of FWHM with the luminosity of the sources. 

Although the comparison focuses on only one source of each category, we cannot draw a clear conclusion about the observed trend. More observations of various sources in different stages of low-mass star formation with high spatial and angular resolution are needed to draw a reliable picture of the chemical evolution during the solar-type star formation process.

\acknowledgments
{This paper is based on the observations carried out as part of the Large Program ASAI (project number 012-12) using IRAM 30 m telescope. IRAM is supported by INSU/CNRS (France), MPG (Germany), and IGN (spain). BB gratefully acknowledges the DST, India's Government, for providing financial assistance through the DST-INSPIRE Fellowship [IF170046] scheme. S.K.M.  acknowledges Institute of Astronomy Space and Earth Science for carrying out part of this work. R.G acknowledges support from the Higher Education Department of the Government of West Bengal. P.G. acknowledges support from a Chalmers Cosmic Origins postdoctoral fellowship. This work was supported by the India-Japan Science Cooperative Program between DST and JSPS, grant No. JPJSBP120207703 and DST/INT/JSPS/P-319/2020. T.S. acknowledges the support from JSPS KAKENHI grant Nos. JP20H05845A and JP21H01145.}


\clearpage
\appendix

\restartappendixnumbering



\section{Observed transitions}
\begin{table*}
\centering
\tiny{
\caption{Observed transitions toward some sources.}\label{tab:observation_1}
\begin{tabular}{|l|l|l|l|l|l|l|l|l|l|l|}
\hline
Species&Tag (Database)&Source&Frequency&E$_{up}$&Quantum No.&A$_{ij}$&V$_{LSR}$&T$_{peak}$&FWHM&$\int$T$_{mb}$.dv\\
&&&(GHz)&(K)&&(s$^{-1}$)&(km.s$^{-1}$)&(K)&(km.s$^{-1}$)&(K.km.s$^{-1}$)\\
\hline
CH$_3$OH&32003(JPL)&L1544&84.521172&40.4&$5_{-1,0}$ - $4_{0,0}$&	$1.97\times10^{-6}$&7.240$\pm$     0.010&   0.017$\pm$     0.001&   0.441$\pm$     0.024&   0.008$\pm$     0.001\\
(Methanol)
&&&96.744545&20.1&$2_{0,0}$ - $1_{0,0}$&$3.41\times10^{-6}$&7.160$\pm$     0.020&   0.109$\pm$     0.001&   0.358$\pm$     0.004&   0.042$\pm$     0.001\\
&&&96.755501&28.0&$2_{1,0}$ - $1_{1,0}$&$2.62\times10^{-6}$&7.070$\pm$     0.010&   0.013$\pm$     0.001&   0.416$\pm$     0.029&   0.006$\pm$     0.001 \\
&&&97.582798&21.5&$2_{1,-0}$ - $1_{1,-0}$&$2.6\times10^{-6}$&7.180$\pm$     0.010&   0.019$\pm$     0.001&   0.428$\pm$     0.021&   0.009$\pm$     0.001\\
\cline{3-11}
&&B1-b&96.739358&12.5&$2_{-1,0}$ - $1_{-1,0}$&$2.56\times10^{-6}$&6.641$\pm$     0.002&   1.243$\pm$     0.003&   1.355$\pm$     0.005&   1.793$\pm$     0.010\\
&&&	96.741371&6.9&$2_{0,+0}$ - $1_{0,+0}$&$3.41\times10^{-6}$&6.615$\pm$     0.001&   1.694$\pm$     0.003&   1.365$\pm$     0.003&   2.462$\pm$     0.010\\
&&&	96.744545&20.1&	$2_{0,0}$ - $1_{0,0}$&$3.41\times10^{-6}$&6.606$\pm$     0.007&   0.306$\pm$     0.003&   1.365$\pm$     0.018&   0.445$\pm$     0.010\\
&&&	96.755501&28.0&	$2_{1,0}$ - $1_{1,0}$&$2.62\times10^{-6}$&6.569$\pm$     0.033&   0.066$\pm$     0.003&   1.310$\pm$     0.084&   0.092$\pm$     0.010\\
&&&	108.893945&13.1&$0_{0,0}$ - $0_{1,-1}$ &$1.47\times10^{-5}$&6.565$\pm$0.005&   0.377$\pm$     0.003&   1.371$\pm$     0.013&   0.550$\pm$     0.009\\
&&&	145.093754&27.0&$3_{0,0}$ - $2_{0,0}$&$1.23\times10^{-5}$&6.671$\pm$     0.003&   0.479$\pm$     0.003&   0.950 $\pm$    0.008&   0.484$\pm$     0.007\\
&&&	145.103185&13.9&$3_{0,+0}$ - $2_{0,+0}$ &$1.23\times10^{-5}$&6.718$\pm$     0.001&   2.316$\pm$     0.003&   1.005$\pm$     0.002&   2.478$\pm$     0.007\\
&&&	157.270832&15.4&$1_{0,0}$ - $1_{-1,0}$&$2.21\times10^{-5}$&6.513$\pm$0.002&   0.896$\pm$     0.003&   1.012$\pm$  0.004&   0.965$\pm$     0.007\\
&&&	157.276019&20.1&$2_{0,0}$ - $2_{-1,0}$&$2.18\times10^{-5}$&6.533$\pm$0.002&   0.583$\pm$     0.003&   0.978$\pm$     0.006&   0.607$\pm$     0.006\\
&&&	165.050175&23.3&$1_{1,0}$ - $1_{0,0}$&$2.35\times10^{-5}$&6.476$\pm$0.004&   0.334$\pm$     0.003&   0.878$\pm$     0.010&   0.312$\pm$     0.006\\
&&&	165.061130&28.0&$2_{1,0}$ - $2_{0,0}$&$2.34\times10^{-5}$&6.474$\pm$0.005&   0.314$\pm$     0.003&   0.905$\pm$     0.011&   0.302$\pm$     0.006\\
&&&	165.099240&34.9&$3_{1,0}$ - $3_{0,0}$&$2.33\times10^{-5}$&6.524$\pm$0.009&   0.189$\pm$     0.002&   1.201$\pm$     0.023&   0.242$\pm$     0.008\\
&&&	170.060592&36.2&$3_{2,0}$ - $2_{1,0}$&2.55$\times10^{-5}$&6.683$\pm$     0.003&   0.510$\pm$     0.003&   0.742$\pm$     0.006&   0.403$\pm$     0.006\\
&&&	213.427061&23.3&$1_{1,0}$ - $0_{0,0}$&$3.37\times10^{-5}$&6.580$\pm$     0.002&   0.354$\pm$     0.002&   0.903$\pm$     0.005&   0.341$\pm$     0.004\\
&&&	254.015377&20.0&$2_{0,0}$ - $1_{-1,0}$&$1.90\times10^{-5}$&6.621$\pm$     0.002&   0.427$\pm$     0.001&   1.035$\pm$     0.005&   0.471$\pm$     0.004\\
&&&	261.805675&28.0&$2_{1,0}$ - $1_{0,0}$&$5.57\times10^{-5}$&6.623$\pm$     0.002&   0.354$\pm$    0.001&   1.005$\pm$     0.005&   0.379$\pm$     0.003\\
\cline{3-11}
&&IRAS4A&96.755501	&28.0	&	$2_{1,0}$ - $1_{1,0}$&$2.62\times10^{-6}$&7.170$\pm$     0.029&   0.143$\pm$     0.002&   2.600$\pm$     0.118&   0.397$\pm$     0.025\\
&&&108.893945	&13.1	&	$0_{0,0}$ - $1_{-1,0}$&	$1.47\times10^{-6}$&7.402$\pm$     0.011&   0.335$\pm$     0.002&   2.446$\pm$     0.046&   0.871$\pm$     0.022\\
&&&143.865795	&28.3	&	$3_{1,+0}$ - $2_{1,+0}$&$1.07\times10^{-5}$&7.279$\pm$     0.010&   0.313$\pm$     0.002&   2.697$\pm$     0.042&   0.899$\pm$     0.019\\
&&&157.178987	&47.9	&	$5_{0,0}$ - $5_{-1,0}$&	$2.04\times10^{-5}$&7.312$\pm$     0.009&   0.286$\pm$     0.002&   2.352$\pm$     0.034&   0.717$\pm$     0.014\\
&&&165.050175	&23.4	&	$1_{1,0}$ - $1_{0,0}$&	$2.35\times10^{-5}$&7.335$\pm$     0.008&   0.263$\pm$     0.002&   2.259$\pm$     0.028&   0.632$\pm$     0.012\\
&&&165.061130&28.0	&	$2_{1,0}$ - $2_{0,0}$&	$2.34\times10^{-5}$&7.242$\pm$     0.008&   0.319$\pm$     0.002&   2.547$\pm$     0.027&   0.865$\pm$     0.013\\
&&&165.099240&35.0	&	$3_{1,0}$ - $3_{0,0}$&	$2.33\times10^{-5}$&7.195$\pm$     0.007&   0.330$\pm$     0.002&   2.553$\pm$     0.026&   0.897$\pm$     0.013\\
&&&213.427061	&23.4	&	$1_{1,0}$ - $0_{0,0}$&	$3.37\times10^{-5}$&7.472$\pm$     0.007&   0.313$\pm$     0.001&   2.254$\pm$     0.023&   0.752$\pm$     0.011\\
&&&230.027047	&39.8	&	$3_{-2,0}$ - $4_{-1,0}$&$1.49\times10^{-5}$&7.345$\pm$     0.019&   0.095$\pm$     0.001&   2.316$\pm$     0.063&   0.233$\pm$     0.010\\
&&&254.015377	&20.1	&	$2_{0,0}$ - $1_{-1,0}$&$1.90\times10^{-5}$&7.073$\pm$     0.005&   0.329$\pm$     0.001&   2.030$\pm$     0.015&   0.711$\pm$     0.008\\
&&&261.805675	&28.0	&	$2_{1,0}$ - $1_{0,0}$&$5.57\times10^{-5}$&7.095$\pm$     0.004&   0.420$\pm$     0.001&   2.275$\pm$     0.014&   1.016$\pm$     0.009	\\
&&&155.320895	&140.6	&	$10_{0,0}$ - $10_{-1,0}$&$1.55\times10^{-5}$&7.289$\pm$     0.077&   0.047$\pm$     0.002&   2.883$\pm$     0.349&   0.145$\pm$     0.022\\
&&&155.997524	&117.5	&	$9_{0,0}$ - $9_{-1,0}$&	$1.67\times10^{-5}$&7.355$\pm$     0.076&   0.060$\pm$     0.002&   4.328$\pm$     0.278&   0.278$\pm$     0.025\\
&&&156.488902	&96.6	&	$8_{8,0}$ - $8_{-1,0}$&	$1.78\times10^{-5}$&7.487$\pm$     0.066&   0.093$\pm$     0.001&   3.158$\pm$     0.253&   0.313$\pm$     0.030\\
&&&156.828517	&78.1	&	$7_{0,0}$ - $7_{-1,0}$&	$1.88\times10^{-5}$&7.305$\pm$     0.031&   0.125$\pm$     0.002&   3.467$\pm$     0.163&   0.460$\pm$     0.027\\
&&&157.048617	&61.8	&	$6_{0,0}$ - $6_{-1,0}$&	$1.96\times10^{-5}$&7.331$\pm$     0.016&   0.192$\pm$     0.002&   2.790$\pm$     0.071&   0.571$\pm$     0.019\\
&&&165.678649	&69.8	&	$6_{1,0}$ - $6_{0,0}$&	$2.30\times10^{-5}$&7.225$\pm$     0.019&   0.148$\pm$     0.002&   2.846$\pm$     0.087&   0.448$\pm$     0.018\\
&&&166.169098	&86.1	&	$7_{1,0}$ - $7_{0,0}$&	$2.28\times10^{-5}$&7.059$\pm$     0.035&   0.119$\pm$     0.001&   3.225$\pm$     0.170&   0.410$\pm$     0.027\\
&&&241.879025	&55.9	&	$5_{1,0}$ - $4_{1,0}$&$5.96\times10^{-5}$&7.350$\pm$     0.006&   0.311$\pm$     0.001&   2.347$\pm$     0.023&   0.777$\pm$     0.011\\
&&&251.738437	&98.5	&	$6_{3,-0}$ - $6_{2,+0}$&$7.46\times10^{-5}$&6.985$\pm$     0.025&   0.113$\pm$     0.001&   3.078$\pm$     0.110&   0.369$\pm$     0.017\\
&&&251.866524	&73.0	&	$4_{3,-0}$ - $4_{2,+0}$&$6.10\times10^{-5}$&6.926$\pm$     0.026&   0.135$\pm$     0.001&   3.119$\pm$     0.090&   0.447$\pm$     0.017\\
&&&251.917065	&63.7	&	$3_{3,+0}$ - $3_{2,-0}$&$4.36\times10^{-5}$&6.867$\pm$     0.032&   0.090$\pm$     0.001&   2.794$\pm$     0.106&   0.267$\pm$     0.014\\
&&&251.984837	&133.4	&	$8_{3,+0}$ - $8_{2,-0}$&$7.99\times10^{-5}$&7.020$\pm$     0.030&   0.073$\pm$     0.001&   2.692$\pm$     0.105&   0.209$\pm$     0.012\\
&&&252.090409	&154.2	&	$9_{3,+0}$ - $9_{2,-0}$&$8.15\times10^{-5}$&6.910$\pm$     0.025&   0.074$\pm$     0.001&   2.341$\pm$     0.080&   0.186$\pm$     0.010\\
\cline{3-11}
&&SVS13A&85.568131&	74.7    &	$6_{-2,0}$ - $7_{-1,0}$&$1.13\times10^{-6}$&8.395$\pm$     0.341&   0.018$\pm$     0.002&   4.570$\pm$     1.399&   0.088$\pm$     0.037\\
&&&	96.755501	&	28.0    &	$2_{1,0}$ - $1_{1,0}$&$2.62\times10^{-6}$&9.336$\pm$     0.460&   0.026$\pm$     0.002&   3.163$\pm$     1.026&   0.088$\pm$     0.036\\
&&&	111.289453	&	102.7	&	$7_{2,+0}$ - $8_{1,+0}$&$2.60\times10^{-6}$&7.288$\pm$     0.078&   0.035$\pm$     0.002&   2.346$\pm$     0.213&   0.088$\pm$     0.013\\
&&&	143.865795	&	28.3	&	$3_{1,+0}$ - $2_{1,+0}$&$1.07\times10^{-5}$&8.528$\pm$     0.022&   0.098$\pm$     0.002&   2.101$\pm$     0.069&   0.219$\pm$     0.011\\
&&&	156.602395	&	21.4	&	$2_{1,+0}$ - $3_{0,+0}$&$8.93\times10^{-5}$&8.250$\pm$     0.032&   0.094$\pm$     0.002&   2.663$\pm$     0.118&   0.267$\pm$     0.016\\
&&&	218.440063	&	45.5	&	$4_{2,0}$ - $3_{1,0}$&$4.69\times10^{-5}$&8.513$\pm$     0.006&   0.246$\pm$     0.001&   1.810$\pm$     0.018&   0.473$\pm$     0.008 \\
&&&	229.758756	&	89.1	&	$8_{-1,0}$ - $7_{0,0}$&$4.19\times10^{-5}$&8.146$\pm$     0.025&   0.155$\pm$     0.001&   3.380$\pm$     0.092&   0.557$\pm$     0.020\\
&&&	241.700159	&	47.9	&	$5_{0,0}$ - $4_{0,0}$&$6.04\times10^{-5}$&8.086$\pm$     0.017&   0.184$\pm$     0.001&   2.744$\pm$     0.053&   0.538$\pm$     0.014\\
&&&	241.791352	&	34.8	&	$5_{0,+0}$ - $4_{0,+0}$&$6.05\times10^{-5}$&8.169$\pm$     0.004&   0.430$\pm$     0.001&   1.680$\pm$     0.010&   0.768$\pm$     0.007\\
&&&	241.879025	&	55.9	&	$5_{1,0}$ - $4_{1,0}$&$5.96\times10^{-5}$&8.051$\pm$     0.029&   0.166$\pm$     0.001&   3.094$\pm$     0.085&   0.547$\pm$     0.019\\
&&&	243.915788	&	49.7	&	$5_{1,-0}$ - $4_{1,-0}$&$5.97\times10^{-5}$&8.444$\pm$     0.012&   0.163$\pm$     0.001&   2.659$\pm$     0.043&   0.461$\pm$     0.011\\
&&&	251.738437	&	98.5	&	$6_{3,-0}$ - $6_{0,+0}$&$7.46\times10^{-5}$&8.463$\pm$     0.037&   0.089$\pm$     0.001&   3.380$\pm$     0.197&   0.320$\pm$     0.023\\
&&&	261.805675	&	28.0	&	$2_{1,0}$ - $1_{0,0}$&$5.57\times10^{-5}$&8.915$\pm$     0.009&   0.154$\pm$     0.001&   1.691$\pm$     0.024&   0.277$\pm$     0.006\\
&&&	266.838148	&	57.1	&$5_{2,0}$ - $4_{1,0}$&$7.74\times10^{-5}$&8.357$\pm$     0.009&   0.197$\pm$     0.001&   2.003$\pm$     0.028&   0.420$\pm$     0.009\\
\hline
\end{tabular}}
\end{table*}

\begin{table*}
\centering
\tiny{
\caption{Observed transitions toward some sources. \label{tab:observation_2}}
\begin{tabular}{|l|l|l|l|l|l|l|l|l|l|l|}
\hline
Species&Tag (Database)&Source&Frequency&E$_{up}$&Quantum No.&A$_{ij}$&V$_{LSR}$&T$_{peak}$&FWHM&$\int$T$_{mb}$.dv\\
&&&(GHz)&(K)&&(s$^{-1}$)&(km.s$^{-1}$)&(K)&(km.s$^{-1}$)&(K.km.s$^{-1}$)\\
\hline
CH$_3$CHO&44003 (JPL)&L1544&93.580909&15.7&5$_{1,5}$ - 4$_{1,4}$, (A)&$2.53\times10^{-5}$&7.140$\pm$0.009&0.022$\pm$0.001&0.438$\pm$0.02&0.010$\pm$0.001\\
(Acetaldehyde)&&&93.595234&15.8&5$_{1,5}$ - 4$_{1,4}$, (E)&$2.53\times10^{-5}$&7.190$\pm$0.007&0.024$\pm$0.001&0.342$\pm$0.017&0.009$\pm$0.001\\
&&&95.947437&13.9&5$_{0,5}$ - 4$_{0,4}$, (E)&$2.84\times10^{-5}$&7.160$\pm$0.005&0.039$\pm$0.001&0.346$\pm$0.011&0.014$\pm$0.001\\
&&&95.963458&13.8&5$_{0,5}$ - 4$_{0,4}$, (A)&$2.84\times10^{-5}$&7.180$\pm$0.006&0.041$\pm$0.001&0.437$\pm$0.013&0.019$\pm$0.001\\
&&&98.863313&16.6&5$_{1,4}$ - 4$_{1,3}$, (E)&$2.99\times10^{-5}$&7.110$\pm$0.008&0.025$\pm$0.001&0.527$\pm$0.019&0.014$\pm$0.001\\
&&&98.900944&16.5&5$_{1,4}$ - 4$_{1,3}$, (A)&$2.99\times10^{-5}$&7.170$\pm$0.008&0.02$\pm$0.001&0.34$\pm$0.02&0.007$\pm$0.001\\
\cline{3-11}
&&B1-b&93.580909&15.7&5$_{1,5}$ - 4$_{1,4}$, (A)&$2.53\times10^{-5}$&6.570$\pm$0.040&0.051$\pm$0.003&1.428$\pm$0.110&0.078$\pm$0.010\\
&&&93.595234&15.8&5$_{1,5}$ - 4$_{1,4}$, (E)&$2.53\times10^{-5}$&6.580$\pm$0.040&0.050$\pm$0.003&1.568$\pm$0.144&0.084$\pm$0.012\\
&&&95.963458&13.8&5$_{0,5}$ - 4$_{0,4}$, (A)&$2.84\times10^{-5}$&6.660$\pm$0.030&0.073$\pm$0.003&1.461$\pm$0.080&0.114$\pm$0.011\\
&&&96.425614&22.9&5$_{2,4}$ - 4$_{2,3}$, (E)&$2.41\times10^{-5}$&6.630$\pm$0.104&0.023$\pm$0.003&1.661$\pm$0.297&0.041$\pm$0.012\\
&&&96.475523&23.0&5$_{2,3}$ - 4$_{2,2}$, (E)&$2.42\times10^{-5}$&6.520$\pm$0.083&0.022$\pm$0.003&0.973$\pm$0.199&0.022$\pm$0.008\\
&&&98.863313&16.6&5$_{1,4}$ - 4$_{1,3}$, (E)&$2.99\times10^{-5}$&6.510$\pm$0.034&0.054$\pm$0.003&1.040$\pm$0.081&0.059$\pm$0.008\\
&&&98.900944&16.5&5$_{1,4}$ - 4$_{1,3}$, (A)&$2.99\times10^{-5}$&6.550$\pm$0.040&0.049$\pm$0.003&1.249$\pm$0.100&0.066$\pm$0.009\\
&&&138.284995&28.9&7$_{1,6}$ - 6$_{1,5}$, (E)&$5.57\times10^{-5}$&6.820$\pm$0.027&0.054$\pm$0.003&0.870$\pm$0.063&0.050$\pm$0.006\\
&&&138.319628&28.8&7$_{1,6}$ - 6$_{1,5}$, (A)&$8.57\times10^{-5}$&6.650$\pm$0.031&0.048$\pm$0.003&0.775$\pm$0.072&0.039$\pm$0.006\\
&&&152.635193&33.1&8$_{0,8}$ - 7$_{0,7}$, (A)&$1.18\times10^{-4}$&6.280$\pm$0.049&0.034$\pm$0.003&0.750$\pm$0.116&0.027$\pm$0.006\\
\cline{3-11}
&&IRAS4A&74.891677&11.2&4$_{1,4}$ - 3$_{1,3}$, (A)&$1.24\times10^{-5}$&6.840$\pm$0.050&0.051$\pm$0.003&   2.641$\pm$0.216&0.144$\pm$0.021\\
&&&74.924133&11.3&4$_{1,4}$ - 3$_{1,3}$, (E)&$1.24\times10^{-5}$&7.030$\pm$0.070&0.050$\pm$0.003&2.375$\pm$0.225&   0.126$\pm$0.019\\
&&&76.878952&9.2&4$_{0,4}$ - 3$_{0,3}$, (A)&$1.43\times10^{-5}$&7.140$\pm$0.060&0.061$\pm$     0.003&   2.332$\pm$     0.185&   0.152$\pm$     0.019\\
&&&77.038601&18.3&4$_{2,3}$ - 3$_{2,2}$, (A)&$1.08\times10^{-5}$&6.960$\pm$0.190&0.022$\pm$     0.003&   2.653$\pm$     0.507&   0.063$\pm$     0.021\\
&&&77.218291&18.3&4$_{2,2}$ - 3$_{2,1}$, (A)&$1.09\times10^{-5}$&6.510$\pm$0.150&0.029$\pm$     0.003&   2.708$\pm$     0.408&   0.082$\pm$     0.021\\
&&&79.099313&11.8&4$_{1,3}$ - 3$_{1,2}$, (E)&$1.46\times10^{-5}$&7.010$\pm$0.070&0.059$\pm$     0.003&   2.299$\pm$     0.207&   0.144$\pm$     0.020\\
&&&93.580909&15.7&5$_{1,5}$ - 4$_{1,4}$, (A)&$2.53\times10^{-5}$&7.320$\pm$0.050&0.065$\pm$     0.003&   2.298$\pm$     0.143&   0.158$\pm$     0.016\\
&&&93.595234&15.8&5$_{1,5}$ - 4$_{1,4}$, (E)&$2.53\times10^{-5}$&7.440$\pm$0.050&0.062$\pm$     0.003&   2.343$\pm$     0.159&   0.155$\pm$     0.017\\
&&&95.947437&13.9&5$_{0,5}$ - 4$_{0,4}$, (E)&$2.84\times10^{-5}$&7.390$\pm$0.030&0.081$\pm$     0.003&   2.277$\pm$     0.102&   0.197$\pm$     0.016\\
&&&96.274252&22.9&5$_{2,4}$ - 4$_{2,3}$, (A)&$2.41\times10^{-5}$&7.330$\pm$0.070&0.037$\pm$     0.003&   2.148$\pm$     0.222&   0.085$\pm$     0.015\\
&&&96.425614&22.9&5$_{2,4}$ - 4$_{2,3}$, (E)&$2.41\times10^{-5}$&7.540$\pm$0.100&0.031$\pm$     0.003&   2.406$\pm$     0.315&   0.078$\pm$     0.017\\
&&&96.475523&23.0&5$_{2,3}$ - 4$_{2,2}$, (E)&$2.42\times10^{-5}$&7.170$\pm$0.090&0.036$\pm$     0.003&   2.357$\pm$     0.248&   0.092$\pm$     0.016\\
&&&98.863313&16.6&5$_{1,4}$ - 4$_{1,3}$, (E)&$2.99\times10^{-5}$&7.360$\pm$0.040&0.076$\pm$     0.003&   2.242$\pm$     0.115&   0.182$\pm$     0.016\\
&&&112.248716&21.1&6$_{1,6}$ - 5$_{1,5}$, (A)&$4.50\times10^{-5}$&7.290$\pm$0.050&0.065$\pm$     0.002&   2.559$\pm$     0.148&   0.178$\pm$     0.017\\
&&&112.254508&21.2&6$_{1,6}$ - 5$_{1,5}$, (E)&$4.50\times10^{-5}$&7.270$\pm$0.060&0.062$\pm$     0.002&   3.061$\pm$     0.199&   0.203$\pm$     0.021\\
&&&133.830491&25.8&7$_{0,7}$ - 6$_{0,6}$, (E)&$7.91\times10^{-5}$&7.330$\pm$0.020&0.084$\pm$     0.002&   2.113$\pm$     0.068&   0.189$\pm$     0.011\\
&&&138.284995&28.9&7$_{1,6}$ - 6$_{1,5}$, (E)&$8.57\times10^{-5}$&7.660$\pm$0.030&0.067$\pm$     0.002&   2.314$\pm$     0.088&   0.165$\pm$     0.011\\
&&&138.319628&28.8&7$_{1,6}$ - 6$_{1,5}$, (A)&$8.57\times10^{-5}$&7.510$\pm$0.030&0.063$\pm$     0.002&   2.541$\pm$     0.102&   0.171$\pm$     0.012\\
&&&152.635193&33.1&8$_{0,8}$ - 7$_{0,7}$, (A)&$1.18\times10^{-4}$&7.480$\pm$0.020&0.077$\pm$     0.002&   2.105$\pm$     0.064&   0.173$\pm$     0.010\\
&&&155.342088&42.5&8$_{2,6}$ - 7$_{2,5}$, (A)&$1.17\times10^{-4}$&7.230$\pm$0.040&0.053$\pm$     0.002&   2.193$\pm$     0.105&   0.123$\pm$     0.010\\
&&&168.093444&42.6&9$_{1,9}$ - 8$_{1,8}$, (A)&$1.57\times10^{-4}$&7.390$\pm$0.030&0.069$\pm$     0.002&   2.622$\pm$     0.095&   0.192$\pm$     0.012\\
&&&208.228560&60.5&11$_{0,11}$ - 10$_{0,10}$, (E)&$3.05\times10^{-4}$&7.360$\pm$0.030&0.054$\pm$     0.002&   2.451$\pm$     0.099&   0.140$\pm$     0.010\\
&&&211.273792&70.0&11$_{2,10}$ - 10$_{2,9}$, (E)&$3.09\times10^{-4}$&7.410$\pm$0.020&0.063$\pm$0.001&1.645$\pm$0.077&   0.110$\pm$     0.008\\
&&&212.257139&81.4&11$_{3,9}$ - 10$_{3,8}$, (A)&$3.00\times10^{-4}$&7.530$\pm$0.020&0.043$\pm$     0.002&   1.306$\pm$     0.059&   0.060$\pm$     0.005\\
&&&214.845041&70.6&11$_{2,9}$ - 10$_{2,8}$, (A)&$3.25\times10^{-4}$&7.470$\pm$0.020&0.055$\pm$     0.001&   1.771$\pm$     0.080&   0.103$\pm$     0.007\\
&&&216.630234&64.8&11$_{1,10}$ - 10$_{1,9}$, (A)&$3.41\times10^{-4}$&7.470$\pm$0.020&0.070$\pm$     0.002&   1.409$\pm$     0.065&   0.105$\pm$     0.007\\
&&&223.660602&72.2&12$_{1,12}$ - 11$_{1,11}$, (A)&$3.77\times10^{-4}$&7.360$\pm$0.020&0.066$\pm$     0.002&   2.027$\pm$     0.061&   0.143$\pm$     0.008\\
&&&242.106020&83.9&13$_{1,13}$ - 12$_{1,12}$, (E)&$4.81\times10^{-4}$&7.130$\pm$0.040&0.051$\pm$     0.001&   2.123$\pm$     0.108&   0.116$\pm$     0.009\\
&&&250.829158&120.3&13$_{4,9}$ - 12$_{4,8}$, (E)&$4.87\times10^{-4}$&7.060$\pm$0.030&0.044$\pm$     0.002&   1.626$\pm$     0.092&   0.077$\pm$     0.007\\
&&&250.934558&104.6&13$_{3,11}$ - 12$_{3,10}$, (A)&$5.10\times10^{-4}$&7.180$\pm$0.030&0.042$\pm$     0.002&   1.497$\pm$     0.093&   0.067$\pm$     0.007\\
&&&251.489286&104.7&13$_{3,10}$ - 12$_{3,9}$, (A)&$5.14\times10^{-4}$&6.850$\pm$0.030&0.055$\pm$     0.001&   1.823$\pm$     0.079&   0.106$\pm$     0.007\\
&&&254.827152&94.1&13$_{2,11}$ - 12$_{2,10}$, (E)&$5.51\times10^{-4}$&7.260$\pm$0.050&0.041$\pm$     0.001&   2.027$\pm$     0.132&   0.089$\pm$     0.009\\
&&&255.326968&88.4&13$_{1,12}$ - 12$_{1,11}$, (E)&$5.64\times10^{-4}$&7.250$\pm$0.030&0.043$\pm$     0.001&   1.884$\pm$     0.115&   0.087$\pm$     0.008\\
&&&262.960097&95.7&14$_{0,14}$ - 13$_{0,13}$, (E)&$6.19\times10^{-4}$&6.730$\pm$0.030&0.056$\pm$     0.001&   1.770$\pm$     0.076&   0.105$\pm$     0.007\\
\cline{3-11}
&&SVS13A&205.170685&61.4&11$_{1,11}$ - 10$_{1,10}$, (A)&$2.90\times10^{-4}$&8.560$\pm$0.030&   0.039$\pm$0.002&   1.550$\pm$     0.094&   0.064$\pm$0.007\\
&&&211.243042&69.9&11$_{2,10}$ - 10$_{2,9}$, (A)&$3.09\times10^{-4}$&8.470$\pm$0.040&   0.040$\pm$     0.001&   1.972$\pm$     0.141&   0.083$\pm$     0.009\\
&&&211.273792&70.0&11$_{2,10}$ - 10$_{2,9}$, (E)&$3.09\times10^{-4}$&8.790$\pm$ 0.050&   0.040$\pm$     0.001&   1.925$\pm$     0.158&   0.081$\pm$     0.010\\
&&&216.581930&64.8&11$_{1,10}$ - 10$_{1,9}$, (E)&$3.41\times10^{-4}$&8.050$\pm$     0.050&   0.044$\pm$     0.001&   1.970$\pm$     0.149&   0.092$\pm$     0.010\\
&&&230.301923&81.0&12$_{2,11}$ - 11$_{2,10}$, (A)&$4.04\times10^{-4}$&8.550$\pm$ 0.030&   0.036$\pm$     0.001&   1.630$\pm$     0.104&   0.062$\pm$     0.007\\
&&&242.106020&83.9&13$_{1,13}$ - 12$_{1,12}$, (E)&$4.81\times10^{-4}$&8.080$\pm$     0.030&   0.053$\pm$     0.001&   1.574$\pm$     0.101&   0.090$\pm$     0.008\\
&&&251.489286&104.7&13$_{3,10}$ - 12$_{3,9}$, (A)&$5.14\times10^{-4}$&8.290$\pm$ 0.030&   0.030$\pm$     0.002&   1.022$\pm$     0.076&   0.032$\pm$     0.004\\
\hline
\hline
\end{tabular}}
\end{table*}
\begin{table*}
\centering
\tiny{
\caption{Observed transitions toward some sources. \label{tab:observation_3}}
\begin{tabular}{|l|l|l|l|l|l|l|l|l|l|l|}
\hline
Species&Tag (Database)&Source&Frequency&E$_{up}$&Quantum No.&A$_{ij}$&V$_{LSR}$&T$_{peak}$&FWHM&$\int$T$_{mb}$.dv\\
&&&(GHz)&(K)&&(s$^{-1}$)&(km.s$^{-1}$)&(K)&(km.s$^{-1}$)&(K.km.s$^{-1}$)\\
\hline
CH$_3$OCHO&60003 (JPL)&B1-b&88.851607&17.9&7$_{1,6}$ - $6_{1,5}$ (A)&$9.82\times10^{-6}$&6.560$\pm$0.217&0.013$\pm$0.003&1.915$\pm$0.635&0.026$\pm$0.014\\
(Methyl formate)&&&90.145723&19.7&$7_{2,5}$ - $6_{2,4}$ (E)&$9.74\times10^{-6}$&6.680$\pm$0.1806&0.013$\pm$0.003&1.585$\pm$0.490&0.022$\pm$0.011\\
&&&90.156473&19.6&$7_{2,5}$ - $6_{2,4}$ (A)&$9.75\times10^{-6}$&6.570$\pm$0.169&0.014$\pm$0.003&1.514$\pm$0.438&0.023 $\pm$0.011\\
&&&100.294604&27.4&$8_{3,5}$ - $7_{3,4}$ (E)&$1.26\times10^{-5}$&6.510$\pm$0.162&0.012$\pm$0.003&1.172$\pm$0.404&0.016$\pm$0.009\\
&&&100.482241&22.8&$8_{1,7}$ - $7_{1,6}$ (E)&$1.43\times10^{-5}$&6.730$\pm$0.131&0.018$\pm$0.003&1.331$\pm$0.348&0.026$\pm$0.011\\
&&&	103.478663&24.6&$8_{2,6}$ - $7_{2,5}$ (A)	&$1.52\times10^{-5}$&6.610$\pm$0.179&0.013$\pm$0.003&1.636$\pm$0.457&0.022$\pm$0.011\\
\cline{3-11}
&&IRAS4A&129.296357&36.4&$10_{2,8}$ - $9_{2,7}$ (E)&$3.06\times10^{-5}$&7.180$\pm$0.060&0.025$\pm$0.002&1.588$\pm$0.161&0.042$\pm$0.007\\
&&&132.928736&40.4&$11_{1,1}$ - $10_{1,9}$ (A)&$3.36\times10^{-5}$&7.000$\pm$0.140&0.016$\pm$0.001&2.210$\pm$0.429&0.039$\pm$0.011\\
&&&135.921949&55.6&$11_{5,7}$ - $10_{5,6}$ (A)&$2.95\times10^{-5}$&7.340$\pm$0.050&0.027$\pm$0.002&1.200$\pm$0.131&0.034$\pm$0.007\\
&&&141.044354&47.5&$12_{2,11}$ - $11_{2,10}$ (A)&	$4.02\times10^{-5}$&7.230$\pm$0.050&0.028$\pm$0.002&1.342$\pm$0.139&0.040$\pm$0.007\\
&&&158.693722&59.6&$13_{3,11}$ - $12_{3,10}$ (E)&$5.61\times10^{-5}$&7.430$\pm$0.090&0.022$\pm$0.002&2.402$\pm$0.306&0.057$\pm$0.011\\
&&&158.704392&59.6&$13_{3,11}$ - $12_{3,10}$ (A)&$5.61\times10^{-5}$&7.220$\pm$0.090&0.023$\pm$0.002&2.153$\pm$0.313&0.052$\pm$0.011\\
&&&200.956372&97.5&$16_{5,11}$ - $15_{5,10}$ (A)&$1.10\times10^{-4}$&7.030$\pm$0.020&0.045$\pm$0.002&1.110$\pm$0.065&0.054$\pm$0.005\\
&&&206.619476&89.2&$16_{3,13}$ - $15_{3,12}$ (A)&$1.28\times10^{-4}$&7.630$\pm$0.030&0.038$\pm$0.002&1.306$\pm$0.095&0.052$\pm$0.006\\
&&&216.216539&109.3&$19_{1,18}$ - $18_{1,17}$ (A)&$1.49\times10^{-4}$&7.460$\pm$0.030&0.030$\pm$0.002&1.322$\pm$0.094&0.042$\pm$0.005\\
&&&228.628876&118.8&$18_{5,13}$ - $17_{5,12}$ (E)&$1.66\times10^{-4}$&7.190$\pm$0.020&0.049$\pm$0.002&1.172$\pm$0.068&0.061$\pm$0.005\\
&&&240.021140&122.3	&$19_{3,16}$ - $18_{3,15}$ (E)&$2.01\times10^{-4}$& 7.140$\pm$0.032&0.036$\pm$0.002&1.325$\pm$0.078&0.051$\pm$0.005\\
&&&247.044146&139.9&$21_{3,19}$ - $20_{3,18}$ (E)&$2.21\times10^{-4}$&7.380$\pm$0.035&0.033$\pm$0.002&0.964$\pm$0.084&0.034$\pm$0.005\\
&&&249.578117&148.7&$20_{6,15}$ - $19_{6,14}$ (E)&$2.14\times10^{-4}$&7.230$\pm$0.032&0.031$\pm$0.002&1.224$\pm$0.075&0.041$\pm$0.005\\
\cline{3-11}
&&SVS13A&100.490682&22.7&$8_{1,7}$ - $7_{1,6}$ (A)&	$1.43\times10^{-5}$&8.830$\pm$0.230&0.015$\pm$0.002&2.474$\pm$0.779&0.040$\pm$0.018\\
&&&164.205978&64.9&$13_{4,9}$ - $12_{4,8}$ (E)&$5.98\times10^{-5}$&8.960$\pm$0.030&0.040$\pm$0.002&1.453$\pm$0.096&0.061$\pm$0.007\\
&&&210.463200&123.0&17$_{7,10}$ - $16_{7,9}$ (A)&$1.17\times10^{-4}$&8.750$\pm$0.030&0.030$\pm$0.002&1.350$\pm$0.092&0.043$\pm$0.005\\
&&&218.280900&99.7&$17_{3,14}$ - $16_{3,13}$ (E)&$1.51\times10^{-4}$&8.620$\pm$0.040&0.037$\pm$0.001&2.472$\pm$0.168&0.098$\pm$0.010\\
&&&222.421489&143.5&$18_{8,10}$ - $17_{8,9}$ (E)&$1.33\times10^{-4}$&8.630$\pm$0.030&0.037$\pm$0.001&1.996$\pm$0.109&0.079$\pm$0.007\\
&&&269.078001&168.8&$24_{2,23}$ - $23_{2,22}$ (E)&$2.90\times10^{-4}$&8.570$\pm$0.020&0.048$\pm$0.001&1.700$\pm$0.062&0.086$\pm$0.006\\
\hline
$\rm{C_2H_5OH}$ v=0&46524 (CDMS)&IRAS4A&129.6657634&23.8&$5_{3,2}$ - $5_{2,3}$,v$_t$=2 - 2&$1.14\times10^{-5}$&6.890$\pm$0.068&0.015$\pm$0.002& 1.140$\pm$0.172&0.019 $\pm$0.005\\
(Ethanol)&&&133.3234312&23.8&$7_{1,7}$ - $6_{0,6}$,v$_t$=2 - 2&$1.83\times10^{-5}$&7.620$\pm$0.080&0.026$\pm$0.002&2.042$\pm$0.231&0.056$\pm$0.010\\
&&&148.3040357&58.15&$11_{1,10}$ - $10_{2,9}$,v$_t$=2 - 2&$1.21\times10^{-5}$&7.580$\pm$0.060&0.017$\pm$0.002&0.981$\pm$0.165&0.018 $\pm$0.005\\
&&&	159.4140526&101.1&$9_{1,8}$ - $8_{1,7}$,v$_t$=1 - 1&$3.65\times10^{-5}$&7.230$\pm$0.050&0.021$\pm$0.002&0.696$\pm$0.125&0.016$\pm$0.005\\
&&&	205.4584717&64.0&$12_{1,12}$ - $11_{0,11}$,v$_t$=2 - 2&$8.04\times10^{-5}$&7.510$\pm$0.050&0.031$\pm$0.002&1.663$\pm$0.132&0.054$\pm$0.007\\
&&&	209.8652009&132.9&$12_{3,9}$ - $11_{3,8}$,v$_t$=0 - 0&$8.01\times10^{-5}$&7.170$\pm$0.030&0.023$\pm$0.003&0.583$\pm$0.076&0.015$\pm$0.003\\
&&&	227.8919226&140.0&$13_{1,12}$ - $12_{1,11}$,v$_t$=1 - 1&$1.12\times10^{-4}$&7.210$\pm$0.020&0.029$\pm$0.002&0.636$\pm$0.062&0.020$\pm$0.004\\
&&&	230.9913834&85.5&$14_{0,14}$ - $13_{1,13}$,v$_t$=2 - 2&$1.20\times10^{-4}$&7.16$\pm$0.02&0.044 $\pm$0.002&0.957$\pm$0.064&0.045$\pm$0.005\\
\cline{3-11}
&&SVS13A&84.5958861&13.4&$4_{2,3}$ - $4_{1,4}$,v$_t$= 2- 2&$3.25\times10^{--6}$&8.640$\pm$0.170&0.012$\pm$0.002&1.353$\pm$0.475&0.017$\pm$0.009\\
&&&130.2463048&19.7&$4_{3,1}$ - $4_{2,2}$,v$_t$=2 - 2&$9.98\times10^{-6}$&8.680$\pm$0.050&0.016$\pm$0.002&0.719$\pm$0.117&0.012$\pm$0.004\\
&&&153.4842039&99.1&$9_{2,8}$ - $8_{1,8}$,v$_t$=0 - 1&$4.17\times10^{-5}$&8.300$\pm$0.060&0.017$\pm$0.002&0.861$\pm$0.150&0.015$\pm$0.005\\
&&&205.4584717&64.0&$12_{1,12}$ - $11_{0,11}$,v$_t$=2 - 2&$8.04\times10^{-5}$&8.770$\pm$0.020&0.037$\pm$0.002&0.884$\pm$0.066&0.035$\pm$0.005\\
&&&244.633959&151.7&$14_{1,13}$ - $13_{1,12}$,v$_t$=1 - 1&$1.42\times10^{-4}$&8.770$\pm$0.030&0.022$\pm$0.002&0.705$\pm$0.091&0.016$\pm$0.004\\
&&&270.4440852&32.6&$5_{4,2}$ - $4_{3,1}$,v$_t$=2 - 2&$1.59\times10^{-4}$&8.720$\pm$0.010&0.042$\pm$0.002&0.811$\pm$0.044&0.036$\pm$0.004\\
\hline
$\rm{HCCCHO}$&54007 (JPL)&L1544&83.775816&20.1&9$_{0,9}$ - 8$_{0,8}$&1.85$\times$10$^{-5}$&7.130 $\pm$0.019&0.011$\pm$0.001&0.393$\pm$0.046&0.004$\pm$0.001\\
(Propynal)&&&99.039070&7.44&4$_{1,4}$ - 3$_{0,3}$&1.13$\times$10$^{-6}$&7.110 $\pm$ 0.020&0.012$\pm$ 0.001& 0.432$\pm$0.045&0.005$\pm$0.001\\
&&&102.298030&29.5&11$_{0,{11}}$ - 10$_{0,{10}}$&3.40$\times$10$^{-5}$&7.080$\pm$ 0.020&0.013$\pm$0.001&0.486$\pm$0.042&0.007$\pm$0.001\\
\hline
$\rm{CH_3OCH_3}$&46514 (CDMS)&L1544&99.324362&10.2&4$_{1,4}$ - 3$_{0,3}$ (EA)&5.53$\times$ 10$^{-6}$&\nodata&\nodata&\nodata&\nodata\\
(Dimethyl ether)&&&99.324364&10.2&4$_{1,4}$ - 3$_{0,3}$ (AE)&5.53$\times$ 10$^{-6}$&\nodata&\nodata&\nodata&\nodata\\
&&&99.325217&10.2&4$_{1,4}$ - 3$_{0,3}$ (EE)&5.53$\times$ 10$^{-6}$&\nodata&\nodata&\nodata&\nodata\\
&&&99.326072&10.2&4$_{1,4}$ - 3$_{0,3}$ (AA)&5.53$\times$ 10$^{-6}$&\nodata&\nodata&\nodata&\nodata\\
\cline{3-11}
&&B1-b&99.324362&10.2&4$_{1,4}$ - 3$_{0,3}$ (EA)&5.53$\times$ 10$^{-6}$&\nodata&\nodata&\nodata&\nodata\\
&&&99.324364&10.2&4$_{1,4}$ - 3$_{0,3}$ (AE)&5.53$\times$ 10$^{-6}$&\nodata&\nodata&\nodata&\nodata\\
&&&99.325217&10.2&4$_{1,4}$ - 3$_{0,3}$ (EE)&5.53$\times$ 10$^{-6}$&\nodata&\nodata&\nodata&\nodata\\
&&&99.326072&10.2&4$_{1,4}$ - 3$_{0,3}$ (AA)&5.53$\times$ 10$^{-6}$&\nodata&\nodata&\nodata&\nodata\\
\cline{3-11}
&&IRAS4A&162.529582&33.0&8$_{1,8}$ - 7$_{0,7}$, (EE)&2.67$\times$10$^{-5}$&7.180$\pm$0.060&0.041$\pm$0.002&2.668$\pm$0.287&0.117$\pm$0.017\\
&&&209.515644&59.3&11$_{1,11}$ - 10$_{0,10}$, (EE)&6.40$\times$10$^{-5}$&7.050$\pm$0.020&0.053$\pm$0.001&1.640$\pm$0.076&0.093$\pm$0.007\\
&&&225.599126&69.5&12$_{1,12}$ - 11$_{0,11}$, (EE)&8.25$\times$10$^{-5}$&7.420$\pm$0.020&0.068$\pm$0.001&1.360$\pm$0.063&0.099$\pm$0.007\\
&&&241.946542&81.1&13$_{1,13}$ - 12$_{0,12}$, (EE)&1.05$\times$10$^{-4}$&7.370$\pm$0.040&0.047$\pm$0.001&2.025$\pm$0.129&0.103$\pm$0.009\\
\hline
$\rm{CH_3CN}$&41001 (JPL)&L1544&91.985314&20.4&5$_{1}$ - 4$_{1}$&6.08$\times$ 10$^{-5}$&7.280$\pm$0.002&0.087$\pm$0.001&0.542 $\pm$ 0.006&0.050$\pm$0.001\\
 (Methyl cyanide)&&&91.987087&13.2&5$_{0}$ - 4$_{0}$&6.33$\times$ 10$^{-5}$&7.282$\pm$0.002&0.127   $\pm$0.001&0.505$\pm$0.004&0.068$\pm$0.001\\  
\cline{3-11}
&&B1-b&91.985314&20.4&5$_{1}$ - 4$_{1}$&6.08$\times$ 10$^{-5}$&6.719$\pm$    0.143&0.018$\pm$0.003&1.731 $\pm$0.388&0.034 $\pm$0.013\\
 &&&91.987087&13.2&5$_{0}$ - 4$_{0}$&6.33$\times$ 10$^{-5}$&6.654$\pm$     0.090&0.029$\pm$0.003&1.846$\pm$0.253&0.058$\pm$0.013\\ 
 \cline{3-11}
 &&IRAS4A&73.588799&16.0&$4_{1}$ - $3_{1}$&2.97$\times$ 10$^{-5}$&6.726 $\pm$   0.141&0.043  $\pm$  0.003&2.758 $\pm$   0.472&0.127 $\pm$   0.029\\
 &&&73.590218&8.8&$4_{0}$ - 3$_{0}$&3.17$\times$ 10$^{-5}$&7.025 $\pm$    0.121&0.055  $\pm$   0.003&2.858$\pm$    0.480&0.169$\pm$    0.036\\
 &&&91.979994&41.8&$5_{2}$ - $4_{2}$&5.32$\times$ 10$^{-5}$&7.071$\pm$    0.170&0.029$\pm$    0.002&2.661 $\pm$   0.853&0.082 $\pm$   0.033\\
 &&&91.985314&20.4&$5_{1}$ - $4_{1}$&6.08$\times$ 10$^{-5}$&6.943$\pm$    0.089&0.054$\pm$    0.002&3.119 $\pm$   0.399&0.179$\pm$    0.031\\
 &&&91.987087&13.2&$5_{0}$ - $4_{0}$&6.33$\times$ 10$^{-5}$&7.083$\pm$    0.056&0.078$\pm$    0.002&2.658 $\pm$   0.239&0.221 $\pm$   0.027\\
 &&& 110.364353&82.8&$6_{3}$ - $5_{3}$&8.33$\times$ 10$^{-5}$&7.297$\pm$    0.051&0.039 $\pm$   0.003&1.737$\pm$    0.131&0.072$\pm$    0.010\\
 &&&110.3813720&25.7&$6_{1}$ - $5_{1}$&1.08$\times$ 10$^{-4}$&7.406 $\pm$   0.055&0.072$\pm$    0.002&3.018 $\pm$   0.235&0.233 $\pm$   0.025 \\
 &&&110.383499&18.5&$6_{0}$ - $5_{0}$&1.11$\times$ 10$^{-4}$&7.299$\pm$    0.056&0.076$\pm$    0.002&3.180$\pm$    0.253&0.258$\pm$    0.028\\
 &&&128.757030&89.0&$7_{3}$ - $6_{3}$&1.46$\times$ 10$^{-4}$&7.052$\pm$    0.135&0.029$\pm$    0.001&2.902$\pm$    0.464&0.089$\pm$    0.018\\
 &&&128.769436&53.3&$7_{2}$ - $6_{2}$&1.64$\times$ 10$^{-4}$&6.812$\pm$    0.072&0.043$\pm$    0.001&3.134 $\pm$   0.266&0.144$\pm$    0.017\\
 &&&128.776881&31.9&$7_{1}$ - $6_{1}$&1.75$\times$ 10$^{-4}$&6.675$\pm$    0.045&0.078 $\pm$   0.001&2.941$\pm$    0.135&0.244$\pm$    0.015\\
 &&&128.779363&24.7&$7_{0}$ - $6_{0}$&1.78$\times$ 10$^{-4}$&6.990$\pm$    0.050&0.092$\pm$    0.001&2.926 $\pm$   0.209&0.285$\pm$    0.024\\
 &&&147.163244&60.4&$8_{2}$ - $7_{2}$&2.52$\times$ 10$^{-4}$&6.898$\pm$    0.062&0.039$\pm$    0.002&2.349$\pm$    0.211&0.098$\pm$    0.013\\
 &&&147.171751&38.9&$8_{1}$ - $7_{1}$&2.64$\times$ 10$^{-4}$&6.854$\pm$    0.051&0.070$\pm$    0.002&3.131$\pm$    0.186&0.232$\pm$    0.019\\
 &&&147.174588&31.8&$8_{0}$ - $7_{0}$&2.69$\times$ 10$^{-4}$&7.147$\pm$    0.031&0.087$\pm$    0.002&2.787$\pm$    0.134&0.258$\pm$    0.017\\
 &&&165.540377&104.0&$9_{3}$ - $8_{3}$&3.42$\times$ 10$^{-4}$&6.815$\pm$    0.113&0.039 $\pm$   0.001&3.176$\pm$    0.399&0.131$\pm$    0.021\\
 &&&165.556321&68.3&$9_{2}$ - $8_{2}$&3.66$\times$ 10$^{-4}$&7.192$\pm$    0.044&0.042$\pm$    0.002&2.204$\pm$    0.134&0.098$\pm$    0.010\\
 &&&165.565891&46.9&$9_{1}$ - $8_{1}$&3.80$\times$ 10$^{-4}$&7.189$\pm$    0.036&0.070$\pm$    0.002&2.912$\pm$    0.141&0.216$\pm$    0.015\\
 &&&165.569081&39.7&$9_{0}$ - $8_{0}$&3.85$\times$ 10$^{-4}$&7.183$\pm$    0.030&0.078$\pm$    0.002&2.656$\pm$    0.122&0.220$\pm$    0.014\\
 &&&202.320442&122.6&$11_{3}$ - $10_{3}$&6.56$\times$ 10$^{-4}$&7.101$\pm$    0.027&0.059 $\pm$   0.002&2.086$\pm$    0.077&0.131$\pm$    0.008\\
 &&&220.709016&133.2&$12_{3}$ - $11_{3}$&8.66$\times$ 10$^{-4}$&7.324 $\pm$   0.051&0.049 $\pm$    0.001&2.570 $\pm$    0.192&0.135 $\pm$    0.014\\
\hline 
\end{tabular}}
\end{table*}
\begin{table*}
\centering
\tiny{
\caption{Observed transitions toward some sources. \label{tab:observation_4}}
\begin{tabular}{|l|l|l|l|l|l|l|l|l|l|l|}
\hline
Species&Tag (Database)&Source&Frequency&E$_{up}$&Quantum No.&A$_{ij}$&V$_{LSR}$&T$_{peak}$&FWHM&$\int$T$_{mb}$.dv\\
&&&(GHz)&(K)&&(s$^{-1}$)&(km.s$^{-1}$)&(K)&(km.s$^{-1}$)&(K.km.s$^{-1}$)\\
\hline
&&SVS13A&91.979994&41.8&$5_{2}$ - $4_{2}$&3.32$\times$ 10$^{-5}$&7.918$\pm$     0.569&0.020 $\pm$    0.002&3.205$\pm$     1.528&0.067 $\pm$    0.038\\
&&&91.985314&20.4&$5_{1}$ - $4_{1}$&6.08$\times$ 10$^{-5}$&8.428$\pm$     0.133&0.029$\pm$     0.002&2.762$\pm$     0.588&0.086$\pm$     0.024\\
&&&91.987087&13.2&$5_{0}$ - $4_{0}$&36.33$\times$ 10$^{-5}$&8.580$\pm$     0.078 &0.035$\pm$     0.002&2.366$\pm$     0.272&0.088$\pm$     0.015\\
&&&110.364353&82.8&$6_{3}$ - $5_{3}$&8.33$\times$ 10$^{-5}$&8.808$\pm$     0.145&0.025$\pm$     0.002&3.050$\pm$     0.596& 0.080$\pm$     0.022\\
&&&110.374989&47.1&$6_{2}$ - $5_{2}$&9.87$\times$ 10$^{-5}$&8.531$\pm$     0.177&0.030$\pm$     0.002&3.366 $\pm$    1.134&0.109 $\pm$    0.043\\
&&&110.381372&25.7&$6_{1}$ - $5_{1}$&1.08$\times$ 10$^{-4}$&8.284$\pm$     0.083&0.042$\pm$     0.002&2.540$\pm$     0.245&0.114$\pm$     0.016\\
&&&110.383499&18.5&$6_{0}$ - $5_{0}$&1.11$\times$ 10$^{-4}$&8.315$\pm$     0.102&0.033$\pm$     0.002&2.649$\pm$     0.332&0.093$\pm$     0.017\\
&&&128.776881&31.8&$7_{1}$ - $6_{1}$&1.75$\times$ 10$^{-4}$&8.392$\pm$     0.034&0.050$\pm$     0.002&1.719$\pm$     0.094&0.092$\pm$     0.008\\
&&&128.779363&24.7&$7_{0}$ - $6_{0}$&1.78$\times$ 10$^{-4}$&8.280$\pm$     0.036&0.050$\pm$     0.001&1.990$\pm$     0.107&0.105$\pm$     0.009\\
&&&147.171751&38.9&$8_{1}$ - $7_{1}$&2.64$\times$ 10$^{-4}$&8.162$\pm$     0.051&0.062 $\pm$    0.002&2.533 $\pm$    0.164&0.167$\pm$     0.015\\
&&&165.565891&46.9&$9_{1}$ - $8_{1}$&3.80$\times$ 10$^{-4}$&8.343$\pm$     0.030&0.091 $\pm$    0.002&2.307$\pm$     0.099&0.223 $\pm$    0.013\\
&&&220.709016&133.2&$12_{3}$ - $11_{3}$&8.66$\times$ 10$^{-4}$&8.576$\pm$     0.033&0.077 $\pm$    0.001&3.238$\pm$     0.136&0.265 $\pm$    0.016\\
&&&220.730260&97.4&$12_{2}$ - $11_{2}$&8.98$\times$ 10$^{-4}$&8.363$\pm$     0.051&0.071$\pm$     0.001&3.900$\pm$     0.236&0.294$\pm$     0.023\\
&&&220.743010&76.0&$12_{1}$ - $11_{1}$&9.18$\times$ 10$^{-4}$&8.300$\pm$     0.028&0.088$\pm$     0.001&2.951$\pm$     0.110&0.278$\pm$     0.014\\
&&&220.747261&68.9&$12_{0}$ - $11_{0}$&9.24$\times$ 10$^{-4}$&8.382$\pm$     0.040&0.090$\pm$     0.001&3.907$\pm$     0.186&0.373$\pm$     0.023\\
&&&239.119504&108.9&$13_{2}$ - $12_{2}$&1.15$\times$ 10$^{-3}$&8.209$\pm$     0.040&0.091$\pm$     0.001&2.952$\pm$     0.139&0.285$\pm$     0.017\\
&&&257.527383&92.7&$14_{0}$ - $13_{0}$&1.48$\times$ 10$^{-3}$&8.426$\pm$     0.059&0.092$\pm$     0.001&4.037$\pm$     0.321&0.394$\pm$     0.036\\

\hline
\end{tabular}}
\end{table*}

\section{Upper limit estimation}
For some species, we were unable to identify suitable and multiple transitions to estimate the column density by rotation diagram and MCMC method. In Table \ref{tab:upper limit}, we note the estimated upper limits of column density for these species. The input parameters used for this estimation are also noted.

\begin{table*}
\centering{
\tiny{
\caption{Estimated upper limit of column density \label{tab:upper limit}}
\begin{tabular}{|c|c|c|c|c|c|c|c|}
\hline
Molecules&Source&Frequency&Quantum No.&E$_{up}$&A$_{ij}$&N$_{tot}$&T$_{k}$ \\
&&(GHz)&&(K)&(s$^{-1}$)&(cm$^{-2}$)&(K)\\
\hline
CH$_3$OCHO&L1544&90.227659&8$_{0,8}$ - 7$_{0,7}$, (E)&20.1&$1.05\times10^{-5}$&$3.7\times10^{12}$&10.0\\
\hline
C$_2$H$_5$OH&B1-b&131.502781&6$_{3,4}$ - 6$_{2,5}$, v$_t$ = $2-2$ &28.9&$1.27\times10^{-5}$&$1.0\times10^{13}$&20.0\\
\hline
HCCCHO&B1-b&93.0432843&10$_{0,10}$ - 9$_{0,9}$&24.6&$2.55\times10^{-5}$&$2.56\times10^{12}$&20.0\\
&IRAS4A&111.53912&12$_{0,12}$ - 11$_{0,11}$&34.8&$4.43\times10^{-5}$&$3.9\times10^{12}$&40.0\\
&SVS13A&236.6916499&21$_{4,17}$ - 22$_{3,20}$&152.0&$4.96\times10^{-6}$&$4.00\times10^{14}$&100.0\\
\hline
\end{tabular}}\\
}
\end{table*}

\section{Rotational diagram}\label{sec:RD}
For the optically thin transitions, upper state column density (N$_u^{thin}$) can be expressed as \citep{gold99},
\begin{equation}
\frac{N_u^{thin}}{g_u}=\frac{3k_B\int{T_{mb}dV}}{8\pi^{3}\nu S\mu^{2}},
\label{eqn:clmn}
\end{equation}
where g$_u$ is the degeneracy of the upper state, k$_B$ is the Boltzmann constant, $\rm{\int T_{mb}dV}$ is the integrated intensity,
$\nu$ is the rest frequency, $\mu$ is the electric dipole moment, and S is the transition line strength. Under the LTE conditions, the total
column density ($N_{total}$) can be written as,
\begin{equation}
\frac{N_u^{thin}}{g_u}=\frac{N_{total}}{Q(T_{rot})}\exp(-E_u/k_BT_{rot}),
\end{equation}
where $T_{rot}$ is the rotational temperature, E$_u$ is the upper state energy, $\rm{Q(T_{rot})}$ is the partition function at rotational
temperature. This can be rearranged as,
\begin{equation}
ln\Bigg(\frac{N_u^{thin}}{g_u}\Bigg)=-\Bigg(\frac{1}{T_{rot}}\Bigg)\Bigg(\frac{E_u}{k}\Bigg)+ln\Bigg(\frac{N_{total}}{Q(T_{rot})}\Bigg).
\end{equation}
There is a linear relationship between the upper state energy and column density at the upper level. The column density and rotational temperature are extracted from the rotational diagram.\\
 The rotational diagram can only be performed when multiple transitions ($>$2) with different up-state energy of a molecule are observed. The estimated rotational temperature (T$_{rot}$) and the column densities are mentioned in table \ref{tab:rotdiag}. In some cases (IRAS4A for CH$_3$OH, CH$_3$CHO and for CH$_3$CN), two temperature components are obtained from the rotational diagram. Details about the components are mentioned in the main text. The error bars (vertical bars) in rotational diagrams are the absolute uncertainty in a log of (N$_u$/g$_u$), which arises from the error of the observed integrated intensity that we measured using a single Gaussian fitting to the observed profile of each transition.

\begin{table}
\centering{
\tiny{
\caption{Results obtained with the rotational diagram analysis.\label{tab:rotdiag}}
\begin{tabular}{|c|c|c|c|c|}
\hline
Molecules&Source&N$_{tot}$&T$_{rot}$&Abundance \\
&&(cm$^{-2}$)&(K)&\\
\hline
CH$_3$OH&L1544&$(7.14^{+0.70}_{-0.60}) \times 10^{12}$&$ (8.60^{+0.40}_{-0.30})$&$(8.02^{+0.80}_{-0.70}) \times 10^{-11}$\\
&B1-b&$(9.73^{+0.05}_{-0.04}) \times 10^{13}$&$(7.80^{+0.20}_{-0.20})$&$(1.23^{+0.01}_{-0.01}) \times 10^{-9}$\\
&IRAS4A&$(1.42^{+0.03}_{-0.02}) \times 10^{14}$&$(15.30^{+0.20}_{-0.20})$&$(4.83^{+0.02}_{-0.02}) \times 10^{-9}$\\
&&$(1.41^{+0.04}_{-0.04}) \times 10^{14}$&$(43.30^{+0.80}_{-0.70})$&$(4.83^{+0.04}_{-0.03}) \times 10^{-9}$\\
&SVS13A&$(1.21^{+0.03}_{-0.03}) \times 10^{14}$&$(65.10^{+2.0}_{-1.9})$&$(1.21^{+0.03}_{-0.03}) \times 10^{-9}$\\
\hline
CH$_3$CHO&L1544&$(1.25^{+0.37}_{-0.28}) \times 10^{12}$  &5.98$^{+0.70}_{-0.57}$&$(1.40^{+0.42}_{-0.32}) \times 10^{-11}$\\
&B1-b&$(3.48^{+0.55}_{-0.47}) \times 10^{12}$&9.09$^{+0.62}_{-0.55}$&$(4.40^{+0.70}_{-0.60}) \times 10^{-11}$\\
&IRAS4A&$(9.47^{+0.56}_{-0.52}) \times 10^{12}$&$(22.09^{+1.03}_{-0.94})$&$(3.30^{+0.19}_{-0.18}) \times 10^{-10}$\\
&&$(9.53^{+1.00}_{-0.95}) \times 10^{12}$&$(64.46^{+5.71}_{-4.85})$&$(3.30^{+0.37}_{-0.33}) \times 10^{-10}$\\
&SVS13A&$(5.89^{+1.60}_{-1.31}) \times 10^{12}$&$ (41.33^{+6.51}_{-4.95})$&$(5.90^{+1.70}_{-1.30}) \times 10^{-11}$\\
\hline
CH$_3$OCHO&L1544&\nodata&\nodata&\nodata\\
&B1-b&$(6.40) \times 10^{12***}$&$(15.50^{***})$&$(8.10) \times 10^{-11***}$\\
&IRAS4A&$(3.49^{+0.50}_{-0.40}) \times 10^{13}$&$(82.30^{+8.40}_{-7.00})$&$(1.20^{+0.4}_{-0.3}) \times 10^{-9}$\\
&SVS13A&$(7.78^{+1.30}_{-1.10}) \times 10^{13}$&$(125.60^{+19.80}_{-15.10})$&$(7.78^{+1.30}_{-1.10}) \times 10^{-10}$\\
\hline
C$_2$H$_5$OH&L1544&\nodata&\nodata&\nodata\\
&B1-b&\nodata&\nodata&\nodata\\
&IRAS4A&$(1.69^{+0.30}_{-0.20}) \times 10^{13}$&$(50.60^{+4.90}_{-4.10})$&$(5.86^{+0.21}_{-0.22}) \times 10^{-10}$\\
&SVS13A&$(1.37^{+0.20}_{-0.10}) \times 10^{13}$&$(59.00^{+7.40}_{-5.90})$&$(1.37^{+0.27}_{-0.17}) \times 10^{-10}$\\
\hline
HCCCHO&L1544&$(3.16^{+0.80}_{-0.70}) \times 10^{12}$&$(5.60^{+0.30}_{-0.30})$&$(3.55^{+0.90}_{-0.70}) \times 10^{-11}$\\
&B1-b&\nodata&\nodata&\nodata\\
&IRAS4A&\nodata&\nodata&\nodata\\
&SVS13A&\nodata&\nodata&\nodata\\
\hline
CH$_3$OCH$_3$&L1544&$1.72\times10^{12}$*&\nodata&$(1.93) \times 10^{-11}$\\
&B1-b&$6.12\times10^{12}$*&\nodata&$(7.74)\times10^{-11}$\\
&IRAS4A&$(1.20^{+0.40}_{-0.30}) \times 10^{13}$&$(61.10^{+20.11}_{-12.10})$&$(4.14^{+1.30}_{-1.00}) \times 10^{-10}$\\
&SVS13A&$1.41\times10^{13}$**&\nodata&$1.41\times10^{-10}$\\
\hline
CH$_3$CN&L1544&$4.85\times10^{11}$*&\nodata&$(5.45) \times 10^{-12}$\\
&B1-b&$4.95\times10^{11}$*&\nodata&$(6.27)\times10^{-12}$\\
&IRAS4A&$(1.78^{+0.17}_{-0.15}) \times 10^{12}$&$(25.78^{+1.9}_{-1.7})$&$(6.10^{+0.59}_{-0.54}) \times 10^{-11}$\\
&&$(2.31^{+0.31}_{-0.28}) \times 10^{12}$&$(61.2^{+5.05}_{-4.34})$&$(8.00^{+1.10}_{-0.96}) \times 10^{-11}$\\
&SVS13A&$(4.57^{+0.23}_{-0.22}) \times 10^{12}$&$(134.69^{+11.8}_{-10.09})$&$(4.6^{+0.23}_{-0.22}) \times 10^{-11}$\\
\hline
\end{tabular}}}\\
{\scriptsize \noindent * Calculated using LTE fitting.\\
** Calculated from \cite{bian19} and scaled for 30$\arcsec$ beam.\\
*** We did not include the errors due to the large uncertainty in data points.\\
The hydrogen column density (N$_{H_2}$) in L1544, B1-b, IRAS4A and SVS13A are $8.9\times10^{22}$ cm$^{-2}$ \citep{hily22}, $7.9\times10^{22}$ cm$^{-2}$ \citep{dani13}, $2.9\times10^{22}$ cm$^{-2}$ \citep{mare02}, 
$1.0\times10^{23}$cm$^{-2}$ \citep{lefl98}, respectively.}
\end{table}

\begin{figure*}
\begin{minipage}{0.35\textwidth}
\includegraphics[width=\textwidth]{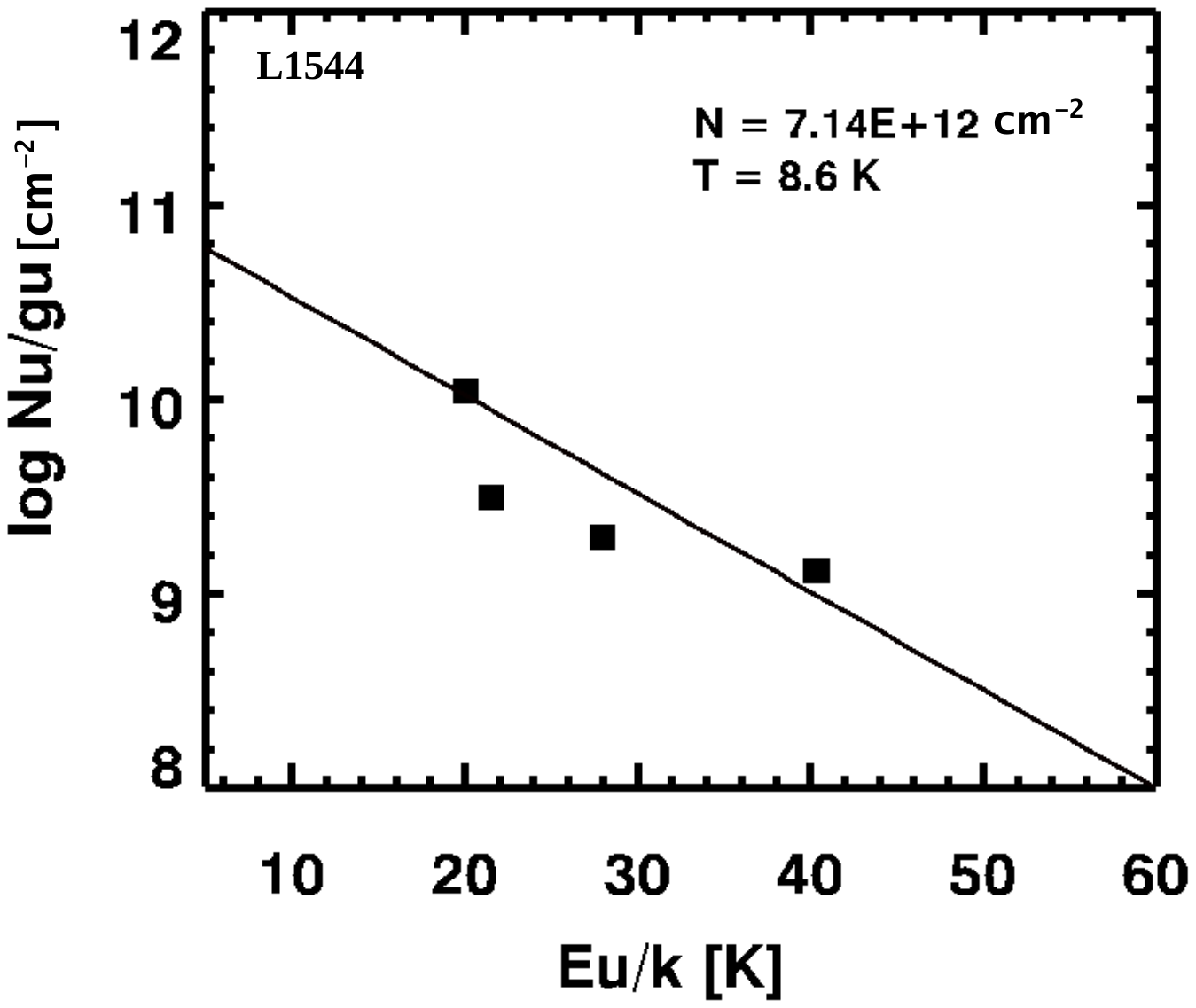}
\end{minipage}
 \begin{minipage}{0.35\textwidth}
 \includegraphics[width=\textwidth]{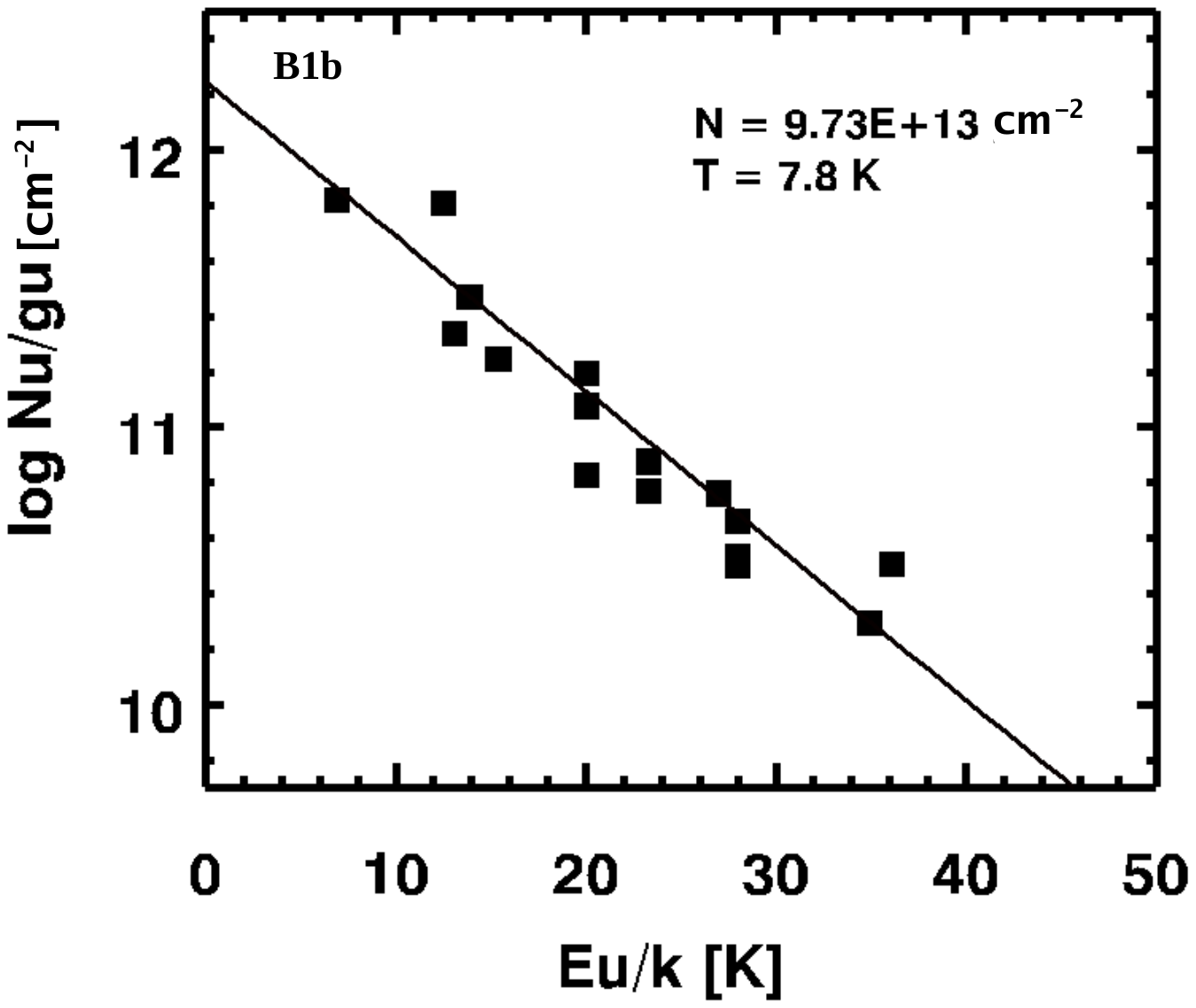}
 \end{minipage}
\begin{minipage}{0.35\textwidth}
 \includegraphics[width=\textwidth]{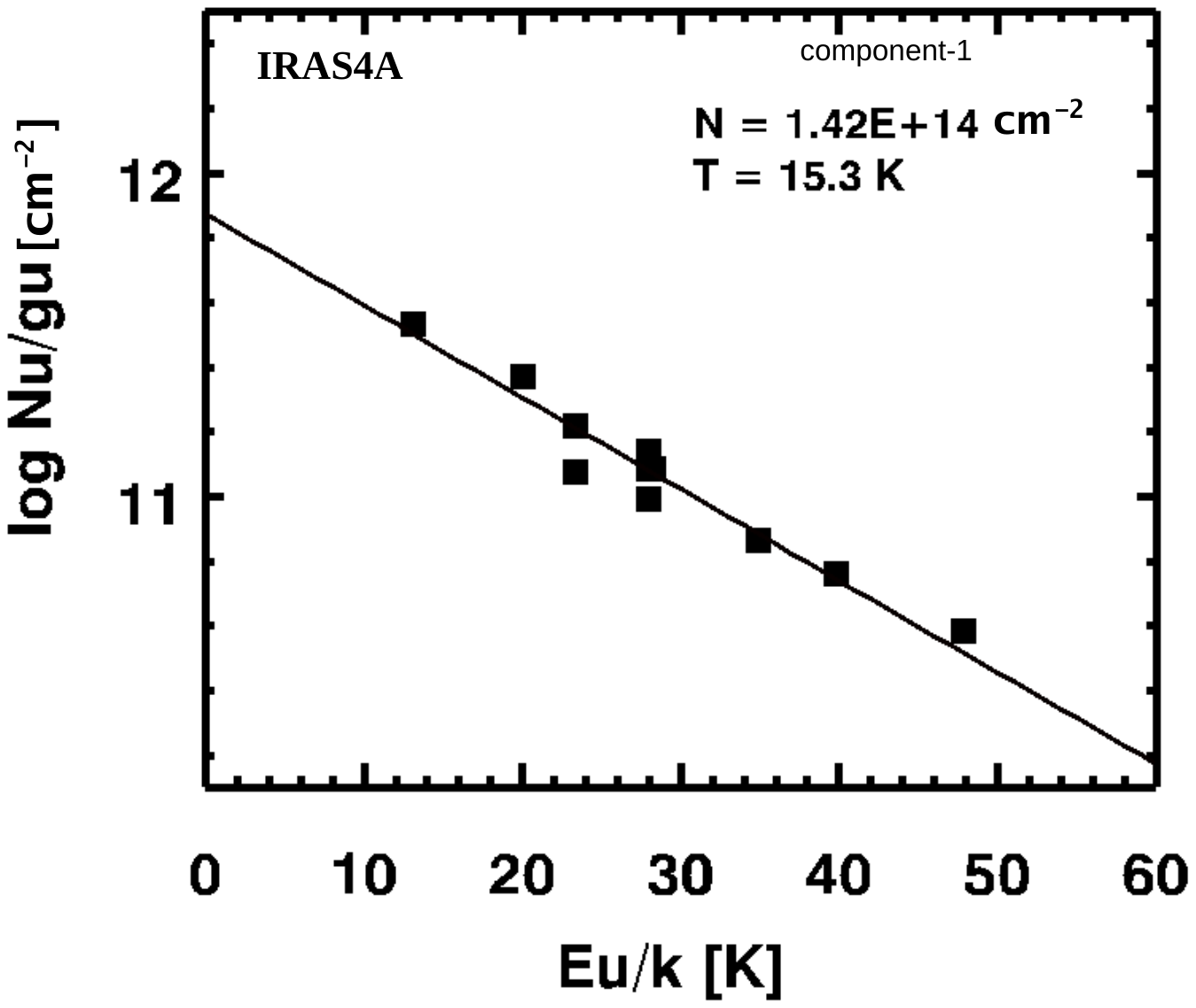}
 \end{minipage}
 \begin{minipage}{0.35\textwidth}
 \includegraphics[width=\textwidth]{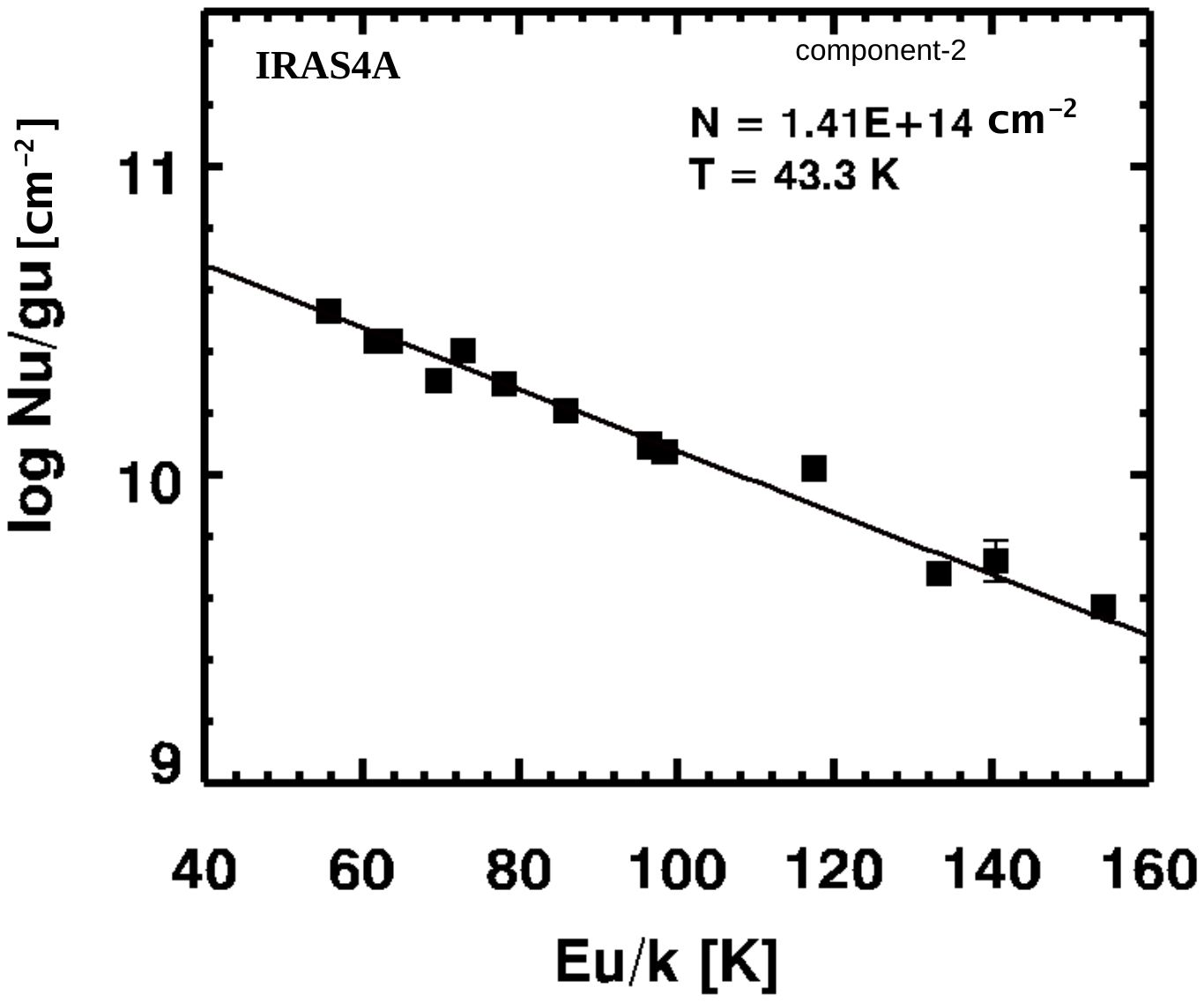}
 \end{minipage}
 \begin{minipage}{0.35\textwidth}
 \includegraphics[width=\textwidth]{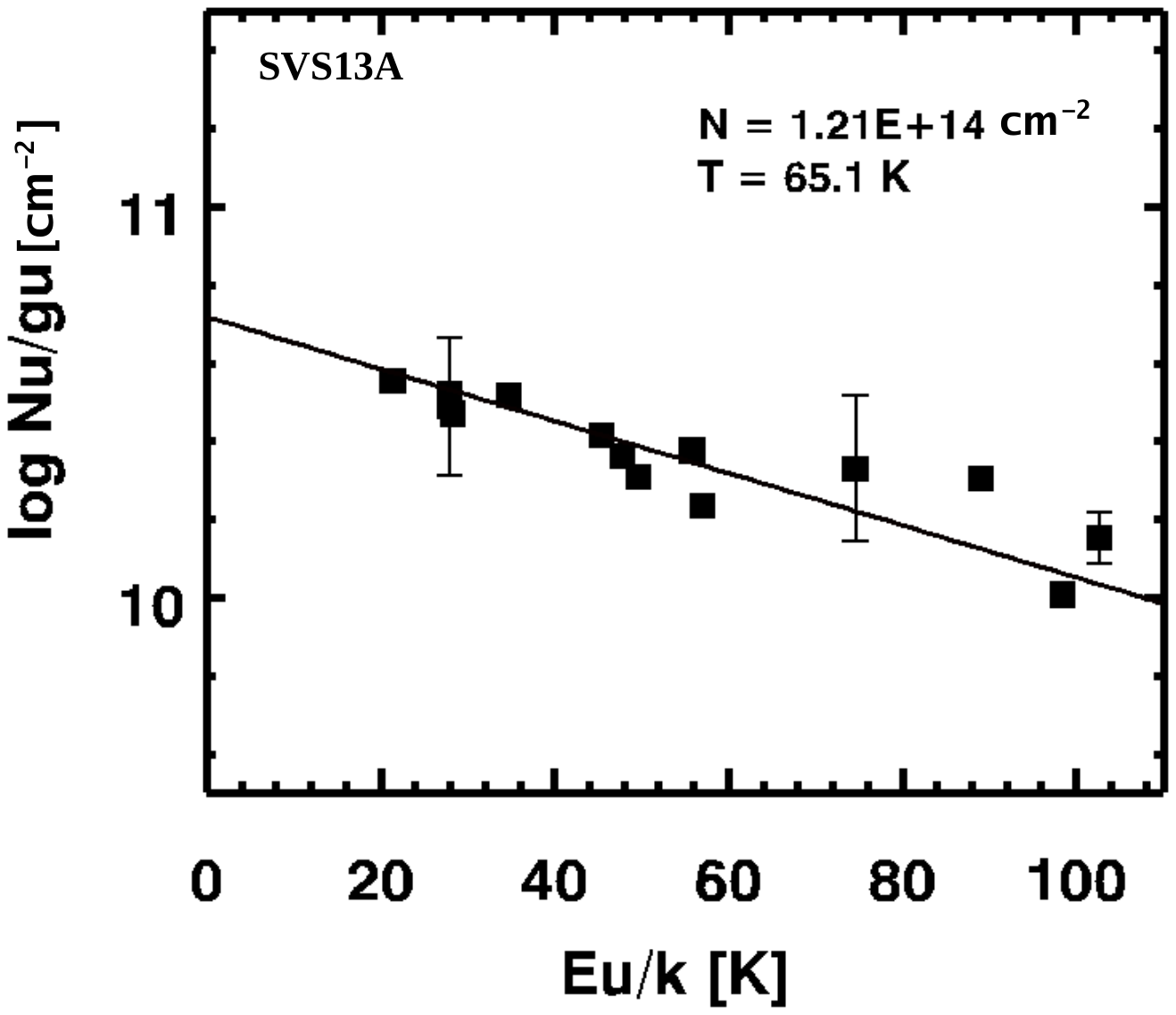}
 \end{minipage}
 \caption{Rotational diagram for CH$_3$OH obtained with various sources. The black points are the position of the data points, the vertical bars are the error bars estimated, and the red lines are the fitted lines to the rotational diagram. The obtained column densities and the excitation temperatures along with the error bars are mentioned in the top right corner of each box. }
\label{fig:rotational_diag_ch3oh_single}
\end{figure*}
\begin{figure*}
\begin{minipage}{0.35\textwidth}
\includegraphics[width=\textwidth]{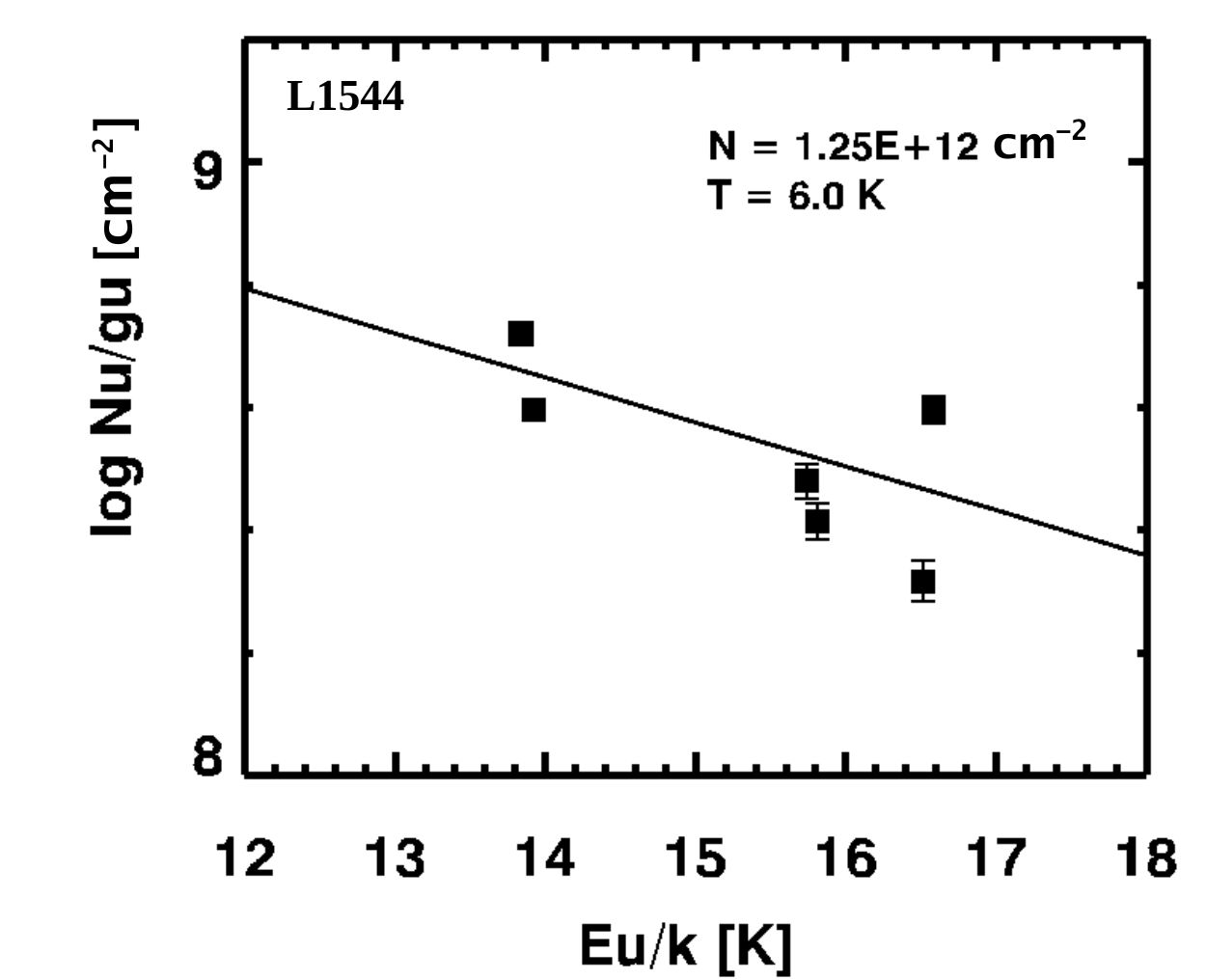}
\end{minipage}
 \begin{minipage}{0.35\textwidth}
 \includegraphics[width=\textwidth]{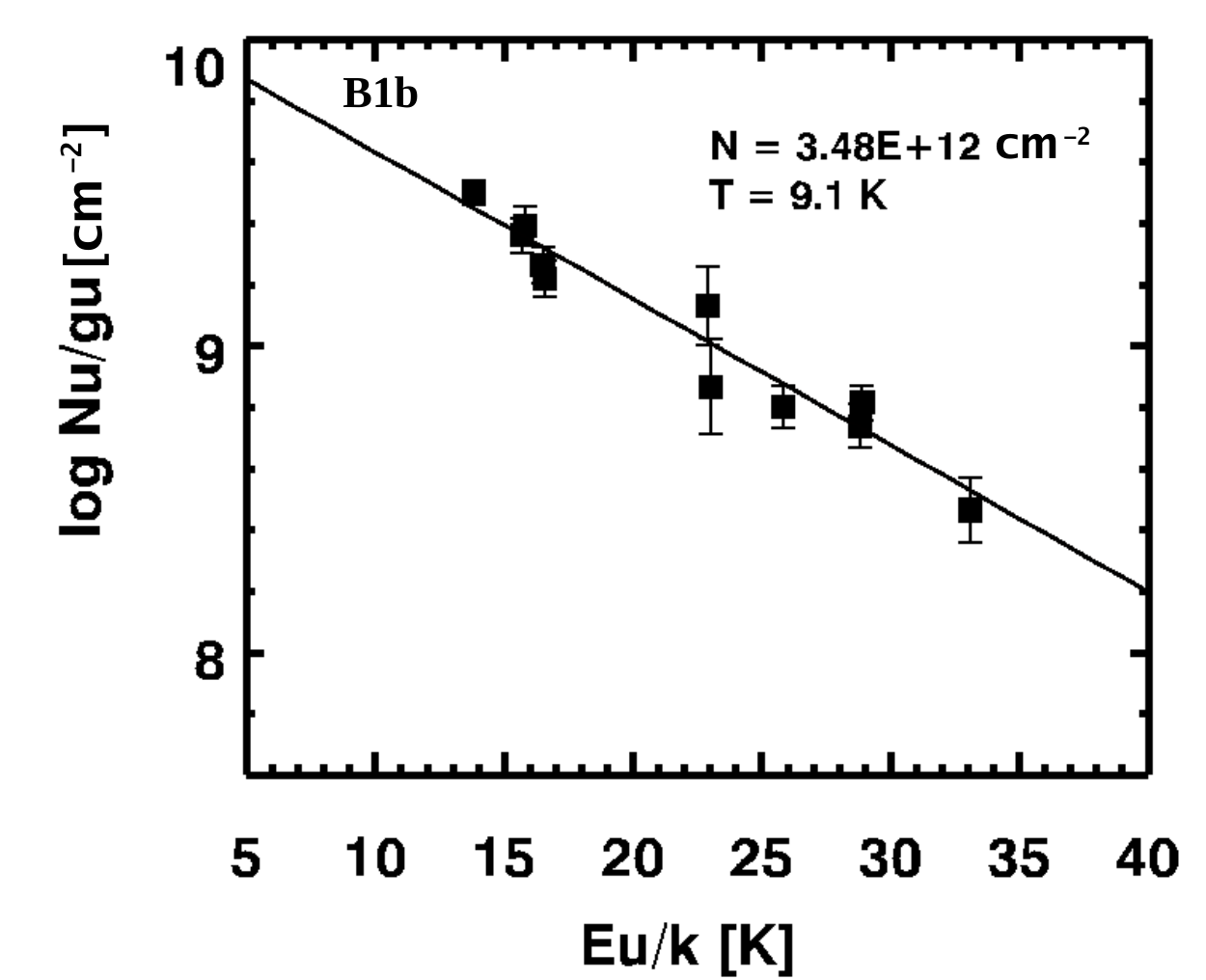}
 \end{minipage}
 \begin{minipage}{0.35\textwidth}
 \includegraphics[width=\textwidth]{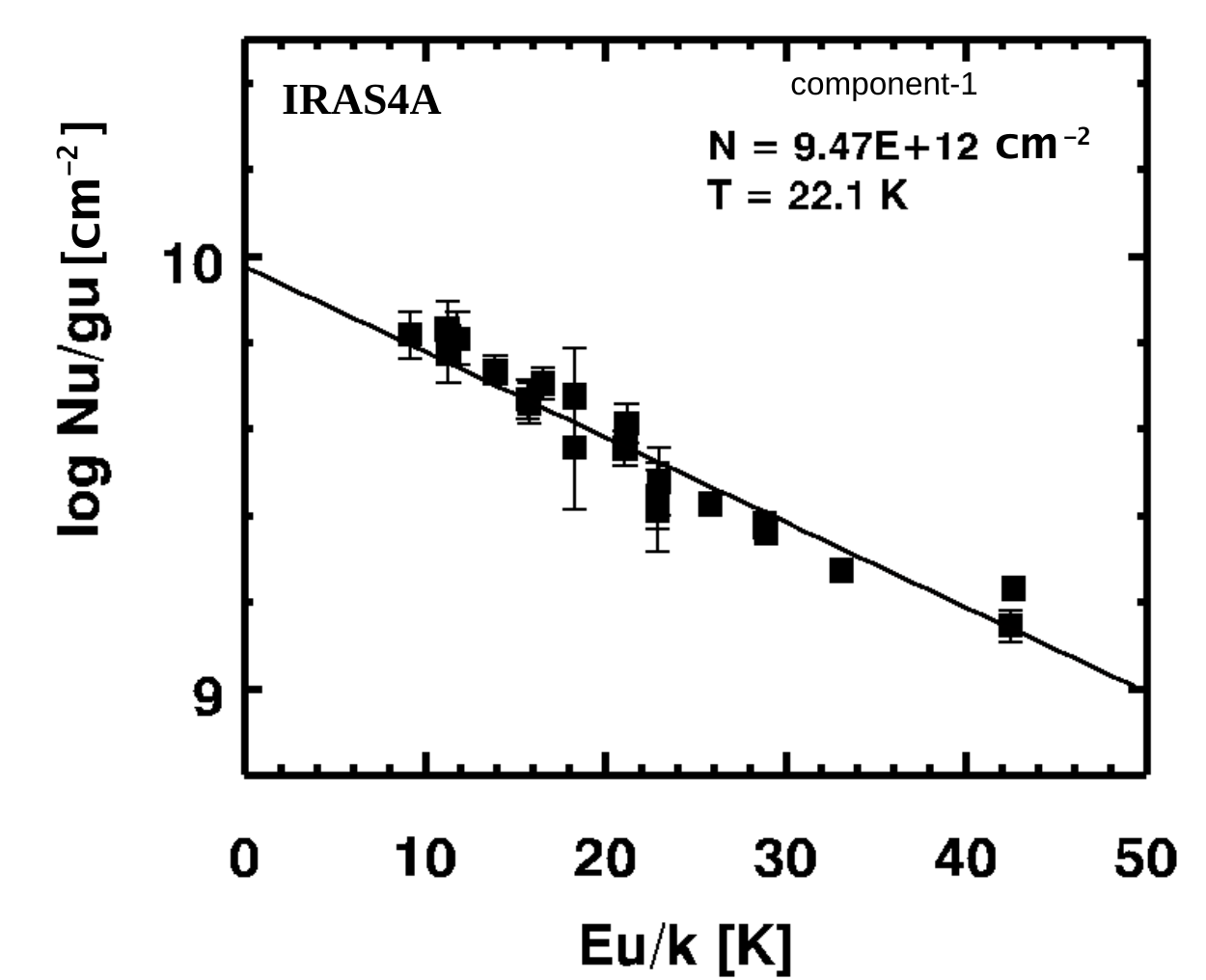}
 \end{minipage}
  \begin{minipage}{0.35\textwidth}
 \includegraphics[width=\textwidth]{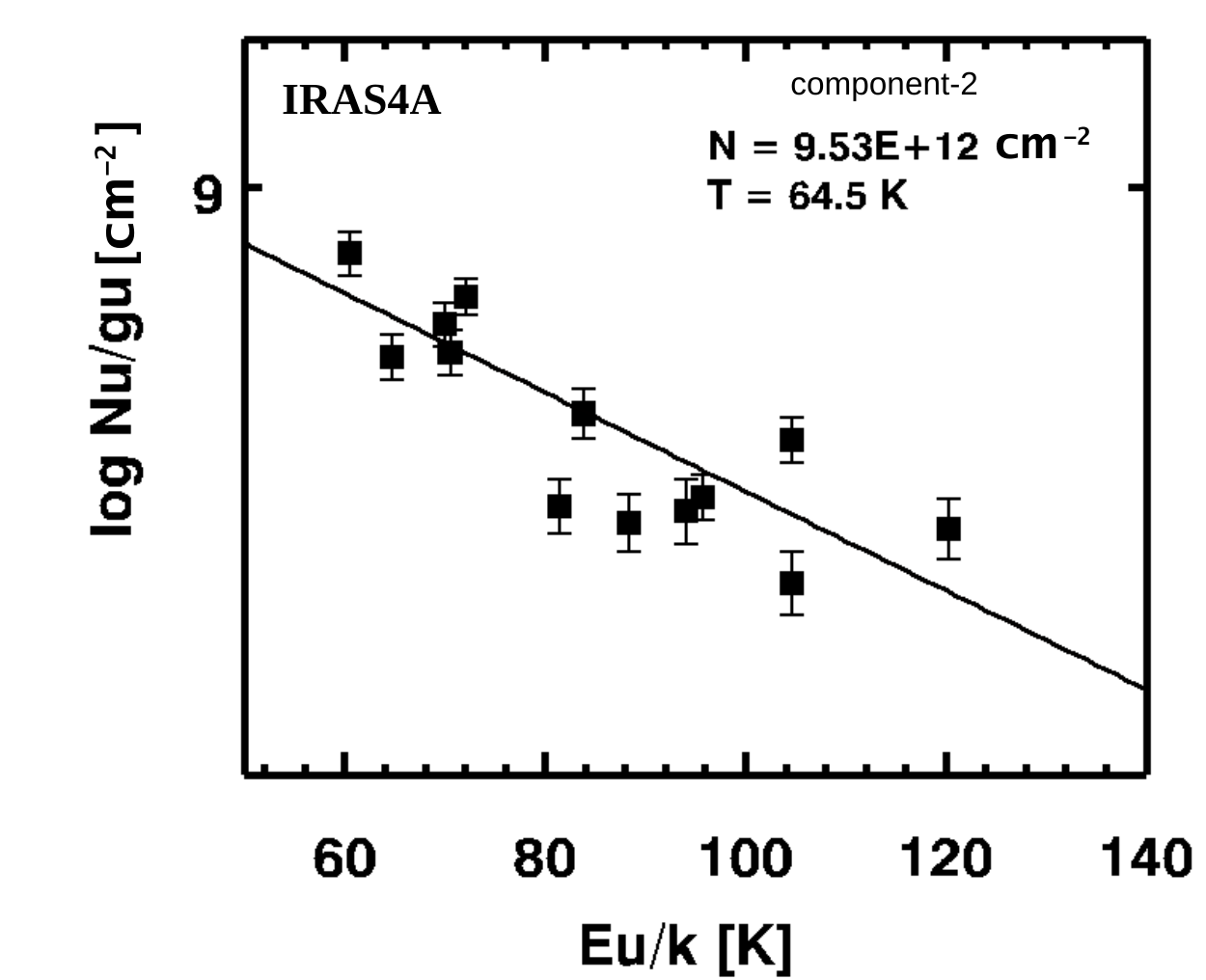}
 \end{minipage}
 \begin{minipage}{0.35\textwidth}
 \includegraphics[width=\textwidth]{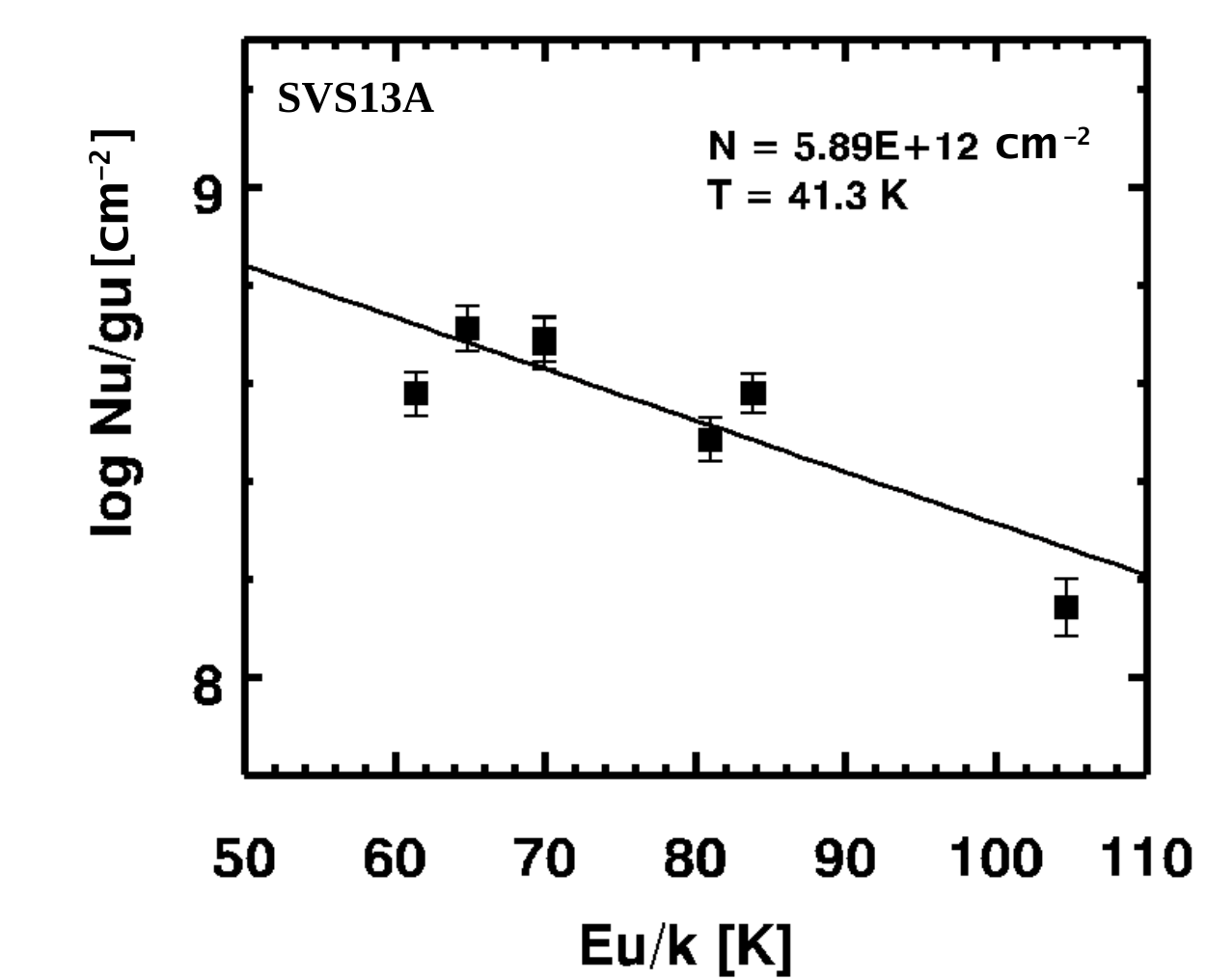}
 \end{minipage}
 \caption{Rotational diagram for CH$_3$CHO obtained for various sources. The symbols represent the same as those depicted in Figure \ref{fig:rotational_diag_ch3oh_single}.}
\label{fig:rotational_diag_ch3cho}
\end{figure*}
\begin{figure*}
 \begin{minipage}{0.35\textwidth}
 \includegraphics[width=\textwidth]{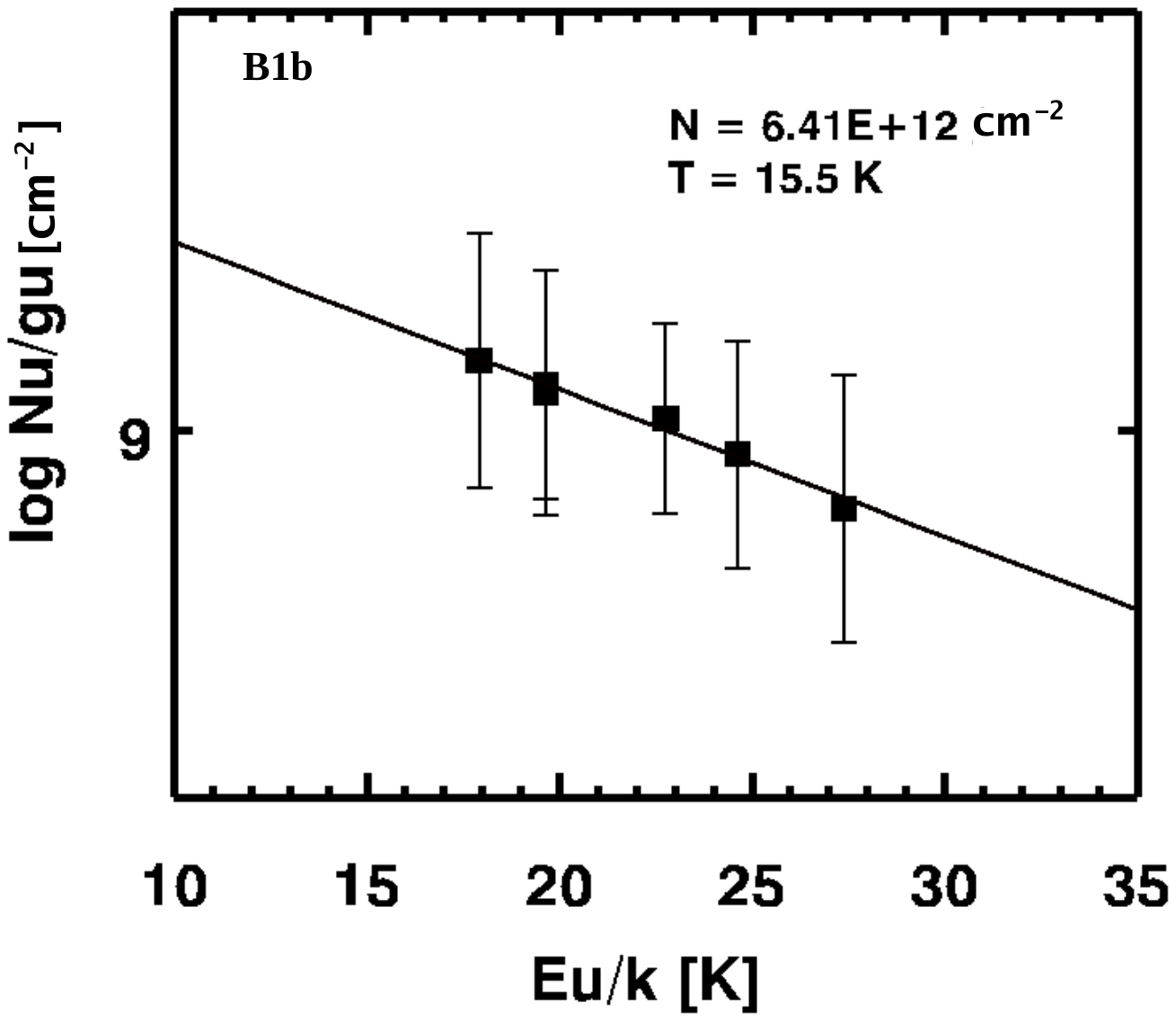}
 \end{minipage}
 \begin{minipage}{0.35\textwidth}
 \includegraphics[width=\textwidth]{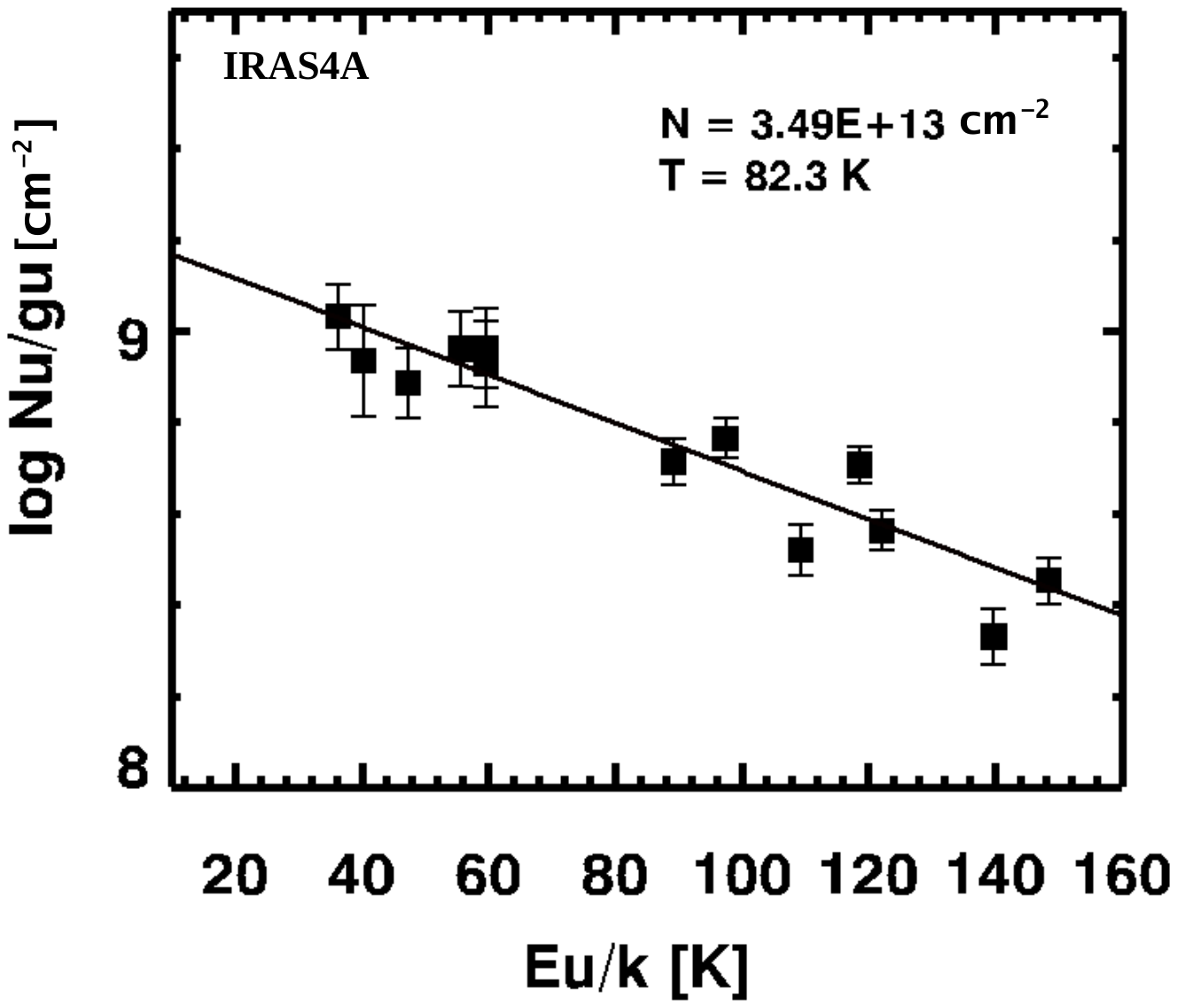}
 \end{minipage}
 \begin{minipage}{0.35\textwidth}
 \includegraphics[width=\textwidth]{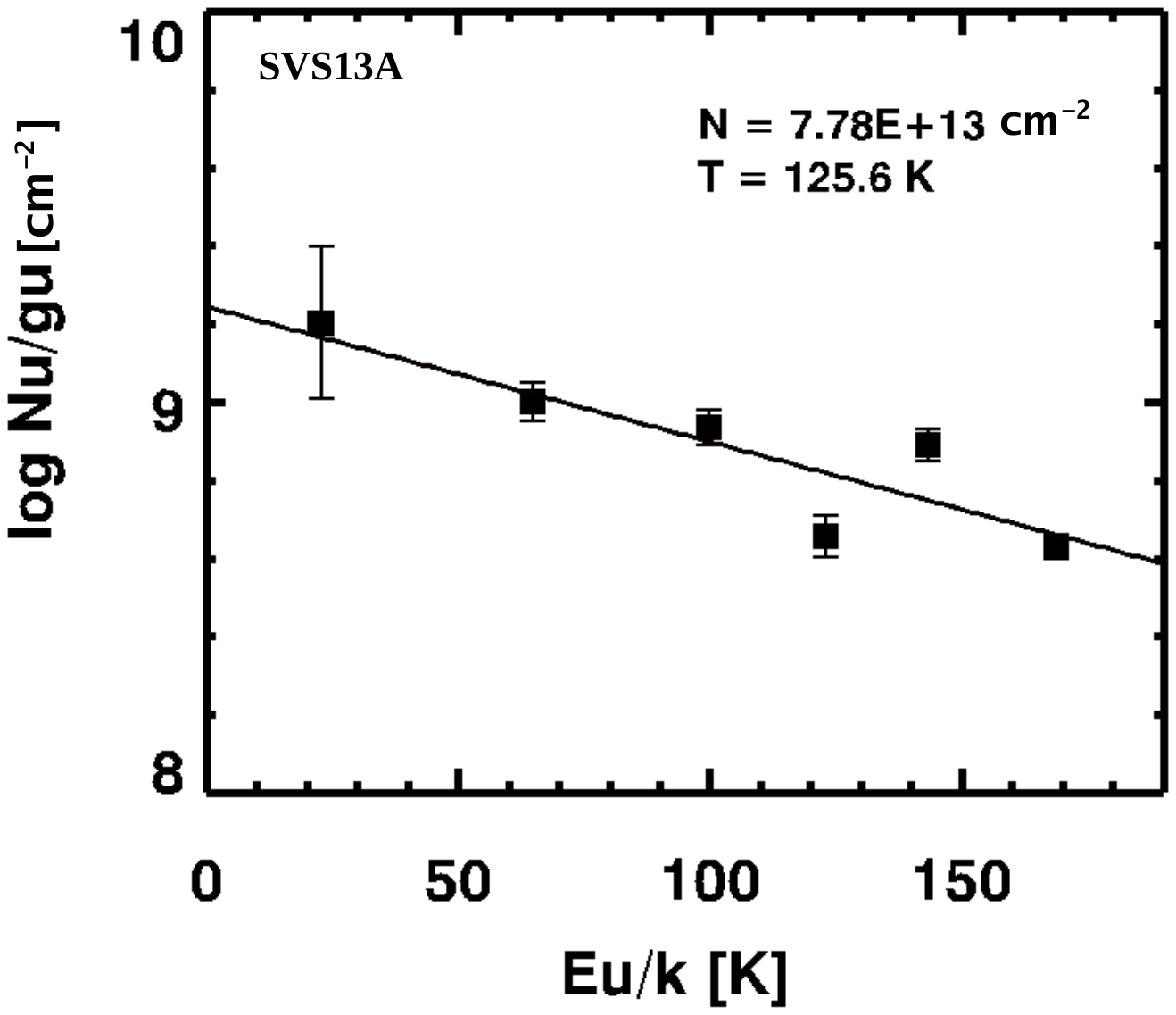}
 \end{minipage}
 \caption{Rotational diagram for CH$_3$OCHO obtained for various sources. The symbols represent the same as those depicted in Figure \ref{fig:rotational_diag_ch3oh_single}.}
\label{fig:rotational_diag_hcooch3}
\end{figure*}
\begin{figure*}
 \begin{minipage}{0.35\textwidth}
\includegraphics[width=\textwidth]{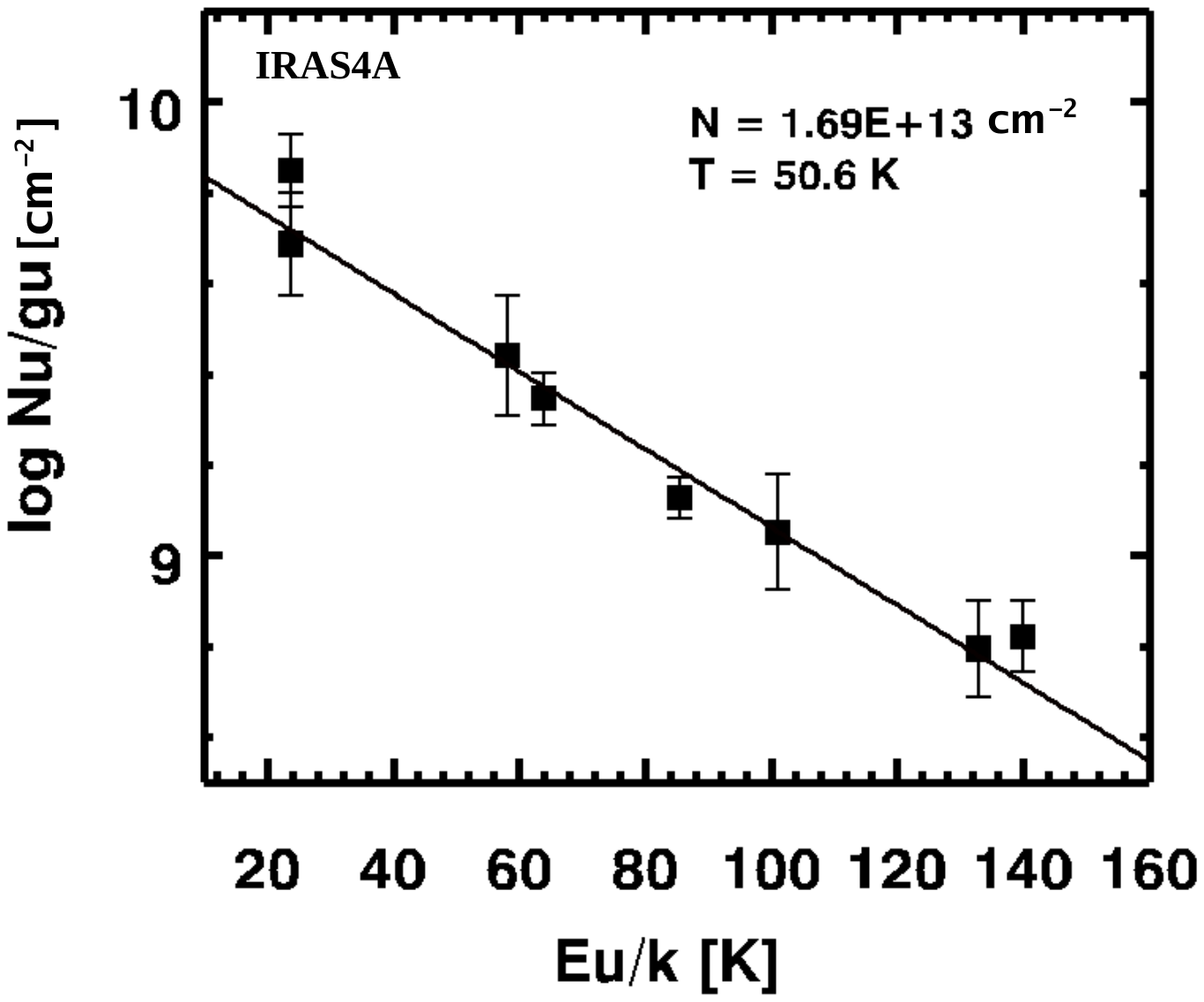}
\end{minipage}
\begin{minipage}{0.35\textwidth}
\includegraphics[width=\textwidth]{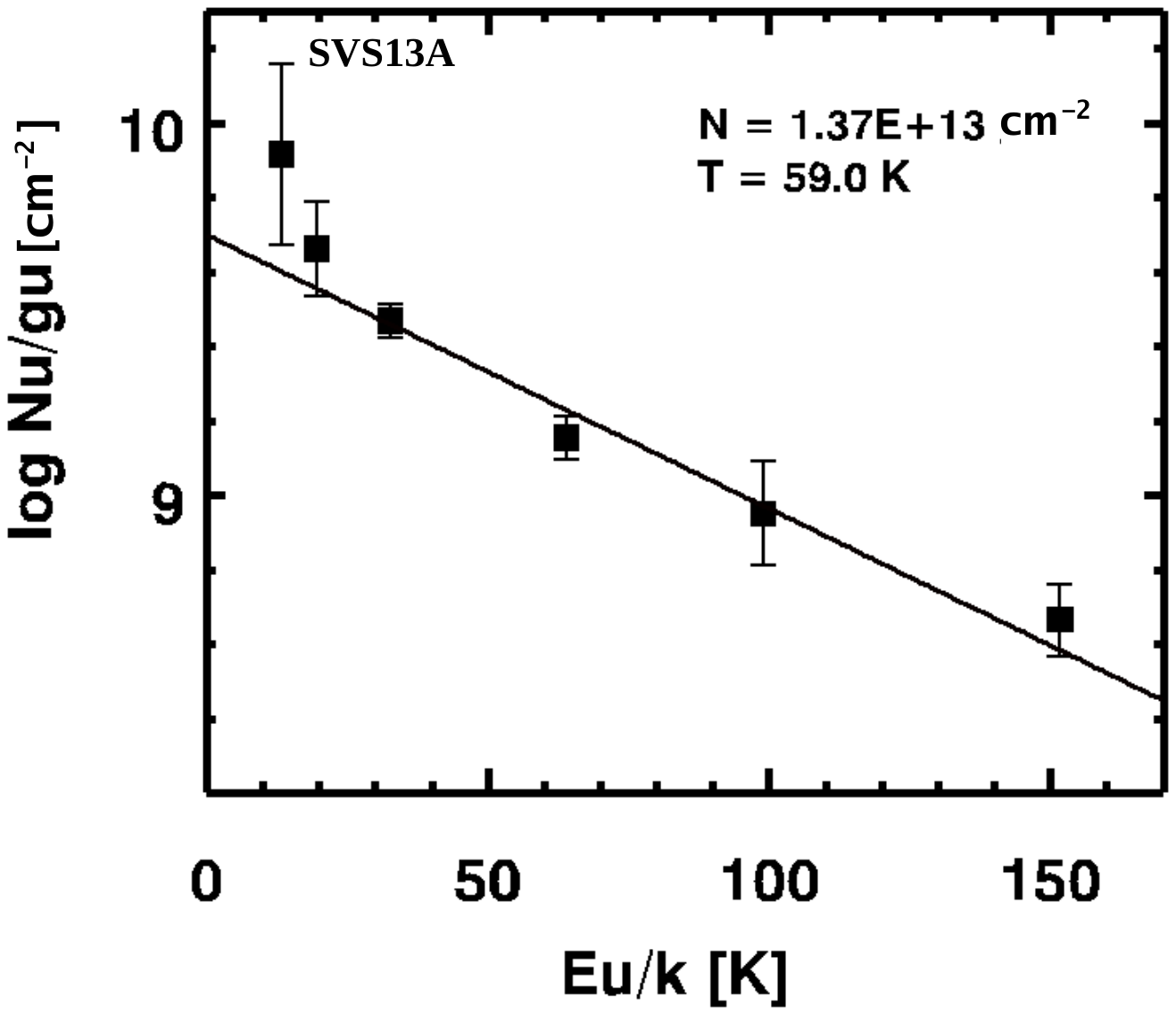}
\end{minipage}
\caption{Rotational diagram for C$_2$H$_5$OH obtained for various sources. Here, we consider the source size and beam size are same. The symbols represent the same as those depicted in Figure \ref{fig:rotational_diag_ch3oh_single}.}
\label{fig:rotational_diag_C2H5OH}
\end{figure*}
\begin{figure*}
\hskip 5cm
\includegraphics[width=8cm]{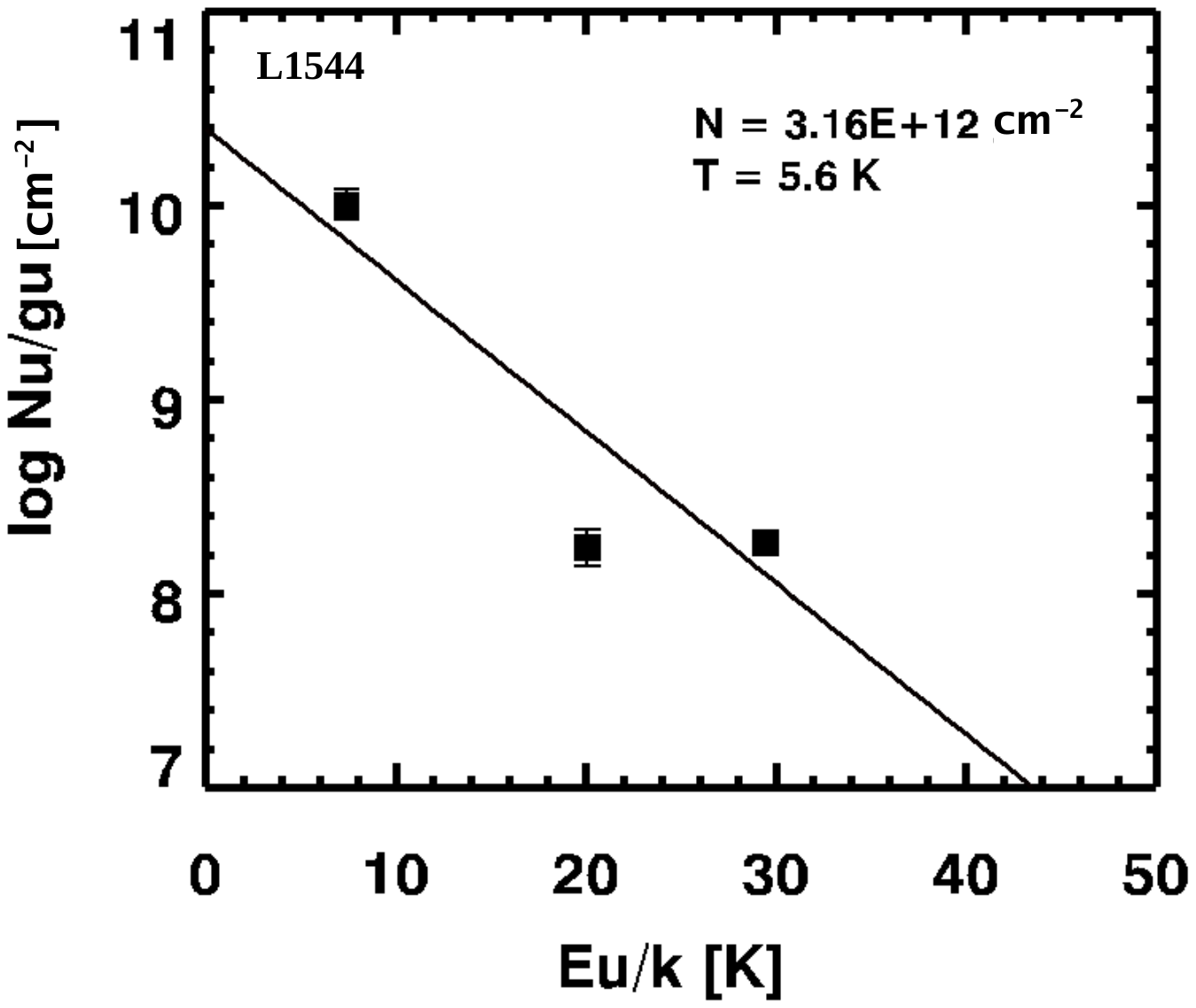}
\caption{Rotational diagram for HCCCHO obtained for one source. Here, we consider the source size and beam size are same. The symbols represent the same as those depicted in Figure \ref{fig:rotational_diag_ch3oh_single}.}
\label{fig:rotational_diag_HCCCHO}
\end{figure*}

\begin{figure*}
\hskip 5cm
\begin{minipage}{0.35\textwidth}
\includegraphics[width=\textwidth]{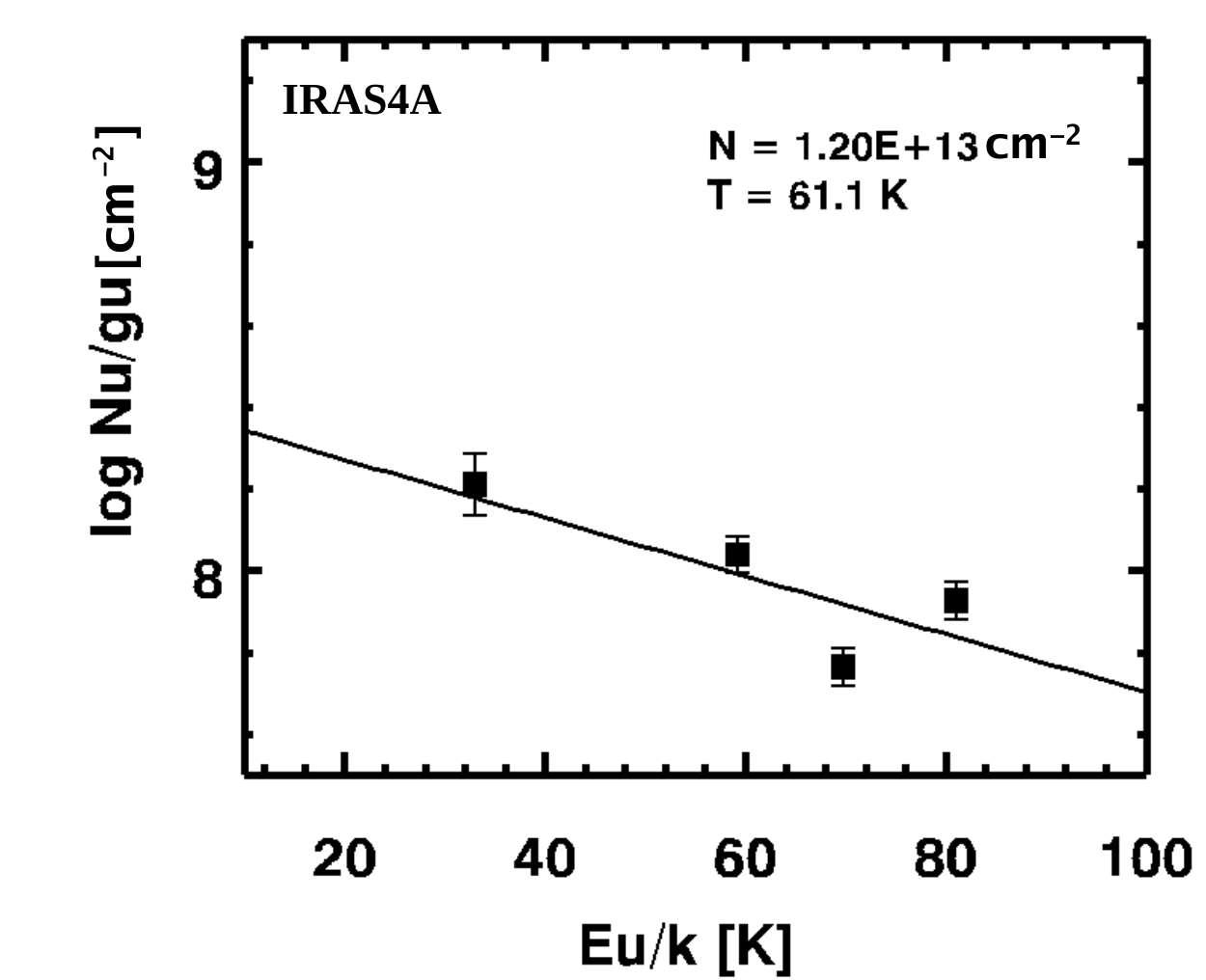}
\end{minipage}
\caption{Rotational diagram for CH$_3$OCH$_3$, v = 0 obtained towards IRAS4A. The symbols represent the same as those depicted in Figure \ref{fig:rotational_diag_ch3oh_single}.}
\label{fig:rotational_diag_CH3OCH3}
\end{figure*}
\begin{figure*}
 \begin{minipage}{0.35\textwidth}
\includegraphics[width=\textwidth]{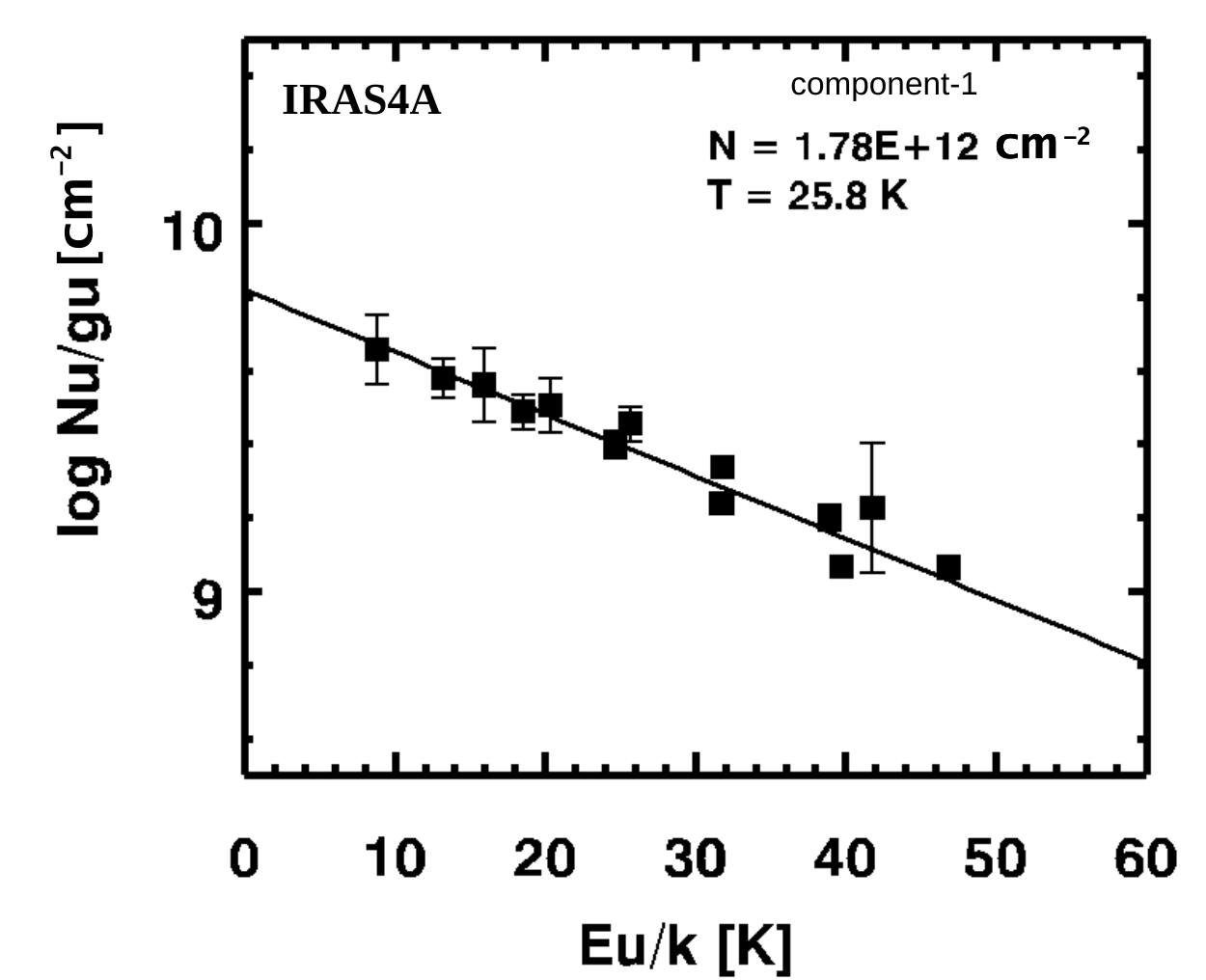}
\end{minipage}
\begin{minipage}{0.35\textwidth}
\includegraphics[width=\textwidth]{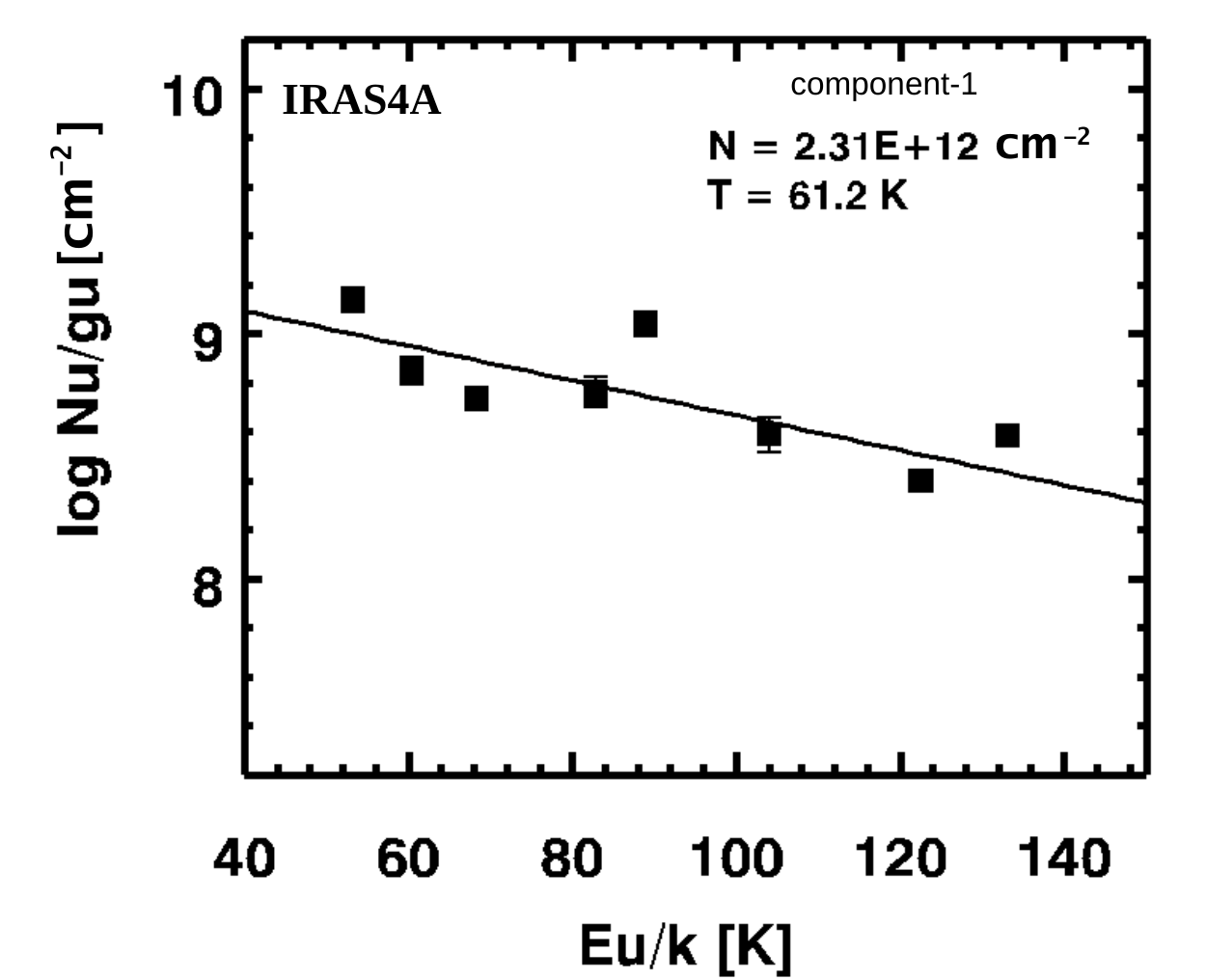}
\end{minipage}
\begin{minipage}{0.35\textwidth}
\includegraphics[width=\textwidth]{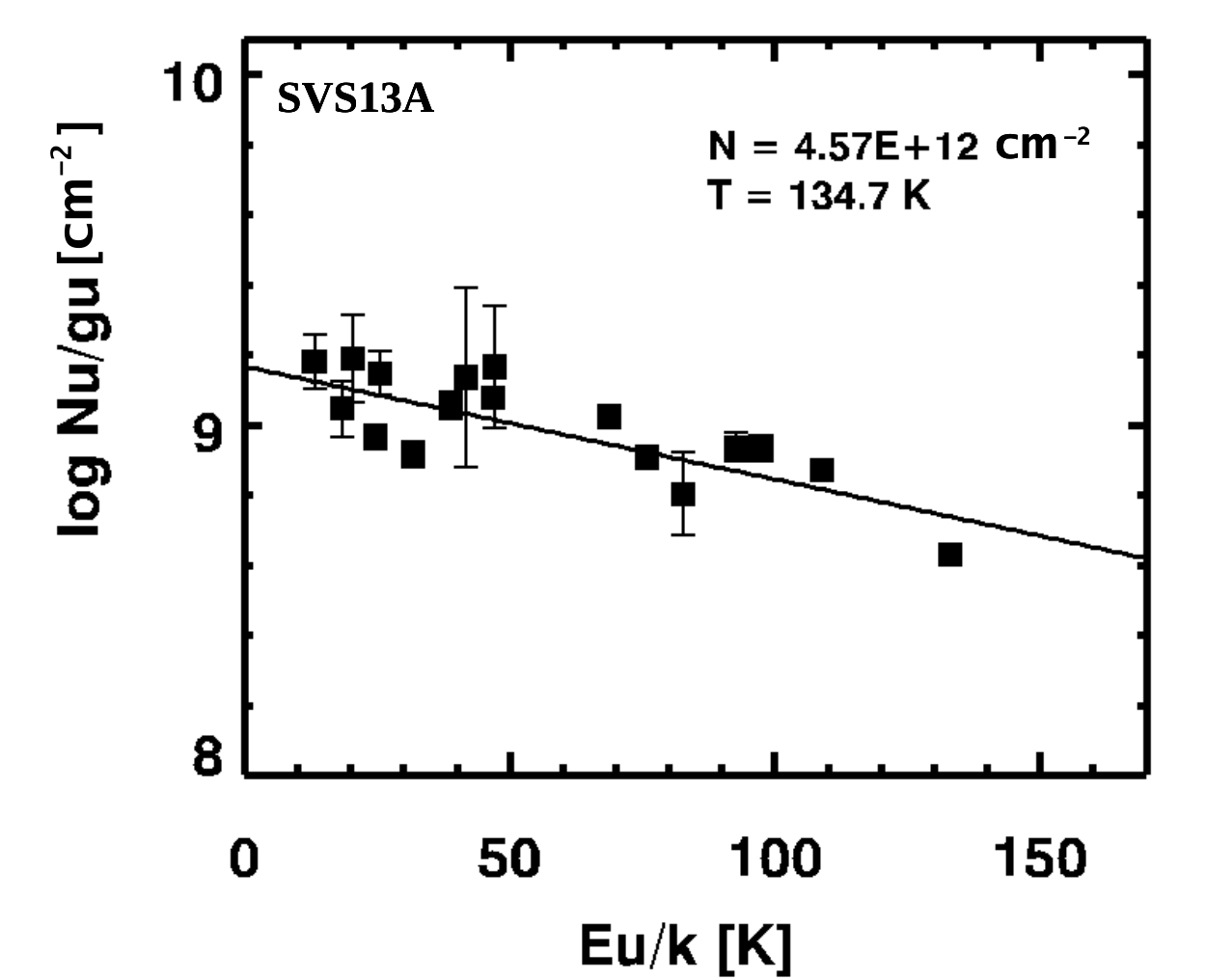}
\end{minipage}
\caption{Rotational diagram for CH$_3$CN obtained for various sources. The symbols represent the same as those depicted in Figure \ref{fig:rotational_diag_ch3oh_single}.}
\label{fig:rotational_diag_CH3CN}
\end{figure*}

\clearpage
\setcounter{figure}{0}    
\setcounter{table}{0}    
\section{Monte Carlo Markov chain method}\label{sec:mcmc}
\begin{figure*}
\centering
\includegraphics[width=6cm, height=16cm, angle=270]{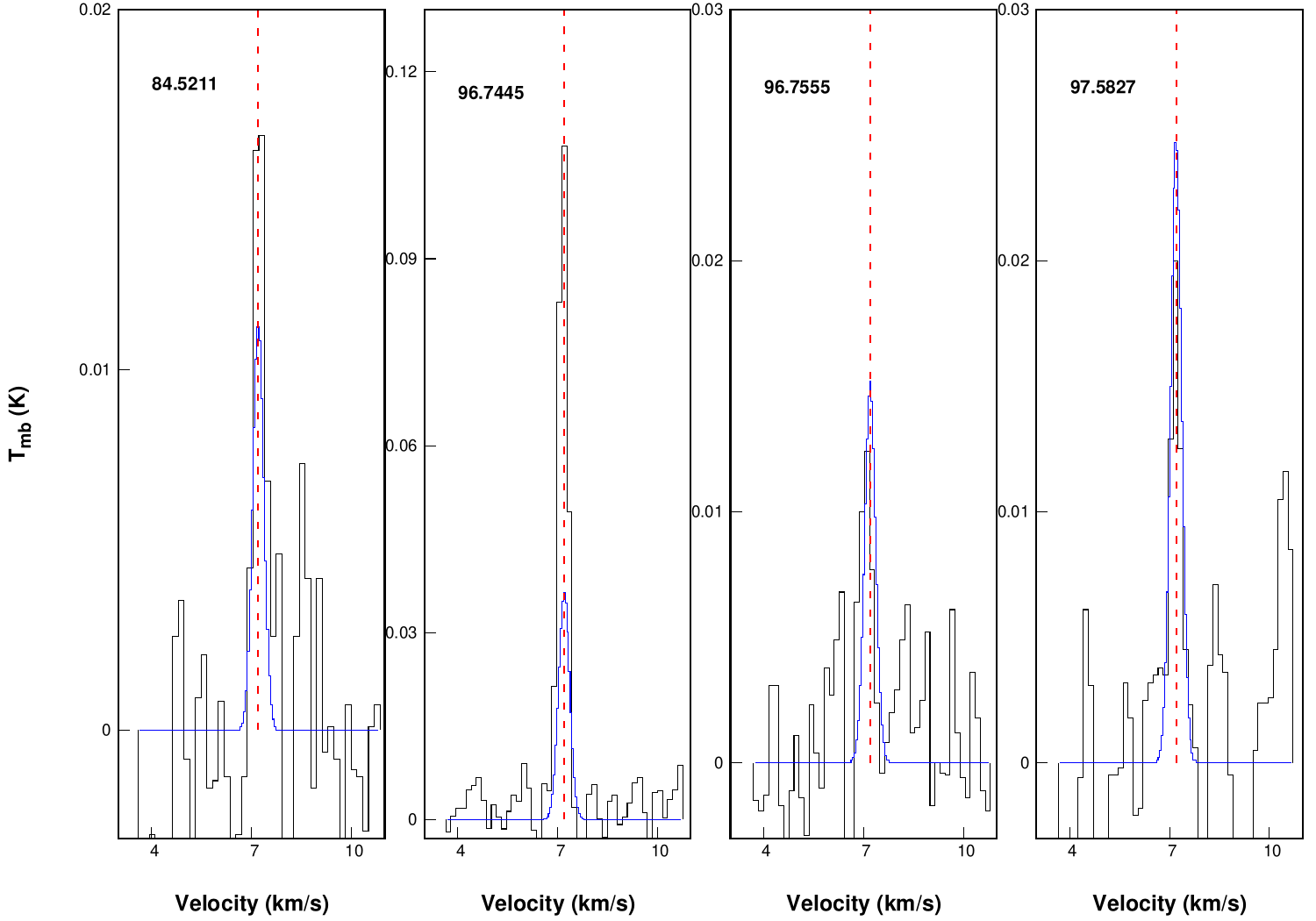}
\caption{MCMC fitting of the observed transitions of $\rm{CH_3OH}$ in L1544. Black lines represent the observed spectra, and blue represents the modeled spectral profile. The vertical red dashed line represents the V$_{lsr}$}.
\label{fig:ch3oh_l1544mcmc}
\end{figure*}

\begin{figure*}
\centering
\includegraphics[width=11cm, height=15cm, angle=270]{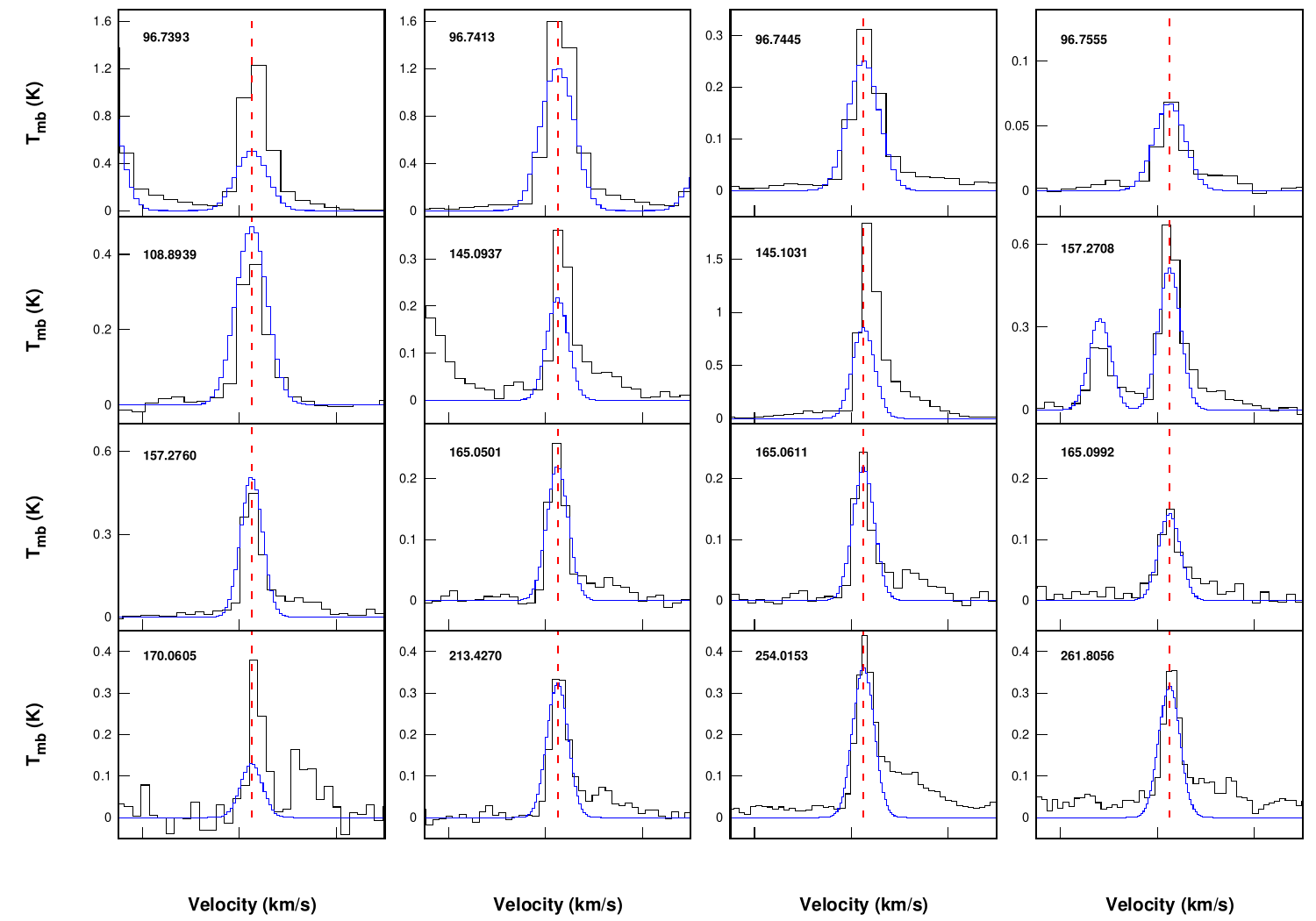}
\caption{Same as Figure \ref{fig:ch3oh_l1544mcmc} for $\rm{CH_3OH}$ in B1-b.}
\label{fig:ch3oh_barnardmcmc}
\end{figure*}

\begin{figure*}
\centering
\includegraphics[width=11cm, height=15cm, angle=270]{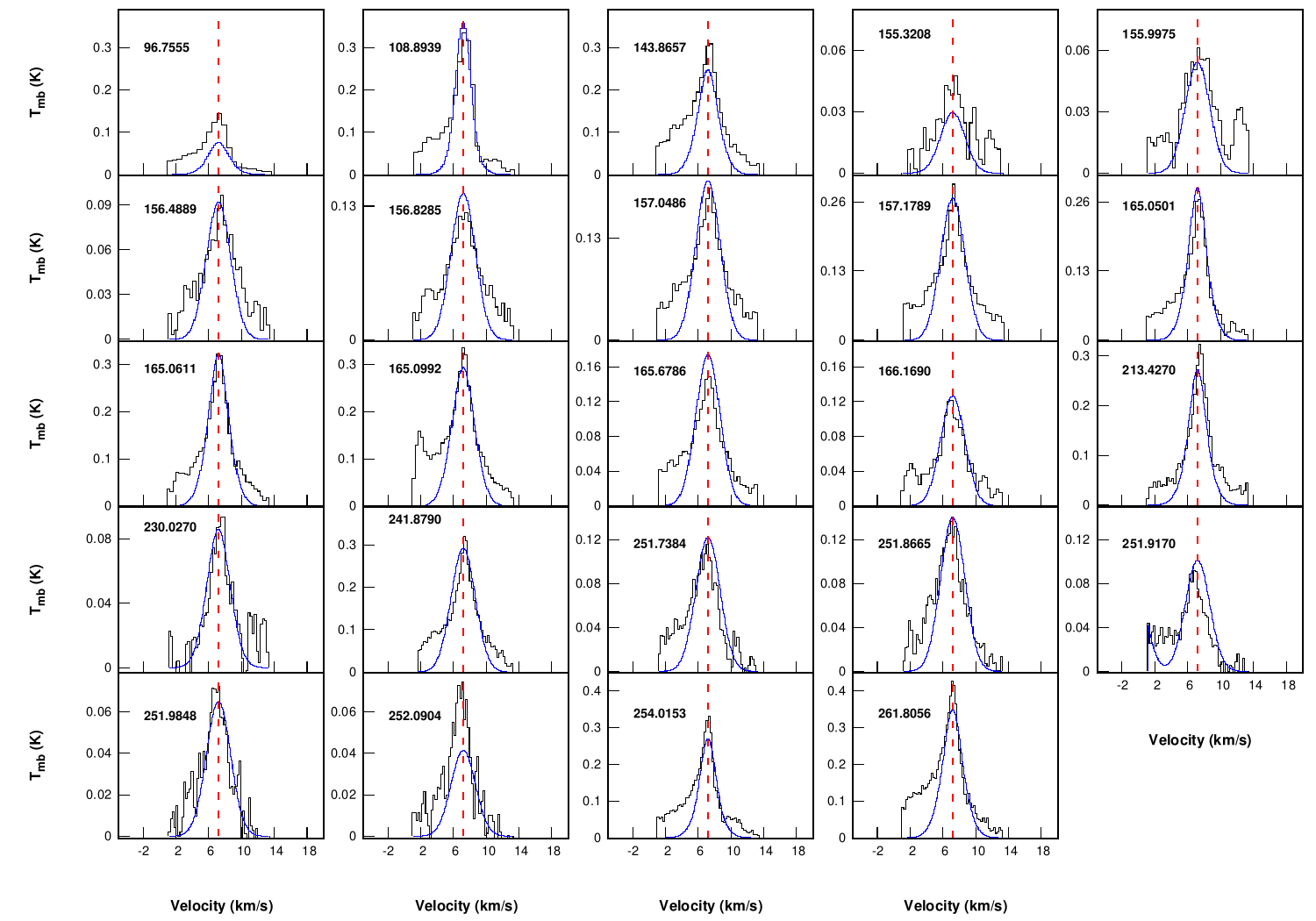}
\caption{Same as Figure \ref{fig:ch3oh_l1544mcmc} for $\rm{CH_3OH}$ in IRAS4A.}
\label{fig:ch3oh_irasmcmc}
\end{figure*}

\begin{figure*}
\centering
\includegraphics[width=11cm, height=15cm, angle=270]{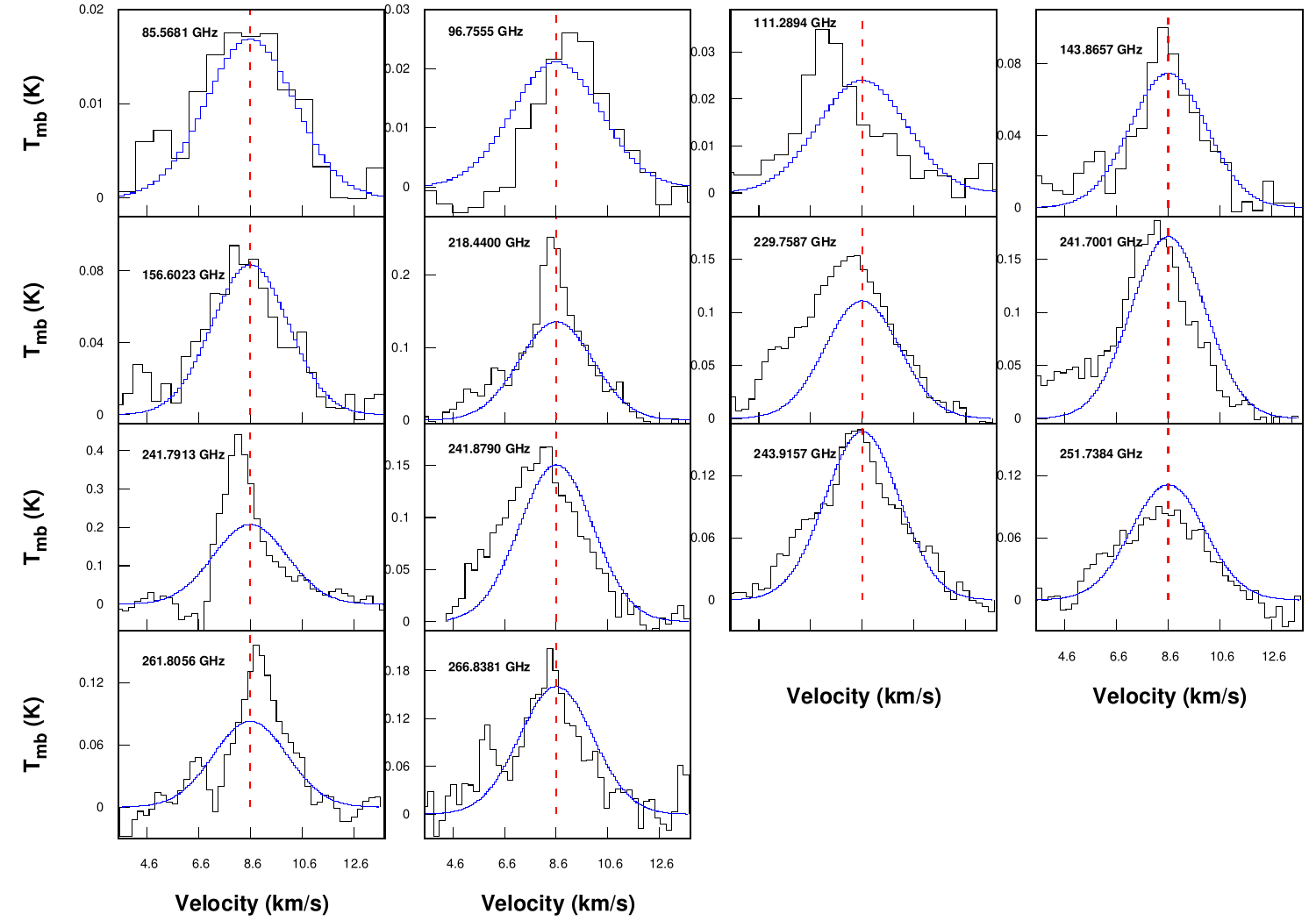}
\caption{Same as Figure \ref{fig:ch3oh_l1544mcmc} for $\rm{CH_3OH}$ in SVS13A.}
\label{fig:ch3oh_svsmcmc}
\end{figure*}

\begin{figure*}
\centering
\hskip -1.67cm
\includegraphics[width=17cm, height=19cm, angle=270]{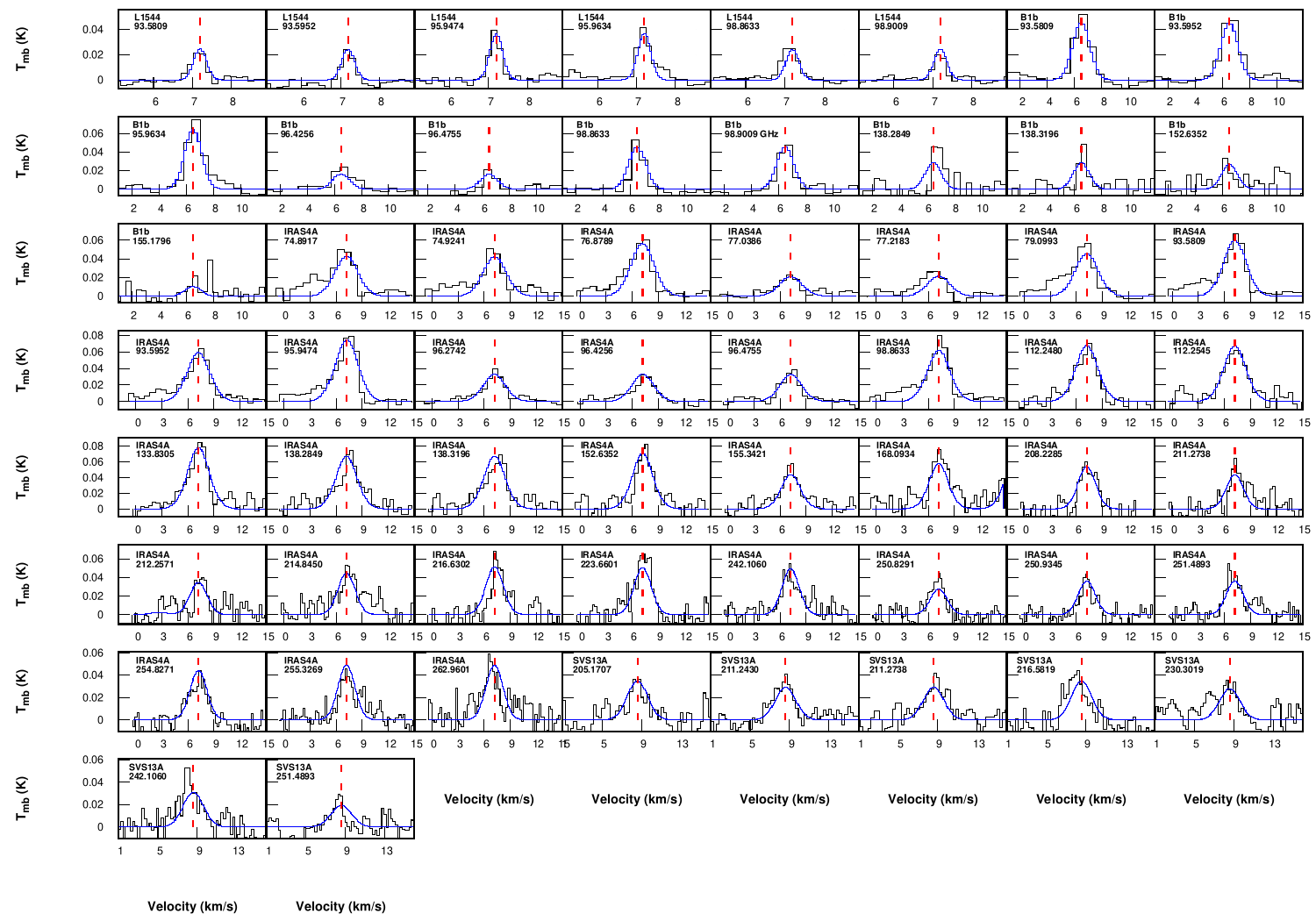}
\caption{Same as Figure \ref{fig:ch3oh_l1544mcmc} for $\rm{CH_3CHO}$.}
\label{fig:ch3cho_mcmc}
\end{figure*}

\begin{figure*}
\centering
\includegraphics[width=15cm, height=18cm, angle=270]{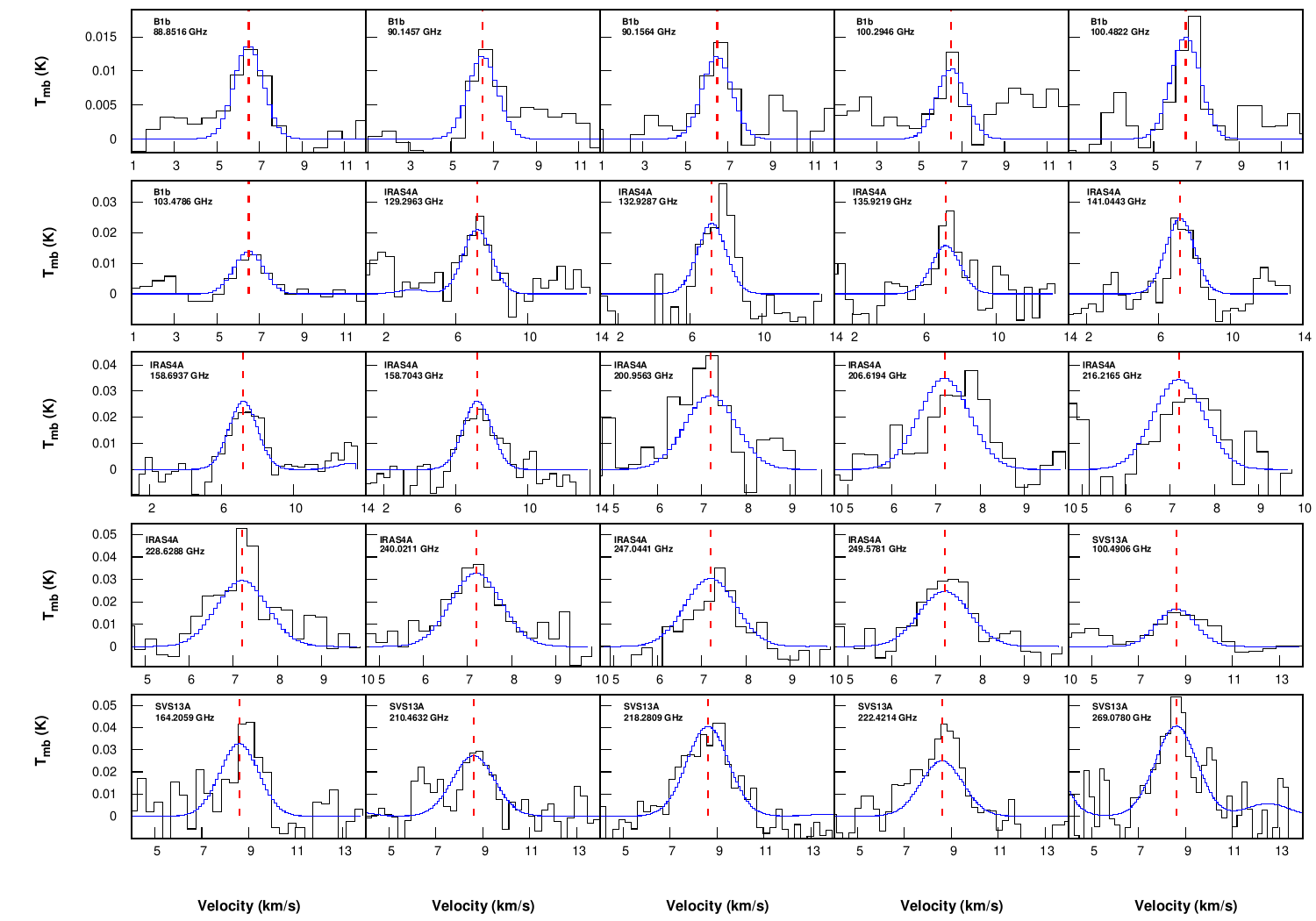}
\caption{Same as Figure \ref{fig:ch3oh_l1544mcmc} for $\rm{CH_3OCHO}$.}
\label{fig:ch3ocho_mcmc}
\end{figure*}

\begin{figure*}
\centering
\includegraphics[width=11cm, height=15cm, angle=270]{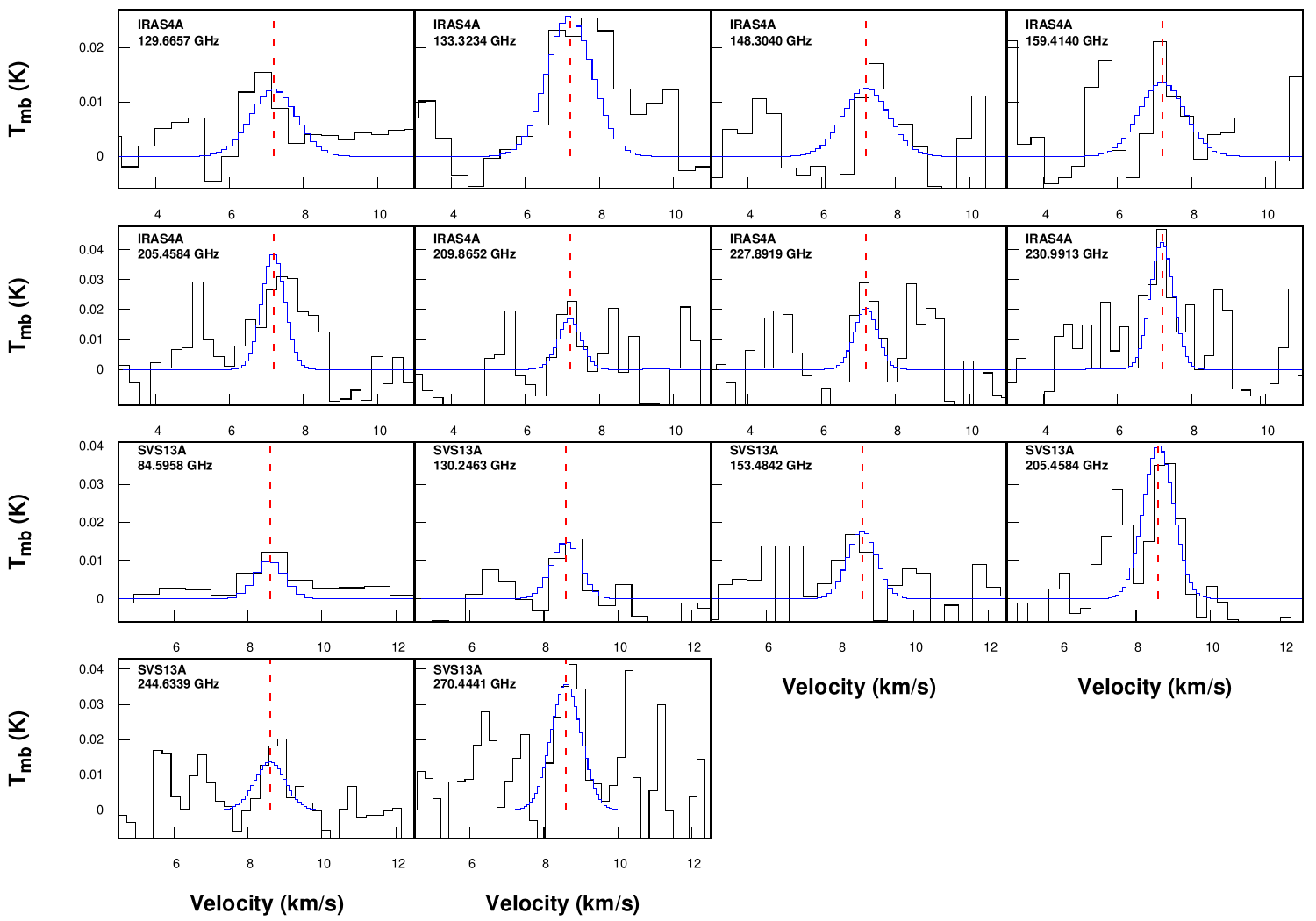}
\caption{Same as Figure \ref{fig:ch3oh_l1544mcmc} for $\rm{C_2H_5OH}$.}
\label{fig:c2h5oh_mcmc}
\end{figure*}

\begin{figure*}
\centering
\includegraphics[width=7cm, height=15cm, angle=270]{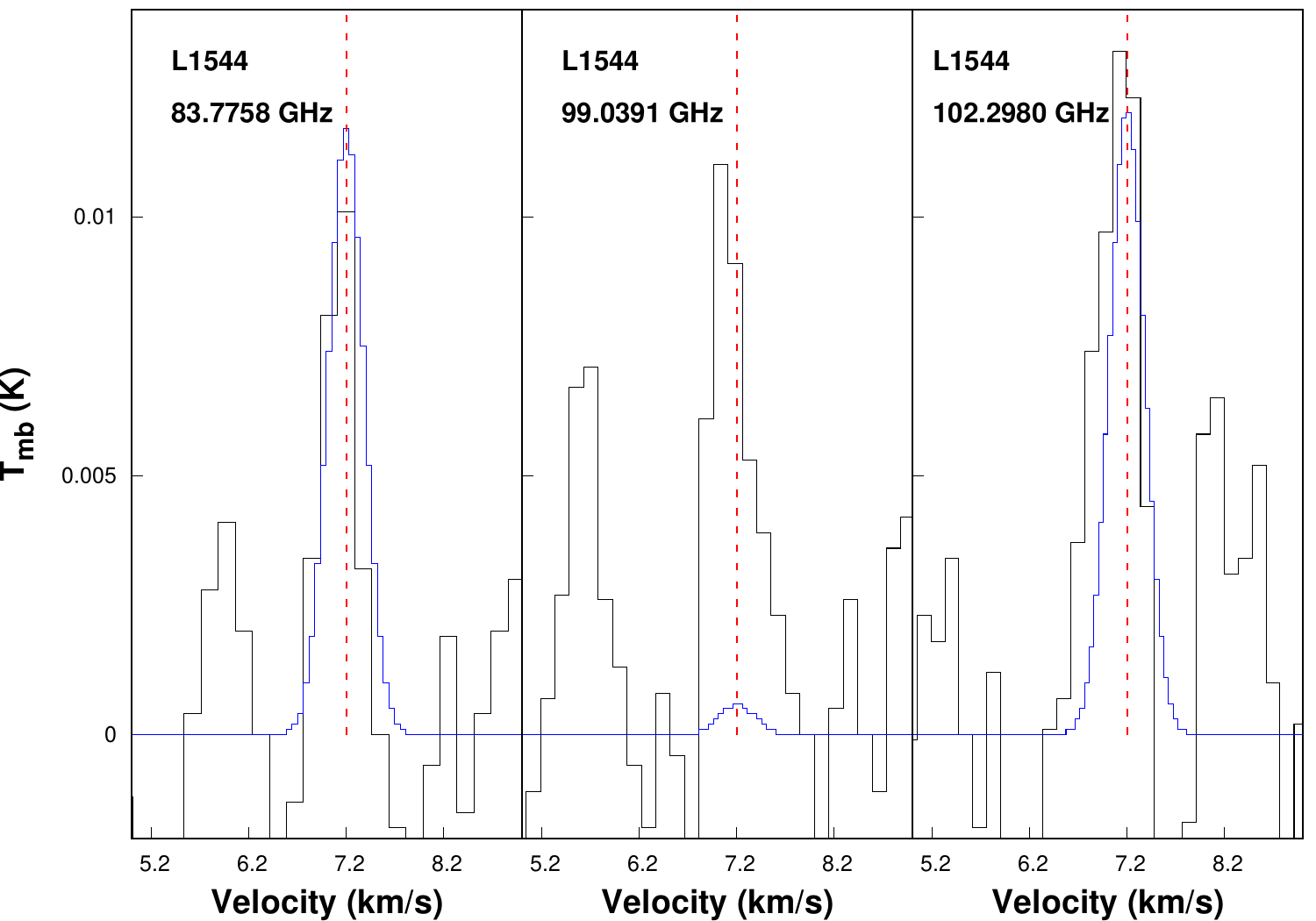}
\caption{Same as Figure \ref{fig:ch3oh_l1544mcmc} for $\rm{HCCCHO}$.}
\label{fig:hcccho_mcmc}
\end{figure*}

\begin{figure*}
\begin{minipage}{0.32\textwidth}
\includegraphics[width=\textwidth,angle=270]{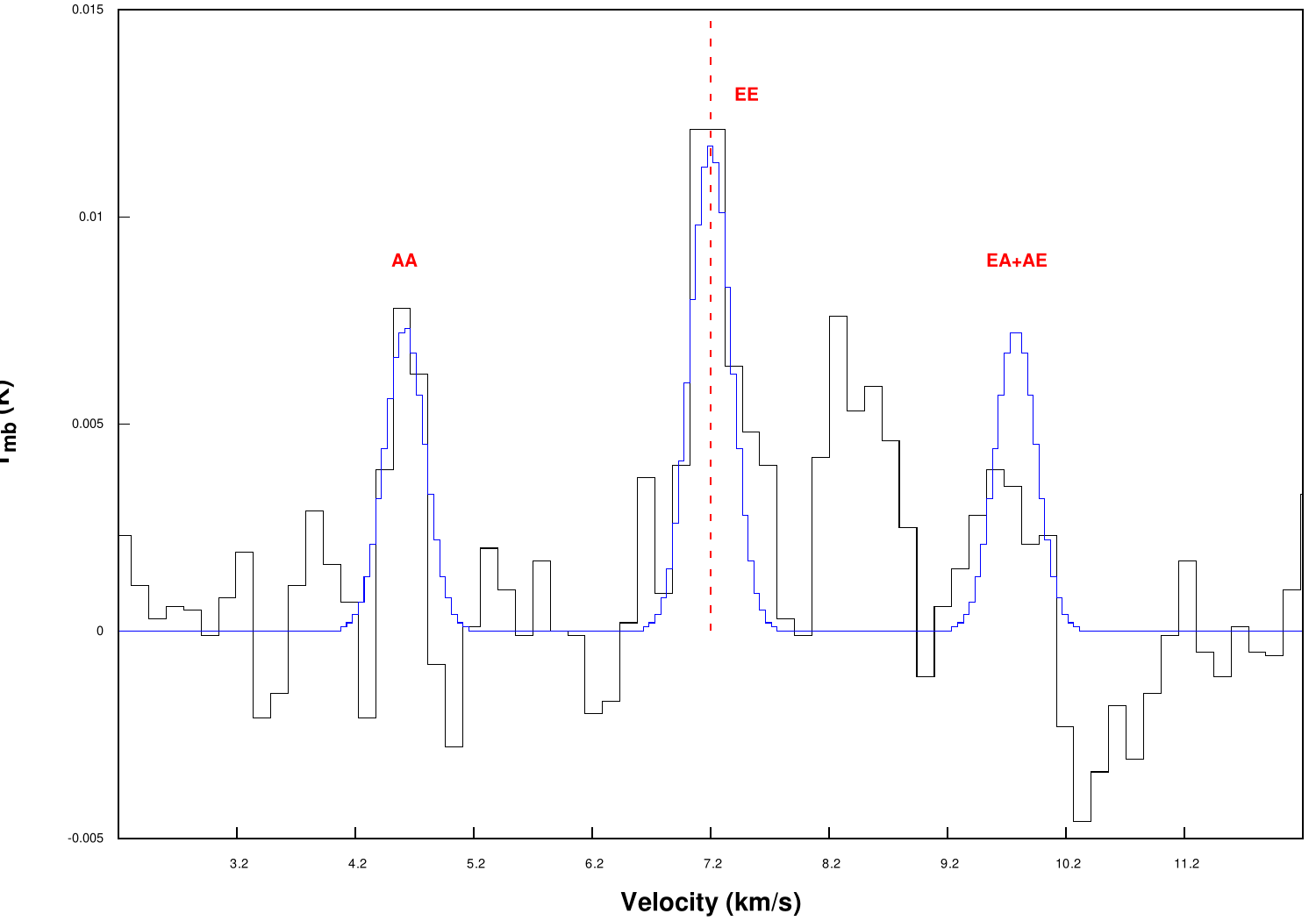}
\end{minipage}
\hskip 3.00 cm
\begin{minipage}{0.32\textwidth}
\includegraphics[width=\textwidth,angle=270]{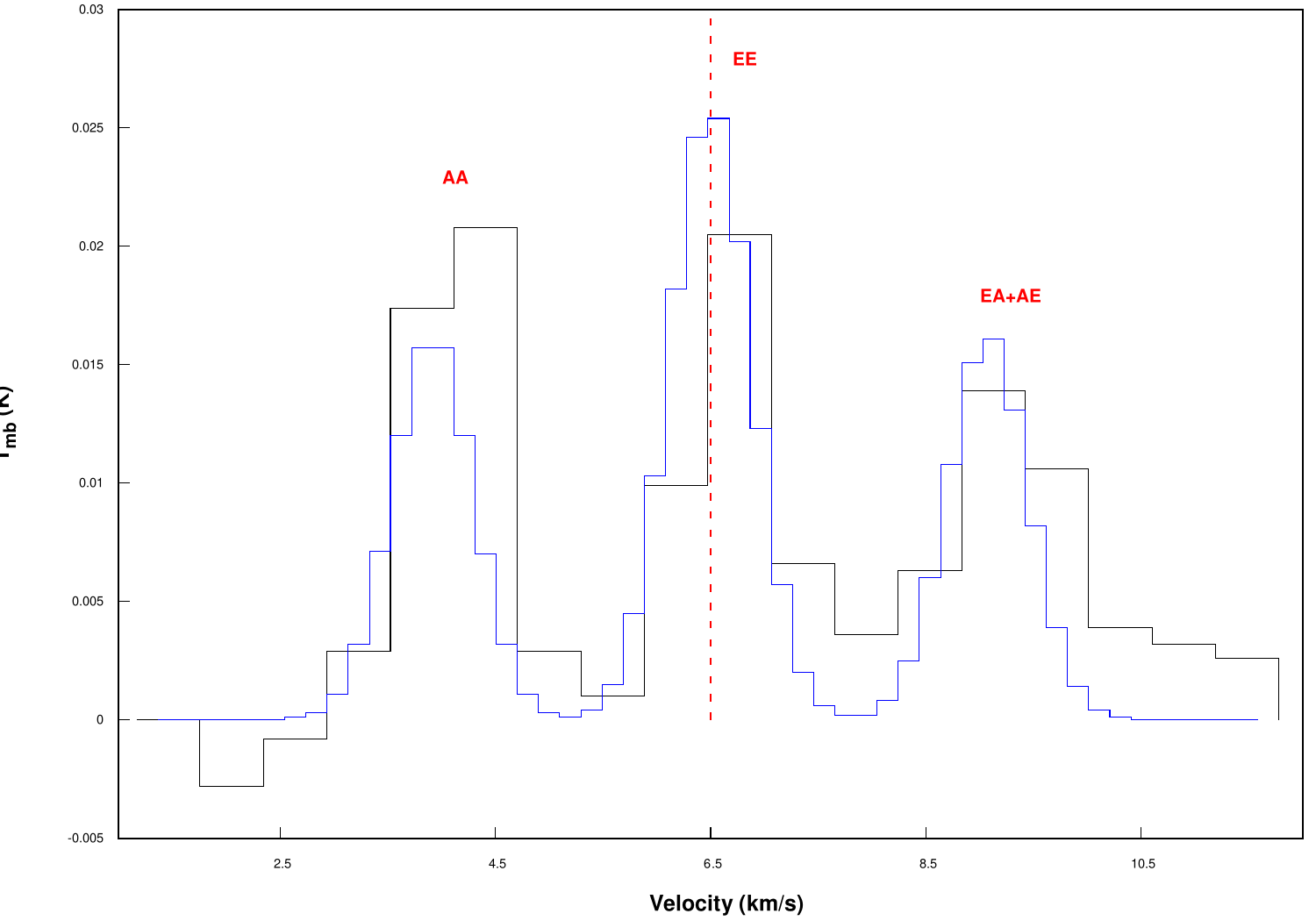}%
\end{minipage}
\newline
\hskip 6.0 cm
\begin{minipage}{0.32\textwidth}
\includegraphics[width=\textwidth,height=14.0 cm,angle=270]{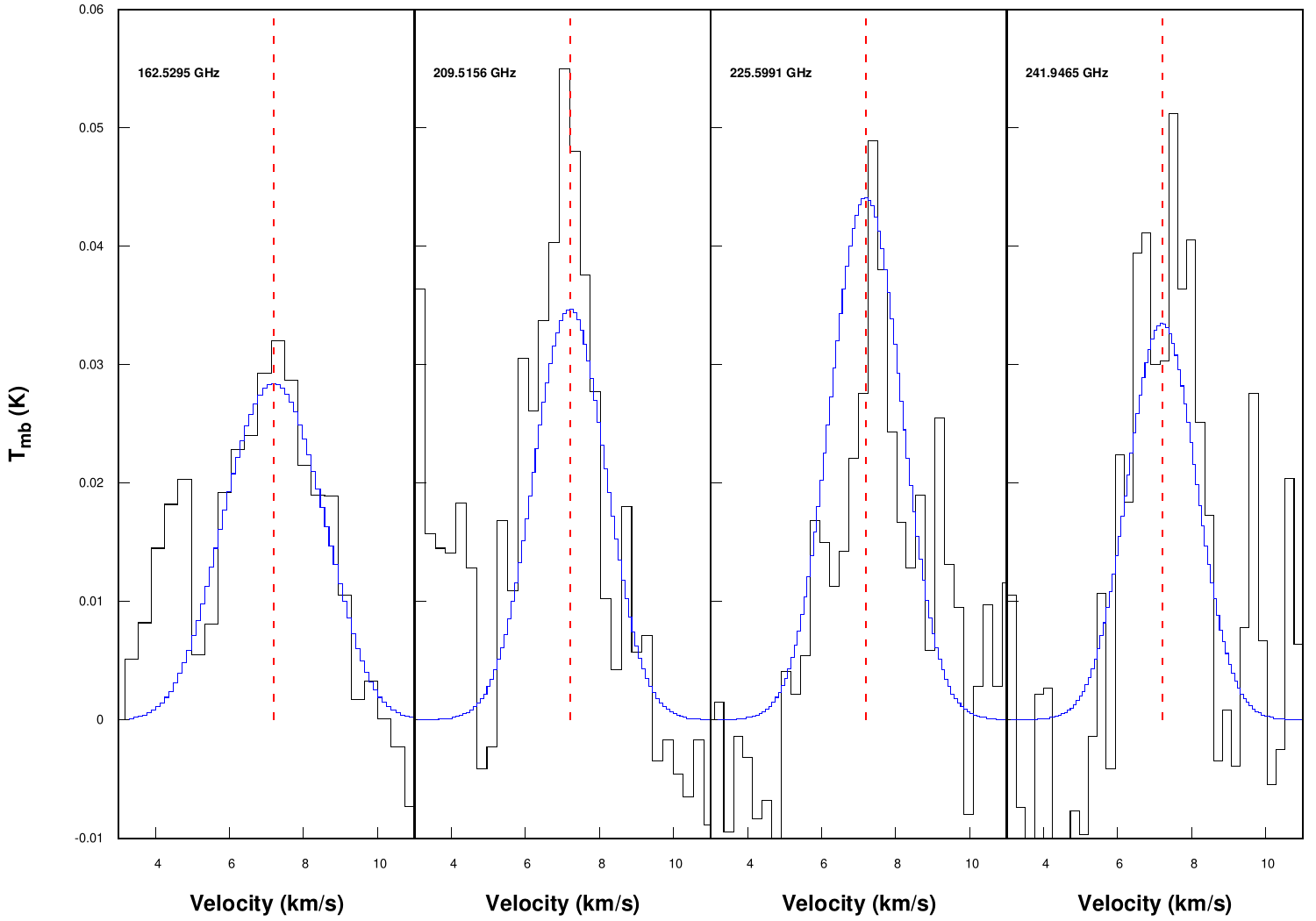}
\end{minipage}
\caption{Same as Figure \ref{fig:ch3oh_l1544mcmc} for $\rm{CH_3OCH_3}$ in L1544 (left), B1-b (right) and IRAS4A (bottom).}
\label{fig:ch3och3_mcmc}
\end{figure*}

The Markov Chain Monte Carlo (MCMC) is one of the most important and popular concept in Bayesian Statistics. Understanding the physical conditions in star-forming regions is a challenge nowadays as the chemistry in the star-forming regions are complicated and varies non-linearly with time and other physical conditions. In astrochemistry we use many trial and error grid-based analysis using simple statistics. But when the parameter space are large, heterogeneous and complex this approach fails. To interpret the astrochemical conditions a new method using Bayesian approach MCMC is implemented by an iterative process to solve it using simulation. The MCMC method is an interactive process that goes through all line parameters (e.g., molecular column density, excitation temperature, source size, line width) with a random walk and heads into the solution's space and the final solution is obtained by $\chi^2$ minimization. As well as rotational diagram method MCMC can also be used to obtain the best suitable physical conditions only for those species for which multiple or at-least more than two transitions are observed with different upstate energies. The $\chi^2$ is defined as,
\begin{equation}
{\rm {\chi_i}^2=\sum_{j=1}^{N_i} \frac{  (I_{obs,ij}-I_{model,ij})^2}{rms^2_i+(cal_i\times I_{obs,ij})^2}},
\end{equation}
where, ${\rm I_{obs,ij}}$ and ${\rm I_{model,ij}}$  are observed and modeled intensity in the channel j of transition i respectively,
rms$_i$ is the RMS of the spectrum i, and cal$_i$ is the calibration error. Here, the MCMC method is only used to fit the observed transitions of those species for which multiple transitions are observed. Here we employed the MCMC method from the CASSIS scripting interface to fit the observed lines of those seven COMs ($\rm{CH_3OH, CH_3CHO, CH_3OCHO, C_2H_5OH, HCCCHO, CH_3OCH_3, CH_3CN}$), which we have considered towards different star-forming regions. Since we obtained a beam average column density with the rotational diagram analysis, we ignored the beam filling factor for the MCMC calculation.
In some cases (CH$_3$OH iand CH$_3$CHO in IRAS4A), we employ a two-component MCMC fitting to have a better match with the observation. The components are considered based on their temperature. Details are described in section \ref{sec:results}. We also kept the LSR velocity constant at 7.2, 6.5, 7.2, 2.6, and 8.4 km/s, respectively, for L1544, B1-b, IRAS4A, L1157-mm, and SVS13A. For the fitting, we varied FWHM, column density, and excitation temperature (T$_{ex}$) and obtained the best fit parameters from the LTE fitting using MCMC method. 
All the parameters along with the best fitted values of the physical parameters are noted in Table \ref{table:mcmc_lte}. Spectral fit for all the transitions arose from various molecules are shown in Figure \ref{fig:ch3oh_l1544mcmc}, \ref{fig:ch3oh_barnardmcmc}, \ref{fig:ch3oh_irasmcmc}, \ref{fig:ch3oh_svsmcmc}, \ref{fig:ch3cho_mcmc}, \ref{fig:ch3ocho_mcmc}, \ref{fig:c2h5oh_mcmc}, \ref{fig:hcccho_mcmc}, and \ref{fig:ch3och3_mcmc}.
\begin{table*}
\tiny{
\caption{Summary of the best fitted line parameters obtained by using MCMC method. \label{table:mcmc_lte}}
\centering
\begin{tabular}{|c|c|c|c|c|c|c|c|c|c|c|}
\hline
\hline
Species&Source&Frequency&FWHM range&Best fitted&Column density&Best fitted column&T$_{ex}$ range&Best fitted&Best fitted&Optical depth \\
&&(GHz)&used (km.s$^{-1}$)&FWHM (Km.s$^{-1}$)&range used (cm$^{-2}$)&density (cm$^{-2}$) &used (K)& T$_{ex}$ (K)&V$_{LSR}$ (km.s$^{-1}$)& ($\tau$)\\
\hline
\hline
CH$_3$OH&L1544&84.5211&&&&&&&&$1.98\times10^{-3}$\\
&&96.7445&0.3-0.5&0.39$\pm$0.03&$1.0\times10^{11}-1.0\times10^{14}$&(4.60$\pm$0.92)$\times10^{12}$&3.5-17.0&12.9$\pm$3.07&7.2&$5.86\times10^{-3}$\\
&&96.7555&&&&&&&&$2.44\times10^{-3}$\\
&&97.5828&&&&&&&&$3.97\times10^{-3}$\\
\cline{2-11}
&B1-b&96.7393&&&&&&&&$2.08\times10^{-1}$\\
&&96.7413&&&&&&&&$5.98\times10^{-1}$\\
 &&96.7445&1.0-3.0&1.50$\pm$0.10&$1.0\times10^{13}-1.0\times10^{15}$&$(1.60\pm0.12)\times10^{14}$&5.0-25.0&7.27$\pm$0.86&6.5&$9.83\times10^{-2}$\\
&&96.7555&&&&&&&&$2.55\times10^{-2}$\\
&&108.8939&&&&&&&&$1.83\times10^{-1}$\\
\cline{3-11}
&&145.0937&&&&&&&&$5.11\times10^{-2}$\\
&&145.1031&&&&&&&&$2.18\times10^{-1}$\\
&&157.2708&&&&&&&&$1.25\times10^{-1}$\\
&&157.2760&1.0-3.0&1.05$\pm$0.08&$1.0\times10^{13}-1.0\times10^{15}$&$(5.90\pm0.51)\times10^{13}$&5.0-25.0&9.06$\pm$1.26&6.5&$1.23\times10^{-1}$\\
&&165.0501&&&&&&&&$5.15\times10^{-2}$\\
&&165.0611&&&&&&&&$5.13\times10^{-2}$\\
&&165.0992&&&&&&&&$3.31\times10^{-2}$\\
&&170.0605&&&&&&&&$3.04\times10^{-2}$\\
\cline{3-11}
&&213.4270&&&&&&&&$1.47\times10^{-1}$\\
&&254.0153&1.0-3.0&1.0$\pm$0.09&$1.0\times10^{13}-1.0\times10^{15}$&$(1.20\pm0.32)\times10^{14}$&5.0-25.0&6.77$\pm$0.95&6.5&$1.89\times10^{-1}$\\
&&261.8056&&&&&&&&$1.68\times10^{-1}$\\
\cline{2-11}
&IRAS4A&96.7555&&&&&&&&$1.14\times10^{-2}$/$2.62\times10^{-3}$\\
&&108.8939&&&&&&&&$1.52\times10^{-1}$/$3.57\times10^{-3}$\\
&&143.8657&&&&&&&&$3.50\times10^{-2}$/$6.90\times10^{-3}$\\
&&155.3208&&&&&&&&$2.89\times10^{-10}$/$1.07\times10^{-3}$\\
&&155.9975&&&&&&&&$1.73\times10^{-8}$/$1.99\times10^{-3}$\\
&&156.4889&&&&&&&&$6.71\times10^{-7}$/$3.41\times10^{-3}$\\
&&156.8285&&&&&&&&$1.69\times10^{-5}$/$5.36\times10^{-3}$\\
&&157.0486&&&&&&&&$2.75\times10^{-4}$/$7.70\times10^{-3}$\\
&&157.1789&&&&&&&&$2.88\times10^{-3}$/$1.00\times10^{-2}$\\
&&165.0501&&&&$Component 1$&&&&$6.79\times10^{-2}$/$5.78\times10^{-3}$\\
&&165.0611&1.0-3.5&3.03$\pm$0.27&$1.0\times10^{13}-1.0\times10^{15}$&$(1.80\pm0.29)\times10^{14}$&5.0-30.0&5.62$\pm$1.68&7.2&$4.94\times10^{-2}$/$8.42\times10^{-3}$\\
&&165.0992&&&&&&&&$1.99\times10^{-2}$/$9.62\times10^{-3}$\\
&&165.6786&&&&$Component 2$&&&&$7.34\times10^{-5}$/$6.48\times10^{-3}$\\
&&166.1690&1.0-3.5&3.50$\pm$0.47&$1.0\times10^{13}-1.0\times10^{15}$&$(2.00\pm0.28)\times10^{14}$&30.0-70.0&35.14$\pm$3.86&7.2&$4.65\times10^{-6}$/$4.66\times10^{-3}$\\
&&213.4270&&&&&&&&$7.56\times10^{-2}$/$5.13\times10^{-3}$\\
&&230.0270&&&&&&&&$3.92\times10^{-3}$/$2.88\times10^{-3}$\\
&&241.8790&&&&&&&&$1.37\times10^{-3}$/$1.05\times10^{-2}$\\
&&251.7384&&&&&&&&$9.90\times10^{-7}$/$4.28\times10^{-3}$\\
&&251.8665&&&&&&&&$5.29\times10^{-5}$/$5.01\times10^{-3}$\\
&&251.9170&&&&&&&&$1.54\times10^{-4}$/$3.63\times10^{-3}$\\
&&251.9848&&&&&&&&$4.59\times10^{-9}$/$2.33\times10^{-3}$\\
&&252.0904&&&&&&&&$1.36\times10^{-10}$/$1.49\times10^{-3}$\\
&&254.0153&&&&&&&&$1.13\times10^{-1}$/$3.85\times10^{-3}$\\
&&261.8056&&&&&&&&$7.95\times10^{-2}$/$8.53\times10^{-3}$\\
\cline{2-11}
&SVS13A&85.5681&&&&&&&&$3.08\times10^{-4}$\\
&&96.7555&2.0-4.5&4.04$\pm$0.59&$1.0\times10^{13}-1.0\times10^{15}$&$(3.00\pm2.42)\times10^{14}$&40.0-200.0&96.72$\pm$18.77&8.6&$3.51\times10^{-4}$\\
&&111.2894&&&&&&&&$3.65\times10^{-4}$\\
\cline{3-11}
&&143.8657&&&&&&&&$1.01\times10^{-3}$\\
&&156.6023&2.0-3.5&3.47$\pm$0.39&$1.0\times10^{13}-1.0\times10^{15}$&$(2.80\pm0.62)\times10^{14}$&40.0-100.0&96.70$\pm$14.03&8.6&$1.09\times10^{-3}$\\
\cline{3-11}
&&218.4400&&&&&&&&$2.37\times10^{-3}$\\
&&229.7587&&&&&&&&$1.93\times10^{-3}$\\
&&241.7001&&&&&&&&$2.97\times10^{-3}$\\
&&241.7913&&&&&&&&$3.60\times10^{-3}$\\
&&241.8790&2.0-3.5&3.36$\pm$0.24&$1.0\times10^{13}-1.0\times10^{15}$&$(1.50\pm0.17)\times10^{14}$&40.0-100.0&68.69$\pm$8.41&8.6&$2.61\times10^{-3}$\\
&&243.9157&&&&&&&&$2.81\times10^{-3}$\\
&&251.7384&&&&&&&&$1.92\times10^{-3}$\\
&&261.8056&&&&&&&&$1.43\times10^{-3}$\\
&&266.8381&&&&&&&&$2.76\times10^{-3}$\\
\hline
\end{tabular}}
\end{table*}
\clearpage
{\tiny
\centering
\begin{tabular}{|c|c|c|c|c|c|c|c|c|c|c|}
\hline
\hline
Species&Source&Frequency&FWHM range&Best fitted&Column density&Best fitted column&T$_{ex}$ range&Best fitted&Best fitted&Optical depth \\
&&(GHz)&used (km.s$^{-1}$)&FWHM (Km.s$^{-1}$)&range used (cm$^{-2}$)&density (cm$^{-2}$) &used (K)& T$_{ex}$ (K)&V$_{LSR}$ (km.s$^{-1}$)& ($\tau$)\\
\hline
\hline
CH$_3$CHO&L1544&93.5809&&&&&&&&$1.32\times10^{-2}$\\
&&93.5952&&&&&&&&$1.31\times10^{-2}$\\
 &&95.9474&0.3-0.6  &0.39$\pm$0.09&$1.0\times10^{11}-1.0\times10^{13}$&$(6.10\pm3.55)\times10^{11}$&3.5-20.0&6.04$\pm$4.42&7.2&$1.93\times10^{-2}$\\
&&95.9634&&&&&&&&$1.96\times10^{-2}$\\
&&98.8633&&&&&&&&$1.25\times10^{-2}$\\
&&98.9009&&&&&&&&$1.26\times10^{-1}$\\
\cline{2-11}
&B1-b&93.5809&&&&&&&&$1.65\times10^{-2}$\\
&&93.5952&&&&&&&&$1.64\times10^{-2}$\\
&&95.9634&&&&&&&&$2.30\times10^{-2}$\\
&&96.4256&0.5-2.0&1.49$\pm$0.36&$1.0\times10^{11}-5.0\times10^{13}$&$(4.70\pm2.39)\times10^{12}$&6.0-100.0&7.52$\pm$6.61&6.5&$5.78\times10^{-3}$\\
&&96.4755&&&&&&&&$5.70\times10^{-3}$\\
&&98.8633&&&&&&&&$1.59\times10^{-2}$\\
&&98.9009&&&&&&&&$1.61\times10^{-2}$\\
\cline{3-11}
&&138.2849&0.5-1.5&1.38$\pm$0.24&$1.0\times10^{11}-5.0\times10^{13}$&$(4.10\pm2.84)\times10^{12}$&6.0-100.0&10.81$\pm$3.63&6.5&$5.08\times10^{-3}$\\
&&138.3196&&&&&&&&$5.11\times10^{-3}$\\
&&152.6352&&&&&&&&$4.59\times10^{-3}$\\
&&155.1796&&&&&&&&$1.81\times10^{-3}$\\
\cline{2-11}
&IRAS4A&74.8917&&&&&&&&$8.67\times10^{-3}$/$1.89\times10^{-4}$\\
&&74.9241&&&&&&&&$8.60\times10^{-3}$/$1.88\times10^{-4}$\\
&&76.8789&&&&&&&&$1.14\times10^{-2}$/$2.13\times10^{-4}$\\
&&77.0386&&&&&&&&$3.79\times10^{-3}$/$1.41\times10^{-4}$\\
&&77.2183&&&&&&&&$3.80\times10^{-3}$/$1.41\times10^{-4}$\\
&&79.0993&&&&&&&&$8.75\times10^{-3}$/$1.98\times10^{-4}$\\
&&93.5809&&&&&&&&$9.64\times10^{-3}$/$2.85\times10^{-4}$\\
&&93.5952&&&&&&&&$9.58\times10^{-3}$/$2.85\times10^{-4}$\\
&&95.9474&&&&&&&&$1.22\times10^{-2}$/$3.13\times10^{-4}$\\
&&96.2742&&&&&&&&$4.56\times10^{-3}$/$2.32\times10^{-4}$\\
&&96.4256&&&&&&&&$4.56\times10^{-3}$/$2.31\times10^{-4}$\\
&&96.4755&&&&&&&&$4.51\times10^{-3}$/$2.31\times10^{-4}$\\
&&98.8633&&&&&&&&$9.56\times10^{-3}$/$2.98\times10^{-4}$\\
&&112.248&&&&&&&&$9.02\times10^{-3}$/$3.88\times10^{-4}$\\
&&112.2545&&&&&&&&$8.97\times10^{-3}$/$3.87\times10^{-4}$\\
&&133.8305&&&&&&&&$8.85\times10^{-3}$/$5.22\times10^{-4}$\\
&&138.2849&&&&&&&&$6.88\times10^{-3}$/$5.08\times10^{-4}$\\
&&138.3196&&&&Component 1&&&&$6.92\times10^{-3}$/$5.08\times10^{-4}$\\
&&152.6352&2.0-3.5&3.30$\pm$0.22&$1.0\times10^{12}-1.0\times10^{14}$&$(1.30\pm0.21)\times10^{13}$&8.0-30.0&11.07$\pm$3.48&7.2&$6.27\times10^{-3}$/$6.18\times10^{-4}$\\
&&155.3421&&&&Component 2&&&&$2.58\times10^{-3}$/$5.19\times10^{-4}$\\
&&168.0934&2.0-3.5&2.3$\pm$0.32&$1.0\times10^{12}-1.0\times10^{14}$&$(1.10\pm0.21)\times10^{13}$&30.0-80.0&71.2$\pm$5.17&7.2&$3.36\times10^{-3}$/$6.66\times10^{-4}$\\
&&208.2285&&&&&&&&$1.13\times10^{-3}$/$8.04\times10^{-4}$\\
&&211.2738&&&&&&&&$4.75\times10^{-4}$/$6.92\times10^{-4}$\\
&&212.2571&&&&&&&&$1.63\times10^{-4}$/$5.68\times10^{-4}$\\
&&214.8450&&&&&&&&$4.63\times10^{-4}$/$7.00\times10^{-4}$\\
&&216.6302&&&&&&&&$8.10\times10^{-4}$/$7.85\times10^{-4}$\\
&&223.6601&&&&&&&&$4.77\times10^{-4}$/$7.99\times10^{-4}$\\
&&242.1060&&&&&&&&$2.04\times10^{-4}$/$8.02\times10^{-4}$\\
&&250.8291&&&&&&&&$7.31\times10^{-6}$/$4.55\times10^{-4}$\\
&&250.9345&&&&&&&&$3.16\times10^{-5}$/$5.94\times10^{-4}$\\
&&251.4893&&&&&&&&$3.16\times10^{-5}$/$5.95\times10^{-4}$\\
&&254.8271&&&&&&&&$8.67\times10^{-5}$/$7.22\times10^{-4}$\\
&&255.3269&&&&&&&&$1.47\times10^{-4}$/$7.97\times10^{-4}$\\
&&262.9601&&&&&&&&$8.63\times10^{-5}$/$8.02\times10^{-4}$\\
\cline{2-11}
&&205.1707&&&&&&&&$9.80\times10^{-4}$\\
&SVS13A&211.2430&&&&&&&&$8.18\times10^{-4}$\\
&&211.2738&&&&&&&&$8.17\times10^{-4}$\\
&&216.5819&0.5-2.5&2.45$\pm$0.33&$1.0\times10^{12}-1.0\times10^{14}$&$(7.20\pm4.89)\times10^{12}$&30.0-200.0&45.16$\pm$11.4&8.6&$9.66\times10^{-4}$\\
&&230.3019&&&&&&&&$7.74\times10^{-4}$\\
&&242.1060&&&&&&&&$8.51\times10^{-4}$\\
&&251.4893&&&&&&&&$5.34\times10^{-4}$\\
\hline
\hline
\end{tabular}}


\clearpage
{\tiny
\centering
\begin{tabular}{|c|c|c|c|c|c|c|c|c|c|c|}
\hline
\hline
Species&Source&Frequency&FWHM range&Best fitted&Column density&Best fitted column&T$_{ex}$ range&Best fitted&Best fitted&Optical depth \\
&&(GHz)&used (km.s$^{-1}$)&FWHM (Km.s$^{-1}$)&range used (cm$^{-2}$)&density (cm$^{-2}$) &used (K)& T$_{ex}$ (K)&V$_{LSR}$ (km.s$^{-1}$)& ($\tau$)\\
\hline
\hline
CH$_3$OCHO&B1-b&88.8516&&&&&&&&$1.48\times10^{-3}$\\
&&90.1457&&&&&&&&$1.30\times10^{-3}$\\
&&90.1564&1.0-2.0&1.64$\pm$0.21&$1.0\times10^{11}-1.0\times10^{14}$&$(1.20\pm1.00)\times10^{13}$&15.0-100.0&18.52$\pm$22.81&6.5&$1.30\times10^{-3}$\\
&&100.2946&&&&&&&&$1.03\times10^{-3}$\\
&&100.4822&&&&&&&&$1.49\times10^{-3}$\\
&&103.4786&&&&&&&&$1.35\times10^{-3}$\\
\cline{2-11}
&IRAS4A&129.2963$^{*}$&&&&&&&&$3.92\times10^{-4}$\\
&&132.9287&&&&&&&&$4.24\times10^{-4}$\\
&&135.9219&&&&&&&&$2.90\times10^{-4}$\\
&&141.0443&1.0-3.0&1.86$\pm$0.54&$1.0\times10^{12}-1.0\times10^{14}$&$(3.90\pm1.54)\times10^{13}$&30.0-130.0&73.87$\pm$24.37&7.2&$4.45\times10^{-4}$\\
&&158.6937&&&&&&&&$4.53\times10^{-4}$\\
&&158.7043&&&&&&&&$4.53\times10^{-4}$\\
\cline{3-11}
&&200.9563&&&&&&&&$4.21\times10^{-4}$\\
&&206.6194&&&&&&&&$5.13\times10^{-4}$\\
&&216.2165&&&&&&&&$5.06\times10^{-4}$\\
&&228.6288&1.0-3.0&1.35$\pm$0.50&$1.0\times10^{12}-1.0\times10^{14}$&$(3.40\pm1.72)\times10^{13}$&30.0-130.0&81.32$\pm$18.46&7.2&$4.28\times10^{-4}$\\
&&240.0211&&&&&&&&$4.75\times10^{-4}$\\
&&247.0441&&&&&&&&$4.38\times10^{-4}$\\
&&249.5781&&&&&&&&$3.57\times10^{-4}$\\
\cline{2-11}
&SVS13A&100.4906&1.0-3.0&1.95$\pm$0.53&$1.0\times10^{12}-1.0\times10^{15}$&$(6.40\pm5.17)\times10^{13}$&40.0-200.0&86.94$\pm$33.36&8.6&$3.04\times10^{-4}$\\
&&164.2059&&&&&&&&$4.73\times10^{-4}$\\
\cline{3-11}
&&210.4632&&&&&&&&$2.81\times10^{-4}$\\
&&218.2809&1.0-3.0&2.10$\pm$0.37&$1.0\times10^{12}-1.0\times10^{15}$&$(7.80\pm1.88)\times10^{13}$&40.0-200.0&113.84$\pm$23.41&8.6&$4.15\times10^{-4}$\\
&&222.4214&&&&&&&&$2.55\times10^{-4}$\\
&&269.0780&&&&&&&&$4.06\times10^{-4}$\\
\hline
$\rm{C_2H_5OH}$&IRAS4A&129.6657&&&&&&&&$2.84\times10^{-4}$\\
&&133.3234&0.4-1.5&1.46$\pm$0.26&$1.0\times10^{12}-1.0\times10^{14}$&$(2.60\pm1.61)\times10^{13}$&30.0-100.0&60.14$\pm$16.62&7.2&$5.90\times10^{-4}$\\
&&148.3040&&&&&&&&$2.76\times10^{-4}$\\
&&159.4140&&&&&&&&$2.92\times10^{-4}$\\
\cline{3-11}
&&205.4584&&&&&&&&$5.22\times10^{-4}$\\
&&209.8652&0.4 - 0.8&0.74$\pm$0.06&$1.0\times10^{12}-1.0\times10^{14}$&$(1.60\pm0.59)\times10^{13}$&30.0 - 100.0&88.46$\pm$18.28&7.2&$2.29\times10^{-4}$\\
&&227.8919&&&&&&&&$2.72\times10^{-4}$\\
&&230.9913&&&&&&&&$5.63\times10^{-4}$\\
\cline{2-11}
&SVS13A&84.5958&&&&&&&&$4.02\times10^{-4}$\\
&&130.2463&0.5 - 2.0&0.92$\pm$0.41&$1.0\times10^{12}-1.0\times10^{14}$&$(1.70\pm1.59)\times10^{13}$&10.0-100.0&45.63$\pm$17.28&8.6&$4.64\times10^{-4}$\\
&&153.4842&&&&&&&&$5.24\times10^{-4}$\\
\cline{3-11}
&&205.4584&&&&&&&&$8.26\times10^{-4}$\\
&&244.6339&0.5-2.0&0.97$\pm$0.28&$1.0\times10^{12}-1.0\times10^{14}$&$(1.50\pm0.86)\times10^{13}$&10.0-100.0&59.58$\pm$21.85&8.6&$2.78\times10^{-4}$\\
&&270.4441&&&&&&&&$7.21\times10^{-4}$\\
\hline
$\rm{HCCCHO}$&L1544&83.7758&&&&&&&&$1.43\times10^{-3}$\\
&&99.0391&0.3-0.6&0.43$\pm$0.07&$1.0\times10^{11}-1.0\times10^{13}$&$(4.00\pm1.33)\times10^{11}$&3.5-20.0&17.40$\pm$5.24&7.2&$6.28\times10^{-5}$\\
&&102.2980&&&&&&&&$1.28\times10^{-3}$\\
\hline
$\rm{CH_3OCH_3}$&L1544&99.324362&&&&&&&&$3.66\times10^{-4}$\\
&&99.324364&0.1-1.0&$0.41\pm$0.22&$1.0\times10^{11}-1.0\times10^{13}$&$(2.20\pm1.61)\times10^{12}$&3.5-20.0&$15.47\pm2.71$&7.2&$5.49\times10^{-4}$\\
&&99.325217&&&&&&&&$1.46\times10^{-3}$\\
&&99.326072&&&&&&&&$9.15\times10^{-4}$\\
\cline{2-11}
&B1-b&99.324362&&&&&&&&$1.49\times10^{-3}$\\
&&99.324364&0.5-1.0&$0.9\pm0.12$&$6.0\times10^{11}-6.0\times10^{13}$&$(8.50\pm5.52)\times10^{12}$&5.0-25.0&$9.85\pm4.56$&6.5&$2.23\times10^{-3}$\\
&&99.325217&&&&&&&&$5.95\times10^{-3}$\\
&&99.326072&&&&&&&&$3.72\times10^{-3}$\\
\cline{2-11}
&IRAS4A&162.5295&&&&&&&&$5.38\times10^{-4}$\\
&&209.5156&&&&&&&&$6.02\times10^{-4}$\\
&&225.5991&1.0-3.0&2.11$\pm$0.35&$1.0\times10^{12}-1.0\times10^{14}$&$(2.10\pm1.03)\times10^{13}$&30.0-130.0&45.28$\pm$28.29&7.2&$5.83\times10^{-4}$\\
&&241.9465&&&&&&&&$5.45\times10^{-4}$\\
\hline
$\rm{CH_3CN}$&IRAS4A&73.588799&&&&&&&&$1.97\times10^{-3}$\\
 &&73.590218&&&&&&&&$2.96\times10^{-3}$\\
 &&91.979994&&&&&&&&$8.26\times10^{-4}$\\
 &&91.985314&&&&&&&&$2.61\times10^{-3}$\\
 &&91.987087&&&&&&&&$3.83\times10^{-3}$\\
 &&110.364353&&&&&&&&$3.08\times10^{-4}$\\
 &&110.381372&&&&&&&&$3.03\times10^{-3}$\\
 &&110.383499&&&&&&&&$4.38\times10^{-3}$\\
 &&128.757030&&&&&&&&$3.48\times10^{-4}$\\
 &&128.769436&&&&&&&&$1.07\times10^{-3}$\\
 &&128.776881&&&&component1&&&&$3.16\times10^{-3}$\\
&&128.779363&1.5-3.0&$2.9\pm0.19$&$1.0\times10^{11}-1.0\times10^{13}$&$(9.9\pm3.15)\times10^{11}$&20.0-50.0&$21.02\pm3.24$&7.2&$4.54\times10^{-3}$\\
 &&147.163244&&&&component2&&&&$1.04\times10^{-3}$\\
 &&147.171751&1.5-3.0&$2.9\pm0.12$&$1.0\times10^{11}-1.0\times10^{13}$&$(1.7\pm0.38)\times10^{12}$&50.0-80.0&$70.11\pm5.32$&7.2&$3.04\times10^{-3}$\\
 &&147.174588&&&&&&&&$4.34\times10^{-3}$\\
 &&165.540377&&&&&&&&$3.21\times10^{-4}$\\
 &&165.556321&&&&&&&&$9.38\times10^{-4}$\\
 &&165.565891&&&&&&&&$2.70\times10^{-3}$\\
 &&165.569081&&&&&&&&$3.84\times10^{-3}$\\
 &&202.320442&&&&&&&&$2.16\times10^{-4}$\\
 &&220.709016&&&&&&&&$1.61\times10^{-4}$\\
 \hline
\end{tabular}}
\clearpage
{\tiny
\centering
\begin{tabular}{|c|c|c|c|c|c|c|c|c|c|c|}
\hline
\hline
Species&Source&Frequency&FWHM range&Best fitted&Column density&Best fitted column&T$_{ex}$ range&Best fitted&Best fitted&Optical depth \\
&&(GHz)&used (km.s$^{-1}$)&FWHM (Km.s$^{-1}$)&range used (cm$^{-2}$)&density (cm$^{-2}$) &used (K)& T$_{ex}$ (K)&V$_{LSR}$ (km.s$^{-1}$)& ($\tau$)\\
\hline
&SVS13A&91.979994&&&&&&&&$2.705\times10^{-4}$\\
&&91.985314&&&&&&&&$3.996\times10^{-4}$\\
&&91.987087&&&&&&&&$4.535\times10^{-4}$\\
&&110.364353&&&&&&&&$4.281\times10^{-4}$\\
&&110.374989&&&&&&&&$3.890\times10^{-4}$\\
&&110.381372&&&&&&&&$5.499\times10^{-4}$\\
&&110.383499&&&&&&&&$6.162\times10^{-4}$\\
&&128.776881&1.5-3.5&$3.48\pm0.09$&$5.0\times10^{11}-5.0\times10^{13}$&$(2.8\pm0.57)\times10^{12}$&40.0-200.0&$83.55\pm19.34$&8.6&$7.041\times10^{-4}$\\
&&128.779363&&&&&&&&$7.832\times10^{-4}$\\
&&147.171751&&&&&&&&$8.538\times10^{-4}$\\
&&165.565891&&&&&&&&$9.913\times10^{-4}$\\
&&220.709016&&&&&&&&$1.210\times10^{-3}$\\
&&220.730260&&&&&&&&$9.624\times10^{-4}$\\
&&220.743010&&&&&&&&$1.270\times10^{-3}$\\
&&220.747261&&&&&&&&$1.394\times10^{-3}$\\
&&239.119504&&&&&&&&$9.939\times10^{-4}$\\
&&257.527383&&&&&&&&$1.441\times10^{-3}$\\
\hline
\hline
\end{tabular}}

\begin{figure*}
\includegraphics[width=11cm, height=15cm, angle=270]{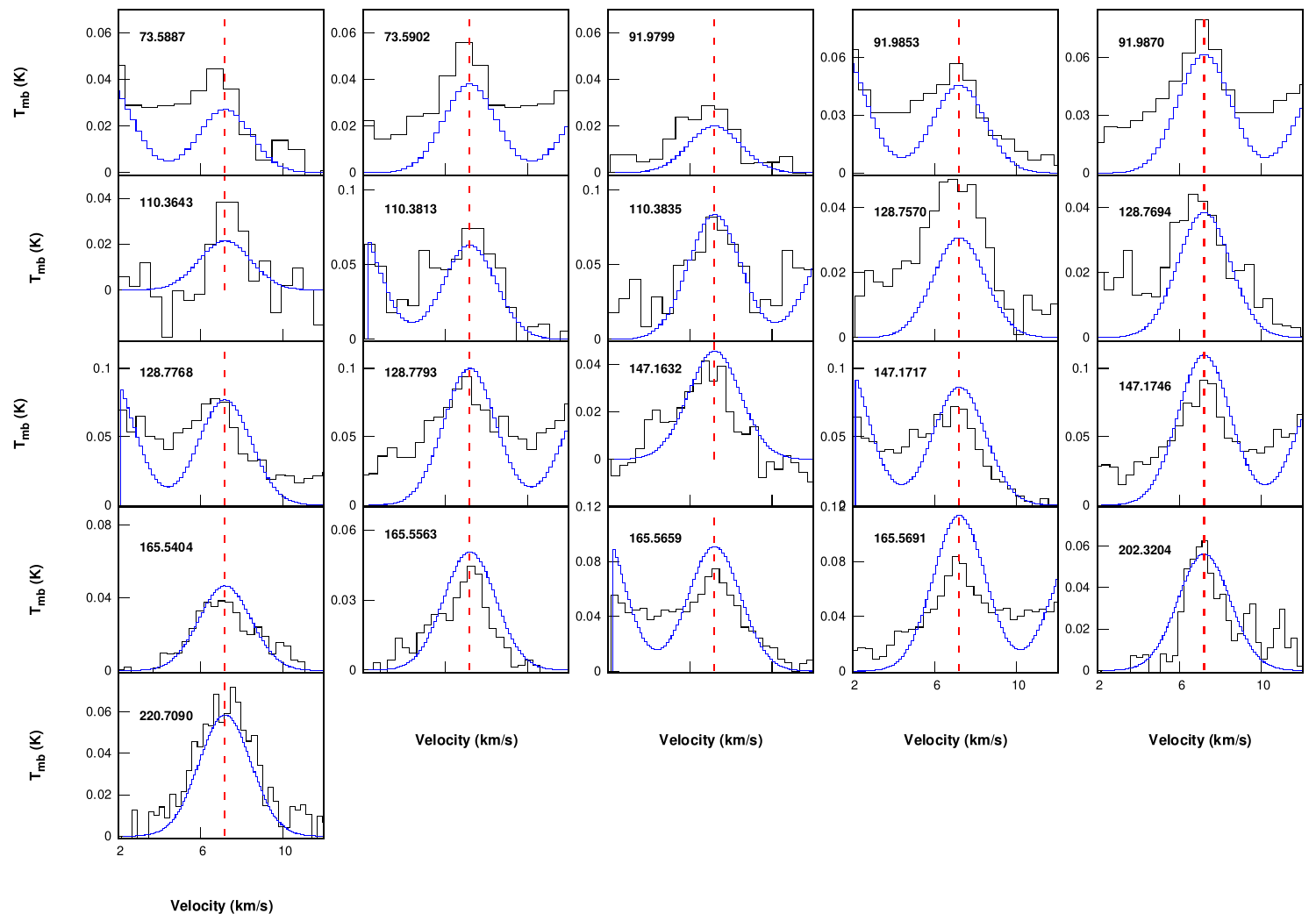}
\caption{Same as Figure \ref{fig:ch3oh_l1544mcmc} for $\rm{CH_3CN}$ in IRAS4A.}
\label{fig:ch3cn_mcmc1}
\end{figure*}

\begin{figure*}
\includegraphics[width=11cm, height=15cm, angle=270]{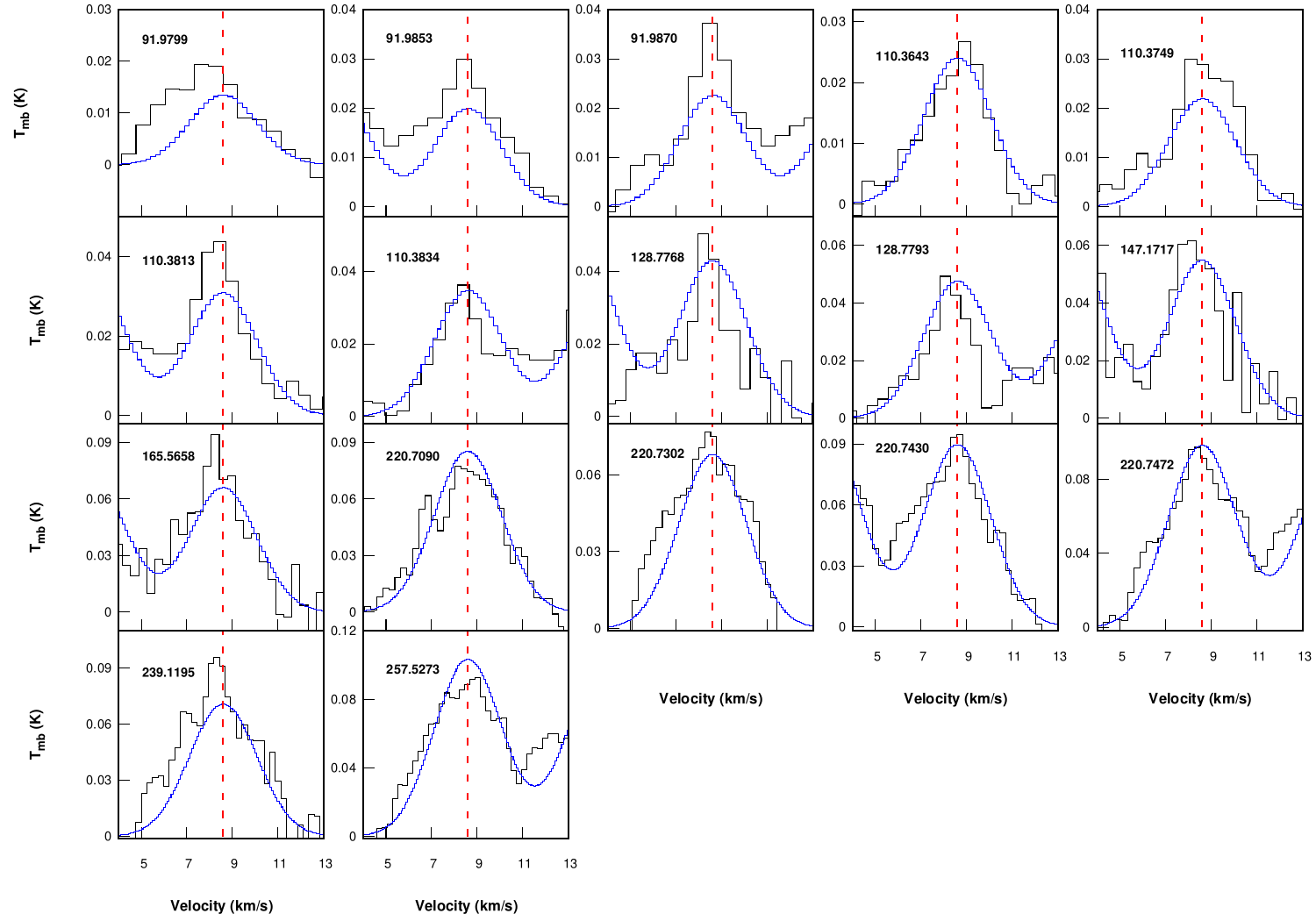}
\caption{ Same as Figure \ref{fig:ch3oh_l1544mcmc} for $\rm{CH_3CN}$ in SVS13A.}
\label{fig:ch3cn_mcmc2}
\end{figure*}
\clearpage

\section{Beam dilution effect}
\setcounter{figure}{0}    
\setcounter{table}{0}
The present data have a large frequency width, and the beam size changes significantly. Therefore, we need to assume the size of the emitting region for the COMs and apply the appropriate beam dilution factor for each source to see its effect. Figure \ref{fig:beamdil} shows the abundances (obtained from the rotational diagram analysis) at the different evolutionary stages of low-mass star-forming regions by considering the beam dilution effect. Additionally, we note the obtained abundances in Table \ref{tab:beam-filling-col-density}. The obtained intensities in the rotational diagram analysis are scaled by considering the beam dilution factor for each source. For simplicity, we consider an average source size, and the choice of source size for each source is justified below:\\
{\it L1544:} \cite{case19,case22} obtained a compact component of $\sim$ 10$\arcsec$ in L1544. Assuming the emission of the COMs from this region, we consider a source size of 10$\arcsec$ (see Figure \ref{fig:beamdil} and in Table \ref{tab:beam-filling-col-density}).\\
{\it B1-b:} \cite{marc18} employed a source model with two components; an inner hot and compact component (200 K, 0.35$\arcsec$) and an outer and colder component (60 K, 0.6$\arcsec$) to fit the observed line profiles in B1b-S. Additionally, they considered another component mimicking the envelope at $\sim 10$ K. 
In our analysis, we obtain a rotational temperature $\sim$ $10$ K for all molecules in this source, we consider a source size of 10$\arcsec$, which is comparable to the smallest beam size in this data (see Figure \ref{fig:beamdil}).\\ 
{\it IRAS16293-2422:} We consider the abundances obtained directly from the ALMA Protostellar interferometric line survey (PILS) (for details, see Section \ref{sec:INO}). \\
{\it IRAS4A:} \cite{lope15} considered a source size of 0.5$\arcsec$  to perform the RD analysis for NH$_2$CHO and HNCO in IRAS4A. We adopt similar source size of $0.5 {\arcsec}$ to consider beam dilution effect ( see Table \ref{tab:beam-filling-col-density} and Figure \ref{fig:beamdil}).\\ 
{\it SVS13A:} \cite{lope15} considered a source size of 1.0$\arcsec$  to perform the RD analysis for NH$_2$CHO and HNCO in SVS13A and \cite{bian19} considered a source size of  0.3$\arcsec$  for various iCOMs . We adopt both sizes to consider the beam dilution effect ( see Table \ref{tab:beam-filling-col-density} and Figure \ref{fig:beamdil}). 

Figure \ref{fig:beamdil} depicts that even with the beam dilution effect, the obtained trend is similar (the abundance is gradually increasing up to class 0 and then decreased) to that obtained without the beam dilution effect shown in Figure \ref{fig:clmdensity}. A very similar trend was also obtained when we used the abundances from various interferometric observations in Figure \ref{fig:inter}.  

 \begin{table*}
     \centering
     \caption{Column density and abundance of the observed species  considering the beam dilution factor.}
     \begin{tabular}{|c|c|c|c|c|c|}
          \hline
     Source&Size&N(H$_2$)&Species&Column density&Abundance\\
     \hline
     \hline
          &&&CH$_3$OH&$5.0 \times 10^{13}$&$7.4 \times 10^{-11}$\\
          &&&CH$_3$CHO&$1.1 \times 10^{13}$&$1.6 \times 10^{-11}$\\
          L1544&10$\arcsec$&$6.8 \times 10^{23c}$&CH$_3$OCHO&$3.7 \times 10^{13*}$&$5.4 \times 10^{-11}$\\
         &&&C$_2$H$_5$OH&\nodata&\nodata\\
          &&&HCCCHO&$2.1 \times 10^{13}$&$3.1 \times 10^{-11}$\\
          &&&CH$_3$OCH$_3$&$1.7 \times 10^{13**}$&$2.5 \times 10^{-11}$\\
          &&&CH$_3$CN& $4.85\times 10^{12**}$&$7.1 \times 10^{-12}$\\
          \hline
          &&&CH$_3$OH&$6.2 \times 10^{14}$&$7.8 \times 10^{-10}$\\
          &&&CH$_3$CHO&$5.1 \times 10^{13}$&$6.4 \times 10^{-11}$\\
          B1-b&10$\arcsec$&$7.9 \times 10^{23d}$&CH$_3$OCHO&$5.4 \times 10^{13}$&$6.8 \times 10^{-11}$\\
          &&&C$_2$H$_5$OH&$1.0 \times 10^{14*}$&$1.2 \times 10^{-10}$\\
          &&&HCCCHO&$2.6 \times 10^{13*}$&$3.2 \times 10^{-11}$\\
          &&&CH$_3$OCH$_3$&$6.0 \times 10^{13**}$&$7.6 \times 10^{-11}$\\
          &&&CH$_3$CN&$4.95\times 10^{12**}$&$6.3 \times 10^{-12}$\\
          \hline
           &&&CH$_3$OH (hot)&$9.7 \times 10^{16}$&$3.9 \times 10^{-8}$\\
          &&&CH$_3$CHO&$4.4 \times 10^{15}$&$1.8 \times 10^{-9}$\\
          IRAS4A&0.5$\arcsec$&$2.5 \times 10^{24}$&CH$_3$OCHO&$2.4 \times 10^{16}$&$9.5 \times 10^{-9}$\\
          &\cite{lope15}&\cite{lope15}&C$_2$H$_5$OH&$1.3 \times 10^{16}$&$5.2 \times 10^{-9}$\\
          &&&HCCCHO&$4.9 \times 10^{15*}$&$2.0 \times 10^{-9}$\\
          &&&CH$_3$OCH$_3$&$6.7 \times 10^{15}$&$2.7 \times 10^{-9}$\\
          &&&CH$_3$CN&$3.55\times 10^{15}$&$1.4 \times 10^{-09}$\\
          \hline
           &&&CH$_3$OH&$3.7 \times 10^{16}$&$3.7 \times 10^{-9}$\\
          &&&CH$_3$CHO&$8.6 \times 10^{14}$&$8.6 \times 10^{-11}$\\
          SVS13A&1$\arcsec$&$1.0 \times 10^{25}$&CH$_3$OCHO&$9.7 \times 10^{15}$&$9.7 \times 10^{-10}$\\
          &\cite{lope15}&\cite{lope15}&C$_2$H$_5$OH&$2.8 \times 10^{15}$&$2.8 \times 10^{-10}$\\
          &&&HCCCHO&$3.9 \times 10^{16*}$&$3.9 \times 10^{-9}$\\
          &&&CH$_3$OCH$_3$&$1.3 \times 10^{16a}$&$1.3 \times 10^{-9}$\\
          &&&CH$_3$CN&$5.16\times 10^{14}$&$5.2 \times 10^{-11}$\\
          \hline
           &&&CH$_3$OH&$4.1 \times 10^{17}$&$1.4 \times 10^{-7}$\\
          &&&CH$_3$CHO&$9.6 \times 10^{15}$&$3.2 \times 10^{-9}$\\
          SVS13A&0.3$\arcsec$&$3.0 \times 10^{24}$&CH$_3$OCHO&$1.1 \times 10^{17}$&$3.6 \times 10^{-8}$\\
          &\cite{bian19}&\cite{chen09}&C$_2$H$_5$OH&$3.1 \times 10^{16}$&$1.0 \times 10^{-8}$\\
          &&&HCCCHO&$4.3 \times 10^{17*}$&$1.4 \times 10^{-7}$\\
          &&&CH$_3$OCH$_3$&$1.4 \times 10^{17b}$&$4.7 \times 10^{-8}$\\
          &&&CH$_3$CN&$5.74\times 10^{15}$&$1.9 \times 10^{-09}$\\
          \hline
     \end{tabular}\\
     { Note: $^*$ indicates upper limit, $^{**}$ indicates LTE derived value, $^{a}$ is scaled value from \cite{bian19}, and $^b$ from \cite{bian19}., $^c$ taken from \cite{hily22} after scaling it for 10$\arcsec$, $^d$ taken from \cite{dani13} after scaling it for 10$\arcsec$.}
     \label{tab:beam-filling-col-density}
 \end{table*}

\begin{figure*}
\centering
\includegraphics[width=14cm, height=8cm]{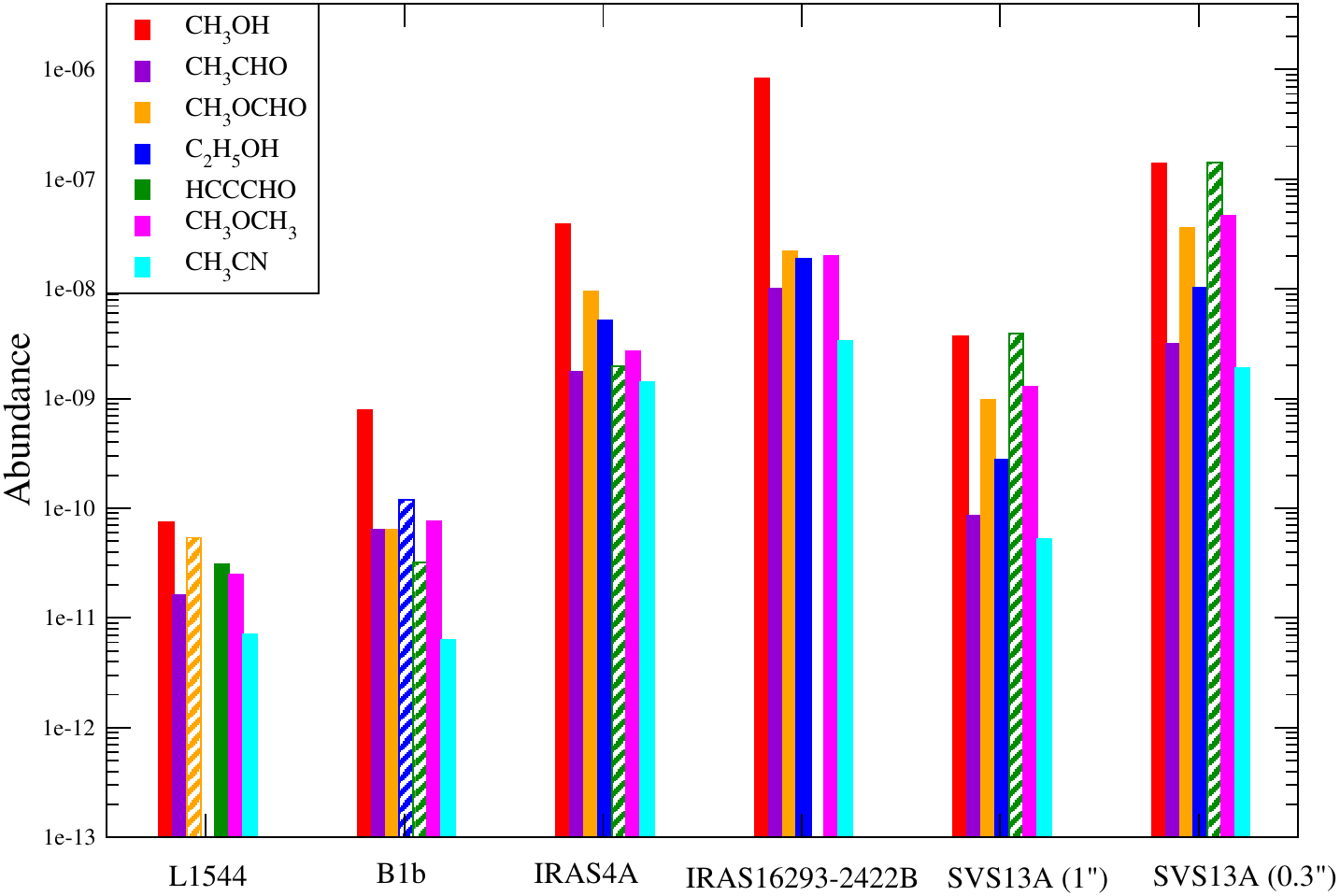}
\caption{Abundances of the COMs at the different evolutionary stages of low-mass star-forming regions. }
\label{fig:beamdil}
\end{figure*}


 \clearpage
\bibliography{asai_apj}{}
\bibliographystyle{aasjournal}

\end{document}